\documentclass{aa}  
\usepackage{graphicx}
\usepackage{natbib}
\usepackage{txfonts}

\newcommand{\degrees}[1]{\ensuremath{#1^\circ}}

\def\NrPulsarsSourceList{191 }
\def\NrConfirmedNewDrifters{26}
\def\NrPulsarsTwoFreq{130}
\def\NrPulsarsTwoFreqAndCoherentOne{11 }
\def\NrPulsarsTwoFreqAndCoherentTwo{11 }
\def\NrPulsarsTwoFreqAndCoherent{22 }
\def\NrPulsarsTwoFreqAndDiffuseTwo{36}
\def\NrPulsarsDiffuseCoherent{four }
\def\NrPulsarsDiffuseNothing{14 }
\def\NrPulsarsDiffuseLong{one }
\def\NrPulsarsTwoFreqAndDiffuse{55 }
\def\NrPulsars{185 }
\def\NrDrifters{76 }

\def\NrNewDrifters{15}
\def\NrPulsarsSNR{84 }
\def\NrDriftersSNR{53 }
\def\NrCandidatesSNR{9 }
\def\DriftPercentageSNR{63}
\def\CandidatePercentageSNR{11}
\def\AgeDriftNonDriftperc{2\% }
\def\AgeCohNondriftperc{0.1\% }
\def\AgeCohNoCohperc{0.9\% }
\def\BNonDriftperc{10\% }
\def\BNonCohpers{3\%}
\def\AgeDriftNonDriftpercCombined{0.02\% }
\def\AgeCohNoCohpercCombined{3\% }
\def\NrDriftersCombinedSample{90}
\def\NrNullers{8 }
\def\NrBursters{8 }

\begin{document}
   \title{The subpulse modulation properties of pulsars at 92 cm}

   \subtitle{and the frequency dependence of subpulse modulation}

\titlerunning{The subpulse modulation properties of pulsars at 92 cm}

   \author{P. Weltevrede
          \inst{1}
	  \and
	  B.~W. Stappers\inst{2,1}
          \and
          R.~T. Edwards\inst{3}
          }

   \offprints{P. Weltevrede}

   \institute{Astronomical Institute ``Anton Pannekoek'',
              University of Amsterdam,
              Kruislaan 403, 1098 SJ Amsterdam, The Netherlands\\
              \email{wltvrede@science.uva.nl}
	 \and
	     Stichting ASTRON, Postbus 2, 7990 AA Dwingeloo, The Netherlands\\
             \email{stappers@astron.nl}
         \and
             CSIRO Australia Telescope National Facility, PO Box 76, Epping NSW 1710, Australia\\
             \email{Russell.Edwards@csiro.au}
             }

   \date{Received ...; accepted ...}

  \abstract
   { A large sample of pulsars has been observed to study their
subpulse modulation at an observing wavelength (when achievable) of
both 21 and 92 cm using the Westerbork Synthesis Radio Telescope.  In
this paper we present the 92-cm data and a comparison is made with the
already published 21-cm results.  }
   { The main goals are to determine what fraction of the pulsars have
drifting subpulses, whether those pulsars share some physical
properties and to find out if subpulse modulation properties are
frequency dependent. }
   { We analysed \NrPulsarsSourceList pulsars at 92 cm searching for
   subpulse modulation using fluctuation spectra. The sample of
   pulsars is as unbiased as possible towards any particular pulsar
   characteristics. }
   { For {\NrNewDrifters} pulsars drifting subpulses are discovered
for the first time and {\NrConfirmedNewDrifters} of the new drifters
found in the 21-cm data are confirmed. We discovered nulling for
\NrNullers sources and \NrBursters pulsars are found to intermittently
emit single pulses that have pulse energies similar to giant
pulses. Another pulsar was shown to exhibit a subpulse phase step.  It
is estimated that at least half of the total population of pulsars
have drifting subpulses when observations with a high enough
signal-to-noise ratio would be available. It could well be that the
drifting subpulse mechanism is an intrinsic property of the emission
mechanism itself, although for some pulsars it is difficult or
impossible to detect. Drifting subpulses are in general found at both
frequencies, although the chance of detecting drifting subpulses is
possibly slightly higher at 92 cm.  It appears that the youngest
pulsars have the most disordered subpulses and the subpulses become
more and more organized into drifting subpulses as the pulsar
ages. The modulation indices measured at the two frequencies are
clearly correlated, although at 92 cm they are on average possibly higher. At 92 cm the modulation index appears to be correlated
with the characteristic age of the pulsar and the complexity
parameters as predicted by three different emission models. The
correlations with the modulation indices are argued to be consistent
with the picture in which the radio emission can be divided in a
drifting subpulse signal plus a quasi-steady signal which becomes, on
average, stronger at high observing frequencies. The measured values
of $P_3$ at the two frequencies are highly correlated, but there is no
evidence for a correlation with other pulsar parameters. }
   {}


   \maketitle

\section{Introduction}

\begin{figure*}[htb]
\rotatebox{270}{\resizebox{0.5\hsize}{!}{\includegraphics[angle=0]{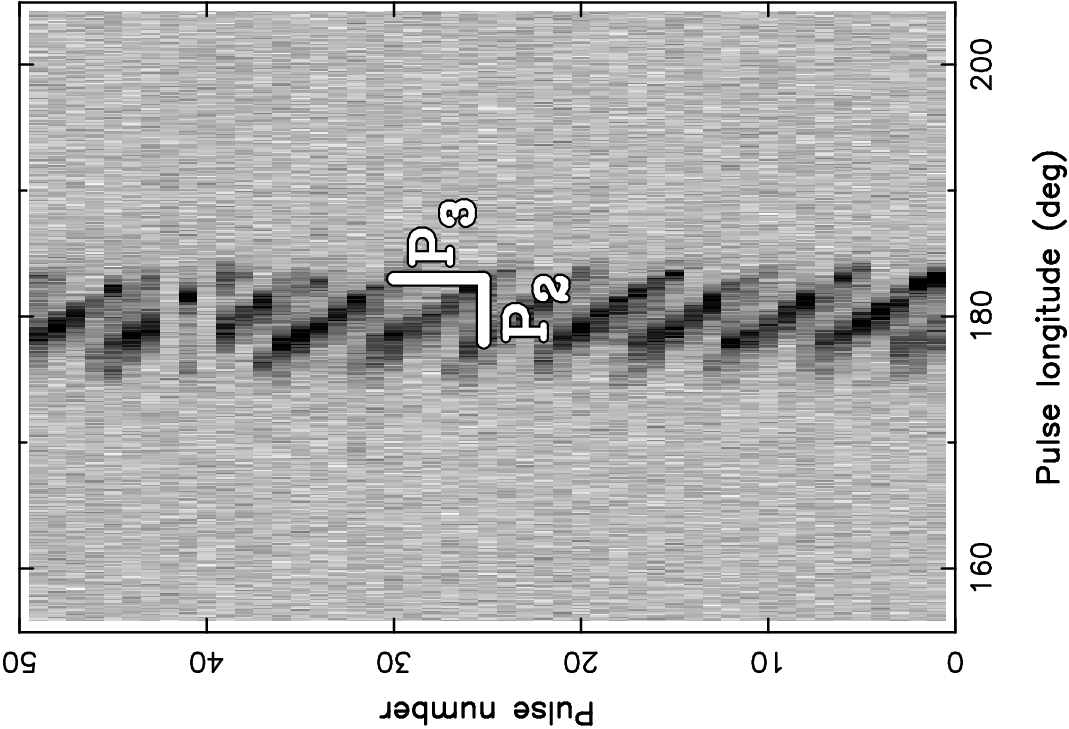}}}
\rotatebox{270}{\resizebox{0.5\hsize}{!}{\includegraphics[angle=0]{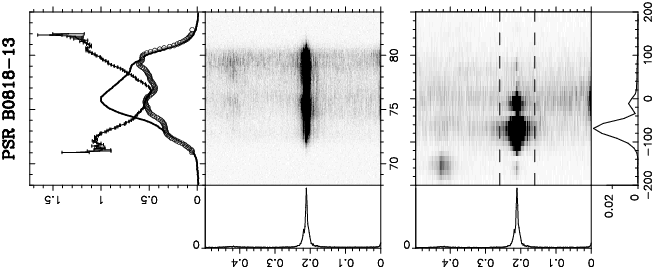}}}
\rotatebox{270}{\resizebox{0.5\hsize}{!}{\includegraphics[angle=0]{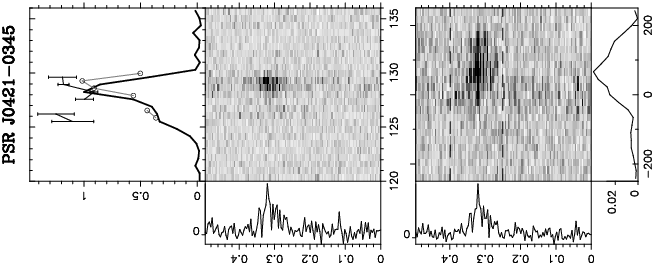}}}
\rotatebox{270}{\resizebox{0.5\hsize}{!}{\includegraphics[angle=0]{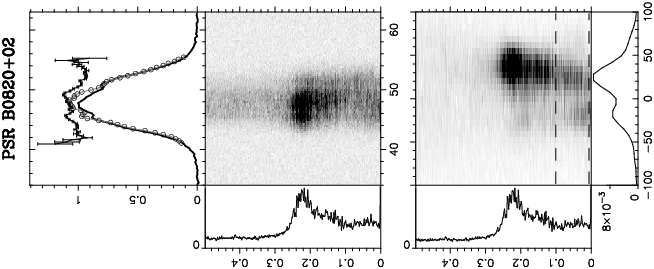}}}
\caption{\label{Classes_fig92}The left panel shows a pulse-stack of
fifty successive pulses of PSR B0818--13. Two successive drift bands
are vertically separated by $P_3$ and horizontally by $P_2$.  The
products of our analysis are shown for three pulsars. The top panel
shows the integrated pulse profile (solid line), the
longitude-resolved modulation index (solid line with error bars) and
the longitude-resolved standard deviation (open circles). Below this
panel the LRFS is shown with on its horizontal axis the pulse
longitude in degrees, which is also the scale for the abscissa of the
plot above. Below the LRFS the 2DFS is plotted and the power in the
2DFS is vertically integrated between the dashed lines, producing the
bottom plots. Both the LRFS and 2DFS are horizontally integrated,
producing the side-panels of the spectra.}
\end{figure*}

Although the pulse profiles of radio pulsars are in general very
stable, the shape of their single pulses are often highly variable
from pulse to pulse. For some pulsars the single pulses are modulated
in a highly organized and fascinating way: they exhibit the phenomenon
of drifting subpulses (\citealt{sspw70,dc68}). An example is shown in
the left panel of Fig. \ref{Classes_fig92}. In this so-called
``pulse-stack'' fifty successive pulses are displayed on top of one
another and a beautiful pattern of diagonal ``drift bands'' emerges.

We have embarked on an extensive observational program to survey a
large sample of pulsars to study their single pulse modulation using
the Westerbork Synthesis Radio Telescope (WSRT) in the
Netherlands. The main goals of this program are to determine what
fraction of the pulsars have drifting subpulses, whether those pulsars
share some physical properties and if subpulse modulation is frequency
dependent. The sample of pulsars studied is selected based only on the
predicted $S/N$ in a reasonable observing time, which makes the
resulting statistics as unbiased as possible towards well-studied
pulsars, pulse profile morphology or any particular pulsar
characteristics. If possible, we observed the pulsars at two
wavelengths around 21 and 92 cm. It should be noted that the actual
central wavelength can be slightly different from observation to
observation.

The results of the pulsars observed at a wavelength of 21 cm are
published in \cite{wes06} (hereafter referred to as paper I). Drifting
subpulses were detected in 68 of 187 analyzed pulsars, 42 of which
were shown to have drifting subpulses for the first time.  In this
paper the results of the 92-cm observations are presented and a
comparison with the results reported in paper I is made. In the next
section the method used to analyse the data is explained concisely,
but for details we refer to paper I. In Sect. \ref{SctIndividual92}
the individual detections are described (which are summarized in Table
\ref{Table92}) and some remarks about the comparison of sources at the
two frequencies are made in Sect. \ref{SctCompTwofreq}. The statistics
of the drifting phenomenon is described in Sect. \ref{Statistics92}
and are discussed in Sect. \ref{SctDisc}. After the summary and
conclusions one can find the plots for all the pulsars observed at 92
cm (Appendix \ref{Figures_ref}).  For some of the pulsars the
Two-Dimensional Fluctuation Spectrum (2DFS; \citealt{es02}) is shown
for two different pulse longitude ranges. The plots of these pulsars
come after the plots of the pulsars with only one 2DFS plot.  Note
that the astro-ph version is missing the appendix due to file size
restrictions. Please download appendices from {\tt
http://www.astron.nl/$\sim$stappers/wiki/doku.php?id=\\resources:publications}.
The plots of the two 21-cm and 92-cm observations can be found side by
side in the PhD. thesis of the main author.\footnote{The thesis of
P. Weltevrede is online available via the following URL: {\tt
http://dare.uva.nl/en/record/217315}}

\section{\label{SctObs92}Observations and data analysis}

\subsection{\label{SctSourceList}The source-list}

All the observations were done using the WSRT in the Netherlands and
therefore the first selection for the source-list is that the pulsars
must have a declination above {\degrees{-30}} (J2000).  For paper I
a source-list was compiled of all pulsars visible with the WSRT for
which a pulse profile could be obtained with an integrated
signal-to-noise ($S/N$) ratio of 130 in an observation of less then
half an hour in duration. Because at 21 cm we discovered drifting
subpulses for many pulsars in observations with a lower $S/N$ a
different criterion was chosen for the 92-cm observations. For this
paper all visible pulsars for which we could obtain a pulse profile
with a $S/N$ ratio of 50 in less than half an hour were included in
the source-list.

The 92-cm frontend at the WSRT was used in combination with the PuMa I
pulsar backend (\citealt{vkv02}), which have the following parameters:
the gain of the telescope $G=1.2$ K/Jy, the number of polarizations of
the antenna $n_p=2$, $T_\mathrm{sys}=125$ K,
$T_\mathrm{sky}\approx100$ K (which is approximately the average of
the entire sky), the bandwidth of the backend $\Delta\nu=10$ MHz and
the digitization efficiency factor $\eta_Q=1$. It should be mentioned
that $T_\mathrm{sky}$ is a strong function of position on the sky,
particularly in galactic latitude. Because the age of the pulsar is
also correlated with the galactic latitude, we are biased against
young pulsars. The same bias also applies for pulsar surveys and is
worse at low frequencies.

The expected observation duration required to get an integrated $S/N$
ratio of at least 50 can be calculated for these parameters (see paper
I for more details):
\begin{equation}
t_\mathrm{obs} \geq \left(\frac{2.1\;\mathrm{Jy}}{S_{400}}\right)^2\frac{w}{P_0-w}.
\end{equation}
Here $S_{400}$ is the flux of the pulsar at the observation wavelength
of 92 cm, $P_0$ the barycentric pulse period and $w$ the FWHM of the
pulse profile. Only catalogued\footnote{{\tt
http://www.atnf.csiro.au/research/pulsar/psrcat/}} pulsars
(\citealt{mhth05}) with the necessary parameters $S_{400}$ and $w$
were included in the source-list.

All the pulsars in the source-list, except the millisecond pulsars,
were observed.  For ten sources (PSRs B0154+61, J1246+22, B1754$-$24,
B1800$-$21, B1804$-$27, B1813$-$26, B1834$-$10, B1900+06, B1907+03 and
B1924+14) the $S/N$ was either too low, there was extreme
radio-frequency-interference (RFI) or the observation failed. Apart
from these sources the results of the remaining \NrPulsarsSourceList
pulsars are presented in this paper.

Similar to paper I, the brightest pulsars were observed for long enough to
obtain at least one thousand pulses and archival data was used when
available. Although for a number of pulsars the data greatly exceed
the minimum required $S/N$ and number of pulses, this does not bias
our statistics too much toward well-studied pulsars as all pulsars in
our source-list were observed long enough to have a good chance to
detect drifting subpulses.

\subsection{Observations and data processing}

The procedure of processing the observations in this paper is
identical to that described in paper I, so we will only summarize it
here. All results in this paper are based on observations done in the
last six years at the WSRT.  The filterbank data were (offline and
incoherently) de-dispersed and the resulting time series were
transformed in a two-dimensional pulse number versus pulse longitude
array (pulse-stack) using the pulse period provided by the TEMPO
software package\footnote{\tt http://pulsar.princeton.edu/tempo/ \rm}.

An example of a pulse-stack is shown in the left panel of
Fig. \ref{Classes_fig92}. In this plot the pulse number is plotted
vertically and the time within the pulses (i.e. the pulse longitude)
horizontally.  The vertical separation (in pulse periods) between the
drift bands is denoted by $P_3$ and the horizontal separation (in
pulse longitude) by $P_2$.  To be able to detect the drifting subpulse
phenomenon in as many pulsars as possible, all the pulse-stacks were
analyzed in a systematic procedure using fluctuation spectra.

Two types of spectra are derived from the pulse-stack: the Longitude
Resolved Fluctuation Spectrum (LRFS; \citealt{bac70b}) and the
Two-Dimensional Fluctuation Spectrum (2DFS; \citealt{es02}).

In Fig. \ref{Classes_fig92} the LRFSs of three pulsars are shown
(middle panels). The units of the vertical axis are in cycles per
period (cpp), which corresponds to $P_0/P_3$ in the case of drifting
subpulses, and the horizontal axis is the pulse longitude in
degrees. The power in the LRFS is horizontally integrated, producing
the side panel.

The top plots of Fig. \ref{Classes_fig92}, which are aligned with the
LRFSs plotted below, show the pulse profiles (solid line, normalized
to the integrated peak intensity), longitude resolved standard
deviation (open dots) and the longitude resolved modulation index
(solid line with error bars). Both the longitude resolved standard
deviation and modulation index are derived from the LRFS, which allows
correction for both (periodic) RFI and interstellar scintillation (see
paper I for details). The modulation index is a measure of the factor
by which the intensity varies from pulse to pulse and could therefore
be an indication for subpulse modulation.

By analyzing the LRFS it can be determined if the subpulse modulation
is disordered or (quasi-)periodic, but the 2DFS is required to find
out if the subpulses are systematically drifting in pulse longitude.
In Fig. \ref{Classes_fig92} the 2DFS is plotted below the LRFS. The unit
of its vertical axis is identical to that of the LRFS and also the
unit of the horizontal axis is cycles per period, which corresponds to
$P_0/P_2$ in the case of drifting. In this paper and paper I we use
the convention that $P_3$ is always positive and the sign of $P_2$
indicates the drift direction. A positive value means that the
subpulses appear later in successive pulses (which is called positive
drifting).

The horizontally integrated power in the 2DFS is shown in the side
panel and the vertically (between the dashed lines) integrated power
is shown in the panel below the 2DFS. These panels are only produced
to make it easier to see by eye what the structure of the feature in
the 2DFS is. In the 2DFS of PSR B0818$-$13 the drift feature is
centered at $P_3\simeq0.21$ cpp $=4.8P_0$ and $P_2\simeq-70$ cpp
$=-0.014P_0$, which agrees well with the values one can measure
directly from the pulse-stack in Fig. \ref{Classes_fig92}. Notice that
besides the primary drift feature also its second harmonic is clearly
detected in the 2DFS (in the top left corner).

\subsection{Analysis of the drift-features}

If there appears to be a drift feature in the 2DFS that is clearly not
produced by RFI its significance and the values $P_2$ and $P_3$ are
determined by calculating the centroid of a rectangular region in the
2DFS containing the feature. 

Any (significant) asymmetry in the 2DFS about the vertical axis
indicates that there exists a preferred drift direction of the
subpulses. For some pulsars the bottom panel of the 2DFS peaks at
zero, but the peak has asymmetric wings which indicates that the
subpulses have a preferred drift direction (see for example PSR
B0823+26 in Fig. \ref{B0823+26}). Drifters (pulsars with drifting
subpulses) are defined in this work as pulsars which have at least one
feature in their 2DFS which has a centroid with a significant finite
$P_2$.  Any structure in the vertical direction in the 2DFS and LRFS
indicates a (quasi-periodic) intensity modulation from pulse to
pulse. An increase of power toward the horizontal axis is called a
low-frequency excess. As discussed in paper I the $P_2$ values are
just a rough measure of the presence of drift, its direction and the
magnitude of the slope in an overall mean sense only.

Following paper I the drifters are classified into three classes. PSR
B0818--13 is a {\em coherent drifter} (class ``Coh'' in Table
\ref{Table92}), because it has a vertically narrow drift feature and
PSR J0421$-$0345 is a {\em diffuse drifter}, because it has a feature
that is vertically broader than 0.05 cpp (Fig. \ref{Classes_fig92}). A
diffuse drifter is indicated by ``Dif'' when its drift feature is
clearly not reaching the alias borders ($P_0/P_3=0$ and
$P_0/P_3=0.5$), as is the case for J0421$-$0345. This is not the case
for PSR B0820+02, which has a feature that is clearly extended toward
$P_0/P_3=0$ and it is therefore indicated by ``Dif$^\ast$'' in Table
\ref{Table92}. Notice also the minimum in the power along the vertical
axis, which is an indication for drift reversals (see paper I for
details and simulations of this effect). If there is no significant
finite value of $P_2$ found, but there is a preferred $P_3$ value, the
modulation is classified as longitude stationary (class ``Lon'' in
table \ref{Table92}). These pulsars are not included in the statistics
because it is not clear if and how this modulation is related to
drifting subpulses.

\section{\label{SctIndividual92}Individual detections}

\subsection{Coherent drifters (class ``Coh'')}

\subsubsection{Known drifters (class ``Coh'')}

{\bf {B0148$-$06}}: both components of the pulse profile of this pulsar
show drifting subpulses with the same $P_3$ value and drift direction
(see Fig. \ref{B0148-06}). The drift bands are clearly visible in the
pulse-stack. This is consistent with results reported by
\cite{bhmm85}, who discovered the drifting subpulses at 645
MHz. Interestingly, the drifting is most prominent in the trailing
component, while in the 21-cm data the most prominent drifting was found in
the leading component. This could be related to the change in the
relative peak intensities of the pulse profile.\\
{\bf {B0320+39}}: this pulsar is known to show very regular drifting
subpulses (\citealt{iks82}), which is confirmed by the very narrow
drift feature in our observation (Fig. \ref{B0320+39}). Notice that
the first harmonic is visible as well $P_0/P_3\simeq0.24$ cpp.
\cite{ikl+93} have shown that drifting at both 102 and 406 MHz occurs
in two distinct pulse longitude intervals and that the energy
contribution in the drifting subpulses is less at higher frequencies,
a trend which continued in the 21-cm data. The drift feature in the
2DFS shows a clear horizontal structure caused by a subpulse phase
step in the drift bands. At both 328 MHz and 1380 MHz the drift bands
of PSR B0320+39 are known to show a phase step (\citealt{esv03} and
\citealt{es03c} respectively). This is the same observation as used by
\cite{es03c}.\\ 
{\bf {B0809+74}}: the subpulse drift is detected with a very high $S/N$
in the 2DFS (Fig. \ref{B0809+74}) and at least two higher harmonics
are detected as well. There is no evidence for a horizontal structure
in the feature in the 2DFS such as detected in the 21-cm data. This is
expected, because the phase step in the drifting subpulses is only
present at high frequencies (\citealt{wbs81} at 1.7 Ghz, \cite{pw86}
at 1420 MHz and \cite{es03c} at 1380 MHz). The drift rate is affected
by nulls (\citealt{th71}) and detailed analysis of this phenomenon
allowed \cite{vsr+03} to conclude that the drift is not aliased for
this pulsar. Notice that the drift-feature is vertically split, which
corresponds with the main drift-mode and an (about 7\%) slower
drift-mode.  This is consistent with the decrease in driftrate
observed by \cite{vkr+02}. In Fig. 12 of \cite{la83} a similar split
of the feature can be seen. \\
{\bf {B0818$-$13}}: this pulsar has a clear drift feature that contains
almost all the power in the 2DFS, as well as a very clear second harmonic
(Fig. \ref{B0818-13}). The first fifty pulses of the observation are
plotted in the left panel of Fig. \ref{Classes_fig92}. The drift feature
has a horizontal structure similar to that observed at 21 cm.
The modulation phase profile is much more complex at this frequency
and seems to consist of at least four distinct pulse longitude ranges
where the drift bands are linear. 
A decrease of the drift rate in the middle of the pulse
profile has been reported at 645 MHz by \cite{bmhm87}.
The longitude resolved modulation index shows a minimum at the
position of the start of the subpulse phase swing (consistent with
\citealt{bmhm87}), something that is also observed for the phase steps
of PSR B0320+39. 
This phase swing seems to be related to the complex polarization
behavior of the single pulses as observed by \cite{edw04}. \cite{la83}
found that the nulling in this pulsar interacts with the drifting
subpulses such that the drift-rate increases after a null. \cite{jv04}
confirmed this effect and argued that the drifting subpulses of this
pulsar are aliased. The feature in the side panel of the LRFS peaks at
$P_3=0.211$ cpp. Interestingly, this feature is not symmetric but
there appears to be an additional overlapping component that peaks at
$P_3=0.219$ cpp. This would correspond with a drift-mode that is 4\%
faster, however this feature is just too weak to be
significant. Nevertheless this feature would be consistent with the
drift-null interaction reported by \cite{jv04} and \cite{la83}.\\
{\bf {B1540$-$06}}: this pulsar has a clear drift feature that contains
almost all the power in the 2DFS (Fig. \ref{B1540-06}). The positive drift
direction is consistent with that reported by \cite{ash82} at
400 MHz and the drifting detected in the 21-cm data. Interestingly,
there is no indication that the trailing part of the pulse profile has
an opposite drift direction, as found in the 21-cm data.\\
{\bf {B1633+24}}: the 2DFS of this pulsar shows a clear drift feature
close to the $P_3=2P_0$ alias border (Fig. \ref{B1633+24}), consistent
with the reported drifting of \cite{hw87} at 430 MHz. This pulsar is
not included in the statistics because it is just too weak to be part
of the source-list. This pulsar was not observed at 21 cm.\\
{\bf  {J1650$-$1654}}: this pulsar shows a weak but clear drift feature
in its 2DFS (Fig. \ref{J1650-1654}), confirming the drifting
discovered in the 21-cm data. In the 21-cm data the feature seems to
show horizontal structure, which could indicate that the drift bands
are curved or show a phase step. At 92 cm the $S/N$ is too low to say
anything more about this.  As observed at 21 cm, there is also a
low-frequency excess.\\
{\bf {B1702$-$19}}: the pulse profile of this pulsar has an interpulse
(\citealt{blh+88}) and we discovered in the 21-cm data that both the
main- and the interpulse show a drift feature and with the same $P_3$
value. At 92 cm there is no significant modulation found in the
interpulse (likely because the interpulse is weaker at this
frequency), but the drift feature in the main pulse is confirmed
(Fig. \ref{B1702-19}). The main- interpulse interaction
of PSR B1702$-$19 is discussed by \cite{wws07}.\\
{\bf  {B1717$-$29}}: a very narrow drift feature is seen in both the
LRFS and the 2DFS of this pulsar (Fig. \ref{B1717-29}), confirming the
drifting discovered at 21 cm and demonstrating that coherent drifting
can be found in low $S/N$ observations. Most of the power is in the
drifting subpulses.\\
{\bf  {B1819$-$22}}: the 2DFS of this pulsar very clearly shows a drift
feature (Fig. \ref{B1819-22}), confirming the drifting discovered at
21 cm. At that frequency the feature was broadened by drift-mode
changes, which is not the case in this observation. The $P_3$ in the
92-cm data is consistent with the slow drift-mode found at 21
cm. Notice also that there is a low-frequency excess, which is much
stronger than at 21 cm. This low-frequency excess is
probably related to ``nulls'', which appear at low frequencies instead
of the fast drift-mode observed in the 21 cm data. Simultaneous
multi-frequency observations will be required to prove this
interpretation and if it turns out to be true, this pulsar would resemble
PSR B0031$-$07 (\citealt{smk05,sms+07}). The drifting subpulses can
be seen by eye in the pulse-stack and this pulsar also appears to show
real nulls (which are most clear at 21 cm), something which has not been
reported for this pulsar before. \\
{\bf  {J1901$-$0906}}: the
trailing component of this pulsar shows a clear and narrow drift
feature in its 2DFS, which is not detected in the leading component
(Fig. \ref{J1901-0906}). However the 2DFS of the leading component
shows a drift feature with a larger $P_3$. This is consistent with the
drifting discovered in the 21-cm data, so it is now shown in two
independent measurements that both components have drifting subpulses
with a different $P_3$.\\
{\bf {B1918+19}}: this pulsar is shown to be a drifter with at least
four drift modes at 430 MHz (\citealt{hw87}). In our observation we
see, besides the strongest feature (the `$B$' mode), also a $P_3 =
5.9\pm0.2P_0$ feature ($P_2 = {\degrees{30}}^{+25}_{-7}$), which is
the `$A$' mode. There is also a low-frequency excess at $P_3 =
70\pm30P_0$ (Fig. \ref{B1918+19}), which is possibly related to the
mode-switching. The drifting subpulses can be seen by eye in the
pulse-stack. Drifting subpulses were not seen in the 21-cm
observation, which could possibly be because of a too low $S/N$.\\
{\bf {B2043$-$04}}: this pulsar has a very clear and sharp drifting
subpulse feature in its 2DFS (Fig. \ref{B2043-04}), which confirms the
drifting discovered in the 21-cm data. Closer inspection shows that
the drift feature appears to be extended toward the alias
border. These apparent variations in the drift-rate could be related
to the small $P_3$ values in combination with intensity fluctuations
of the subpulses. Almost all the power is in the drift subpulses.\\
{\bf {B2303+30}}: this pulsar is known to drift close to the alias
border (e.g. \citealt{so75} at 430 MHz). This is also seen in our
observation as a clear double-peaked feature in the 2DFS exactly at
the alias border (Fig. \ref{B2303+30}). This suggests that the apparent
drift direction changes during the observation because the alias
changes constantly. At both frequencies the same drift direction
dominates. The change of drift direction can clearly be seen by eye in
the pulse-stack and also in the pulse-stacks shown in
\cite{rwr05}. These authors show that, besides this
$P_3\!\!\approx\!\!2P_0$ `$B$' drift mode, there is occasionally also
a $P_3\!\!\approx\!\!3P_0$ `$Q$' drift mode if the $S/N$ conditions
are good. The feature in the 2DFS seems to be extended all the way up
to $P_0/P_3\simeq0.3$, but there is no evidence that this is
associated with a (quantized) drift-mode. There is also a strong
low-frequency excess, probably caused by nulling. This is all
consistent with the 21-cm data.\\
{\bf {B2310+42}}: the two components of this pulsar are clearly drifting
at the alias border (Fig. \ref{B2310+42}). The drift feature in both
components are clearly double peaked, so the alias mode is probably
changing during the observation. The dominant drift direction (at both
frequencies) is consistent with the positive drifting found by
\cite{ash82} at 400 MHz. The low-frequency excess of the leading
component is clearly drifting with two signs as well ($P_2 =
\degrees{240}\pm50$,$P_3 = 15.7\pm0.6P_0$ and $P_2 =
-\degrees{50}\pm30$, $P_3 = 60\pm7P_0$), consistent with our 21-cm
results. At low frequencies the low-frequency excess of the trailing
component shows a preferred drift direction ($P_2 =
-\degrees{130}\pm20$, $P_3 = 19\pm2P_0$) and this feature in the 2DFS
appears to be more detached and higher above the horizontal axis
compared with the 21-cm observation.\\
{\bf {B2315+21}}: drifting with a negative drift direction has been
reported for this pulsar at 430 MHz by \cite{bac81}, which is
confirmed in our spectra (Fig. \ref{B2315+21}). No drifting subpulses
were found in the 21-cm data. Interestingly, the drift feature in the
92-cm data is only found in the leading half of the profile. This
suggests that the non-detection of drifting subpulses in the 21-cm data
could be because of a profile evolution, although the $S/N$ of the
21-cm data is somewhat lower.\\

\subsubsection{New drifters (class ``Coh'')}

{\bf {B0105+65}}: the pulsar shows a clear drift feature in the 2DFS at
the $P_3=2P_0$ alias border that contains most of the power (see
Fig. \ref{B0105+65}). In the LRFS of the 21-cm data there appears to
be a $P_3=2P_0$ modulation as well (although much weaker) and it did
not show up in the 2DFS as a feature with a preferred drift
direction. Although the $S/N$ was slightly lower at 21 cm, this seems
at most only part of the explanation. The number of recorded pulses
was very similar at both frequencies and there seems to be not much
frequency evolution in the pulse-profile. The drift-feature in the
2DFS of the 92-cm data appears to be split into a component left and
right of the vertical axis, suggesting that the subpulses are drifting
in both directions during the observation or that the drift-bands are
highly non-linear. The $P_2$ value in the table is for the centroid of the
whole feature, but it peaks at $P_2 = -{\degrees{7.7}}\pm0.3$.\\
{\bf {J0459$-$0210}}: the 2DFS of pulsar has a very strong drift feature
containing virtually all the power in the 2DFS
(Fig. \ref{J0459-0210}). The feature has a clear horizontal structure,
which is probably because there is a subpulse phase difference between
the two components of the profile. This pulsar was not observed at 21
cm. Notice that the pulse profile is similar to that of
PSR B0320+39 at 103 MHz (\citealt{kl99c}), where the two out-of-phase drift components 
apparently are more separated than at higher frequencies.\\
{\bf {B1730$-$22}}: there is a clear detection of a drift-feature in the
spectra of this pulsar (Fig. \ref{B1730-22}). In the 21-cm data there
is probably modulation at a similar $P_3$ value, but it was very weak.
Possibly the drifting subpulses were not discovered at 21 cm because
of the remarkable profile evolution or because the 21-cm observation was too
short.\\
{\bf {J2302+6028}}: this pulsar shows a strong drift feature in its 2DFS
(Fig. \ref{J2302+6028}). This pulsar was not observed at  21 cm.\\

\subsection{Diffuse drifters (classes ``Dif'' and ``Dif$^\ast$'')}

\subsubsection{Known drifters (classes ``Dif'' and ``Dif$^\ast$'')}

{\bf {B0031$-$07}}: this pulsar shows a broad drifting feature in its
2DFS (Fig. \ref{B0031-07}). Three drift modes have been found for this
pulsar by \cite{htt70} at 145 and 400 MHz. In our observation most 
power in the 2DFS is due to  `$B$'-mode drift ($P_3=6P_0$) at 0.15
cpp. The slope of the drift bands change from band to band
(e.g. \citealt{vj97}), causing the feature to extend vertically in the
2DFS. The `$A$'-mode drift ($P_3=12P_0$) is visible as a downward
extension of the main drift feature to 0.08 cpp. At 0.25 cpp the
`$C$'-mode drift ($P_2 = -{\degrees{19}}^{+2}_{-1}$ and $P_3 =
4.1\pm0.1P_0$) is visible as well. The feature at 0.3 cpp is the
second harmonic of the main drift feature. In the 21-cm data the
`$A$'-mode dominated over the '$B$'-mode, which dominates this 92-cm
observation, and the '$C$'-mode was not detected. This is consistent
with the multi-frequency study of \cite{smk05,sms+07}.\\
{\bf  {B0136+57}$^\ast$}: the drift feature discovered in the 21-cm data
is confirmed (Fig. \ref{B0136+57}) and it is, as observed at 21 cm, strongest
in the leading part of the pulse profile. It appears that the feature
extends toward the horizontal axis and there is a hint of a
low-frequency excess.\\
{\bf  {B0149$-$16}}: the 2DFS of both components of this pulsar show a
drift feature with the same preferred drift direction
(Fig. \ref{B0149-16}). This pulsar was discovered to have drifting
subpulses in the 21-cm data. Probably because of a higher $S/N$ in the
observation at 92 cm, the drifting is detected in both components and
the features are broader than could be found in the 21-cm
data. Therefore this pulsar is classified a diffuse drifter, and not
as a coherent drifter.\\
{\bf {B0301+19}$^\ast$}: both components shows a broad drift feature in
their 2DFS (Fig. \ref{B0301+19}). This pulsar is observed to have
linear drift bands in both components of the pulse-stack
(\citealt{ss73} at 430 MHz), which is confirmed by this
observation. The feature in the trailing component is reported to be
broader than in the leading component (\citealt{brc75}, also at 430
MHz), probably because drifting subpulses appear more erratic in the
trailing component. This is also the case is our observation, as
especially the feature in the trailing component is broad and may even
be extended toward the alias border. The measured value for $P_3$ is
significantly larger for the trailing component. Interestingly, no
drifting is detected in the leading component at 21 cm. The LRFSs at
both frequencies are similar, which suggest that the same two
components are observed at the two frequencies.\\
{\bf {B0329+54}$^\ast$}: the power in the LRFS of this pulsar peaks
toward $P_0/P_3=0$, as is reported by \cite{th71} for low frequencies
(Fig. \ref{B0329+54}). The plotted 2DFSs are of the leading and
trailing part of the central peak. The trailing part of the central
peak shows a broad drift feature in its 2DFS and the subpulses have a
preferred positive drift direction, something that is also reported by
\cite{tmh75} at 400 MHz. The leading peak and leading part central
peak also show  a preferred positive drift direction
($P_2=\degrees{250}\pm150$ and $\degrees{200}\pm150$ and $P_3= 6\pm2$
and $4\pm2$ respectively). The right peak does not show significant
drifting. This is very similar to the 21-cm results, except the
preferred drift direction of the left part of the central peak which
was found to be negative at high frequencies. Possibly this means that
the left part of the central peak just does not have a stable
preferred drift direction (at both frequencies). The value of $P_3$ in
the leading part of the central peak appears to be, as observed at 21 cm, on
average smaller than in the trailing part of the central peak. This is
possibly because the drift feature, if it really is a drift feature,
is much weaker and therefore the spectrum is much flatter. Notice the
very high modulation index  between the leading and central
component of the profile.\\
{\bf  {B0450+55}$^\ast$}: the drift feature of the main component that
was discovered in the 21-cm data is confirmed
(Fig. \ref{B0450+55}). The $S/N$ of the data is unfortunately not
enough to detect any subpulse modulation in the leading component,
which is almost merged with the main component at 92 cm. This
component was found to have an opposite preferred drift direction and
a different $P_3$ value.\\
{\bf  {B0525+21}$^\ast$}: subpulse modulation without apparent drift as
well as some correlation between the subpulses of the two components
of the pulse profile has been detected for this pulsar by \cite{bac73}
at 318 MHz and \cite{tmh75} at 400 MHz. We find that the leading
component shows a broad drift-feature with a preferred negative drift
direction (Fig. \ref{B0525+21}), which confirms the drifting
discovered in the 21-cm data. The feature is possibly extended toward
the $P_3=2P_0$ alias border. In this observation there is no evidence
for a drift feature with opposite drift direction in the trailing
component, as found in the 21-cm data.\\
{\bf {B0628$-$28}$^\ast$}: sporadic drifting subpulses with a positive
drift direction have been reported for this pulsar by \cite{ash82} at
400 MHz, but the $P_2$ and $P_3$ values could not be measured. The
positive drift direction has been confirmed in paper I as a clear
excess of power in the right half of the 2DFS.  At this frequency
(Fig. \ref{B0628-28}) the drifting is also detected and it originates
from the trailing half of the pulse profile (see the bottom panel of
Fig. \ref{B0628-28}). Interestingly, the leading half of the pulse
profile shows a quasi-periodic feature with a significantly longer
period. A similar behavior is revealed in the 21-cm data after
re-analysing the data, however there it is less clear. There
is no indication that $P_3$ is different at the two frequencies. The
drift feature in the 2DFS of the trailing half of the profile is not
separated from either alias border, like observed at 21 cm. \\
{\bf {B0751+32}$^\ast$}: the 2DFS of the leading component of the pulse
profile of this pulsar shows a very broad drift feature with a
negative $P_2$ value (Fig. \ref{B0751+32}). This can be seen in the
bottom plot of the first 2DFS, which shows an excess of power in the
left half. This confirms the drifting found in the 21-cm data and as
reported by \cite{bac81} at 430 MHz. Interestingly, the trailing
component shows a faster periodicity (without measured preferred drift
direction), which was not significant in the 21-cm data (possibly
because a lower $S/N$). Both components also show a strong long period
fluctuation ($P_3=60\pm20P_0$ and $P_3=70\pm20P_0$ respectively) as
was also reported for the 21-cm data. These features are not
significantly offset from the vertical axis.\\
{\bf {B0820+02}$^\ast$}: the positive drifting, as has been reported by
\cite{bac81} at 430 MHz, is clearly detected in the 2DFS
(Fig. \ref{B0820+02}). The drift feature is spread out over the whole
$P_3$ range, especially towards the $P_0/P_3=0$ axis where it is
clearly double peaked (the region between the dashed lines). This
means that the drift direction could be changing during the
observation. The tabulated $P_2$ and $P_3$ values are for the main
feature between $P_0/P_3=0.1$ and 0.3 cpp. In the 21-cm data (which
was much shorter) no drifting was found, probably because of a too low
$S/N$.\\
{\bf  {B0823+26}$^\ast$}: only the pulse longitude range of the main
pulse is shown in Fig. \ref{B0823+26} and the 2DFS of the leading
component of the main pulse shows a clear broad drift
feature. \cite{bac73} found that at 606 MHz this pulsar shows drifting
in bursts, but the drift direction is different for different
bursts. In our observation there seems to exist a clear preferred
subpulse drift direction, confirming the preferred drift direction found
in the 21-cm data.\\
{\bf {B0834+06}$^\ast$}: the 2DFS of both components show a drift
feature at the $P_3=2P_0$ alias border (Fig. \ref{B0834+06}). This
confirms the drifting detected by \cite{sspw70} at a similar
frequency. Both components showed a positive drift with
$P_3\simeq2P_0$ at 21 cm. However at this frequency the leading
component shows a negative drift, which is consistent with the
direction found by \cite{sspw70}. Possibly the relatively short
observation in combination with a changing alias mode is responsible
for the different preferred drift directions found at the two
frequencies. The trailing component does possibly show a positive
drift, although this is not very significant. Besides this feature
there is a very broad drift feature with a positive preferred drift
direction in the leading component ($P_3 = 3.5\pm0.6P_0$ and $P_2 =
\degrees{45}\pm6$). The circulation time of this pulsar ($\hat P_3$)
has been measured by \cite{ad05}.\\
{\bf  {B0919+06}$^\ast$}: the power in the 2DFS (Fig. \ref{B0919+06})
is significantly offset from the vertical axis over the whole $P_3$
range, confirming the preferred negative drift direction discovered in
the 21-cm data. Interestingly, the drift feature is much less clear at
this frequency and there is also no sign of the low-frequency
excess. Notice that the main component seems to be split into two
components. This can also seen clearly in the LRFS, which shows
subpulse modulation in two distinct pulse longitude ranges. No
drifting has been reported for this pulsar by \cite{bac81} at 430
MHz.\\
{\bf {B0950+08}$^\ast$}: drifting has been reported for this pulsar
(e.g. \citealt{bac73} and \citealt{wol80}) with a variable
$P_3\simeq6.5$. This is confirmed in the 2DFS of our observation
(Fig. \ref{B0950+08}) wherein subpulse modulation is seen over the
whole $P_3$ range with a (weak) preferred negative drift direction. In the
21-cm data there was no significant drifting detected, which could be
because the lower number of pulses or because it is only detectable at
low frequencies. The observing frequency of \cite{bac73} was 430 MHz,
but it is not mentioned in \cite{wol80}.\\
{\bf  {B1039$-$19}}: both components of this pulsar show a clear, broad
drift feature in its 2DFS with the same preferred positve drift
direrection (Fig. \ref{B1039-19}). This confirms the drifting
discovered in the 21-cm data.\\
{\bf {B1112+50}$^\ast$}: a broad feature offset from the vertical axis
is detected in the 92-cm data (Fig. \ref{B1112+50}). This pulsar is
known to show nulling and pulse profile mode switching and in one of
these modes (positive) drifting subpulses are reported
(e.g. \citealt{wsw86} at 1412 MHz).  According to \citealt{wsw86} the
leading component is weak when there are no drifting subpulses in the
trailing component. Reconsidering the 21-cm data it seems that the
there was no drift-feature detected at 21 cm because in that
observation the leading component was strong, showing that the
emission was dominated by the mode without drifting subpulses. \\
{\bf {B1133+16}$^\ast$}: the 2DFS of both components of this pulsar
(Fig. \ref{B1133+16}) show a very broad drift feature with a preferred
positive drift direction. The trailing component shows also a long
period drift feature ($P_2 = \degrees{120}\pm50$ and $P_3 =
30\pm8P_0$), consistent with the 21-cm data. At 92 cm also the leading
component shows a long period fluctuation ($P_3=33\pm5P_0$). This
positive drifting is consistent with the drifting found by
\cite{now96} at 430 and 1418 MHz and by \cite{bac73} \& \cite{tmh75}
at low frequencies. There is some indication that the
long period feature is related to nulls which appear to be
quasi-periodical.\\
{\bf {B1237+25}$^\ast$}: the 2DFS of the outer components of the pulse
profile are clearly drifting with opposite drift direction (they are
plotted in Fig. \ref{B1237+25}). The 2DFS of the second component
(which is not plotted) also shows a preferred drift direction ($P_2 =
\degrees{6}\pm3$, $P_3 = 2.77\pm 0.03$). The 92-cm and 21-cm results
are similar, except that the third component shows a clear
low-frequency excess and that the fourth component does not show a
preferred drift direction at low frequencies. Notice also that the
modulation index profiles are very different at the two
frequencies. This is all consistent with \cite{pw86} at 408 and 1420
MHz and \cite{sr05} at 327 MHz. \\
{\bf {B1508+55}$^\ast$}: \cite{th71} report that the subpulse modulation
of this pulsar is unorganized and without a preferred drift direction
or a particular $P_3$ value at 147 MHz. However the 21-cm data
revealed a broad drift feature, which is offset from the vertical axis
over the whole $P_3$ range. The drifting subpulses are confirmed in
the 2DFS of the 92-cm data (Fig. \ref{B1508+55}), where the drift
feature appears much stronger. Moreover, we find that the $P_3$ values
for the leading and trailing components are significantly different
from each other. The central component shows a low frequency excess
without a preferred drift direction, which was also reported by
\cite{th71}. All these details were not found at  21 cm, possibly
because of a lower $S/N$ and a shorter observation duration or because
of the profile evolution with frequency. \\
{\bf {B1530+27}$^\ast$}: the 2DFS of this pulsar shows a preferred
negative drift direction (Fig. \ref{B1530+27}), which is confirmed in
another observation. A negative drift direction was also found by
\citealt{bac81} at 430 MHz. Notice that the feature is extended over
the whole $P_3$ range and that the vertically integrated power in
between the dashed lines appears to be double peaked. This could mean
that the drift direction is constantly changing or that the
drift-bands are highly non-linear. This pulsar was not observed at
21 cm.\\
{\bf  {B1604$-$00}$^\ast$}: the very broad feature has a preferred
positive drift direction (Fig. \ref{B1604-00}), which confirms the
discovery in the 21-cm data. There is a strong low-frequency excess as
well at both frequencies. It is difficult to characterize the
single pulses. It sometimes looks like there is only one component on
at the time while there is also a short term flickering. The component
switching can be more abrupt or more like a drift-band. The component
switching would be consistent with the double peaked low-frequency
excess that appears in the 2DFS.\\
{\bf {B1612+07}$^\ast$}: the power in the 2DFS is offset from the
vertical axis over the whole $P_3$ range (Fig. \ref{B1612+07}), which
is consistent with the negative subpulse drift that has been reported by
\cite{bac81} at 430 MHz for this pulsar. At 92 cm this pulsar has  a
strong low frequency excess. Although the $S/N$ was lower at 21 cm,
this cannot explain why the low-frequency excess was not observed at
21 cm.\\
{\bf  {B1642$-$03}$^\ast$}: drifting is observed to occur in bursts in
this pulsar with both drift directions (\citealt{th71} at 400 MHz) and
also \citealt{tmh75} report that there is no preferred drift direction
at 400 MHz. The 2DFS of our observation (Fig. \ref{B1642-03}) reveals
a clear broad drift feature with a clear preferred negative drift
direction, so this pulsar is classified as a drifter. Because the
observation contains almost 15,000 pulses and the drift feature is
clearly offset from the vertical axis the detected preferred negative
drift direction seems highly significant. Interestingly, compared with
paper I the opposite drift direction seems to dominate and
also the $P_3$ values seems to be significantly different. At both
frequencies the alias border seems to be crossed on both sides,
because the feature is extended over the whole $P_3$ range and seems
double peaked. Notice also that the LRFS are very different at the two
frequencies. While at 21 cm the quasi-periodic feature in the LRFS is
centered in the middle of the main component, at 92 cm it is
centered at the leading edge of the profile.  Because the observations
at the two frequencies were not simultaneously recorded, it is
impossible to prove that these differences are because of a frequency
dependence in the drifting phenomenon rather than a time-dependence.\\
{\bf  {B1738$-$08}$^\ast$}: the drifting subpulses discovered in both
components in the 21-cm data is confirmed (Fig. \ref{B1738-08}). There
is, consistent with the 21-cm data, also a low frequency excess in
both components, possibly because of nulling (not reported in the
literature). The drifting subpulses and the nulls can be seen by eye
in the pulse-stack. The observations are too short to make the
difference in $P_3$ significant.\\
{\bf  {B1753+52}$^\ast$}: the trailing part of the pulse profile shows a
broad drift feature in its 2DFS (second 2DFS in Fig. \ref{B1753+52}),
consistent with the drift feature discovered at 21 cm. There is no
evidence for drifting subpulses in the rest of the profile (first
2DFS). The drifting can be seen by eye in the pulse-stack at both
frequencies.\\
{\bf  {B1857$-$26}}: the components at both edges of the pulse profile
are drifting with the same drift direction, which can be seen in
Fig. \ref{B1857-26} as an excess of power in the 2DFS at positive
$P_2$ values. The drift bands are sometimes visible to the eye in the
pulse-stack. The center part of the pulse profile does not show
drifting in its 2DFS and is therefore not plotted. This confirms the
drifting discovered in the 21-cm data. Notice the dramatic profile
evolution with frequency. Nevertheless the LRFSs at both frequencies
are similar, making it possible to identify which component
corresponds to which at the other frequency.\\
{\bf  {B1900+01}$^\ast$}: drifting was clearly detected over the whole
$P_3$ range in the 21-cm data, which is (although weaker) confirmed in
this observation (Fig. \ref{B1900+01}).\\
{\bf  {B1917+00}$^\ast$}: this pulsar shows a broad drifting component in
its 2DFS, which is visible in the bottom plot as an excess of power at
positive $P_2$ (Fig. \ref{B1917+00}). This observation confirms the
drifting subpulses discovered in the 21-cm data. The drift feature is
smeared out over the whole $P_3$ range.  According to \cite{ran86} a
much longer $P_3\simeq50P_0$ value without a measured $P_2$ was
reported in a preprint by L.A. Nowakowski and T.H. Hankins, but to the
best of our knowledge the paper was never published. Notice also the
high modulation index because of occasional strong pulses.\\
{\bf {B1919+21}}: both components of this pulsar have a strong drift
feature in their 2DFS (Fig. \ref{B1919+21}) and the feature of the
leading component probably shows horizontal structure,  similar to that observed
in the 21-cm data.  The reason for this horizontal structure in the
drift feature is, similar to PSR B0320+39, that there is a subpulse phase
step in the drift bands. This observation confirms the reported phase
step by \cite{pw86} seen at 1420 MHz. There is also a strong
low-frequency excess detected in both components and the leading
component shows a $P_3=2P_0$ flickering. This observation is too short
to derive fluctuation spectra with high resolution. By comparing the
LRFSs at the two frequencies it seems that the three more separated components in the average pulse profile at 21 cm
 merge at lower frequencies.\\
{\bf {B1923+04}$^\ast$}: this pulsar shows in its 2DFS a very broad
drift feature at the $P_3=2P_0$ alias border (Fig. \ref{B1923+04}). It
is clearly double peaked, which could indicate that the drift direction
changes constantly during the observation. This pulsar was not part of
the 21-cm survey. The positive drift is consistent with the findings
of \cite{bac81} at 430 MHz.\\
{\bf {B1929+10}$^\ast$}: the LRFS (Fig. \ref{B1929+10}) shows a very
clear feature with $P_3=11.4\pm0.6P_0$, comparable to what was found
by \cite{nuwk82} at 0.43, 1.7 and 2.7 GHz. \cite{ohs77} suggested,
using 430 MHz data, that this pulsar has drifting
subpulses. \cite{bac73} reported two features in the LRFS of this
pulsar at 606 MHz and the short period feature appeared to have a
negative drift and the long period fluctuations appeared to be
longitude stationary. Also in our data the long period fluctuations do
not have a preferred drift direction, while the shorter period
fluctuations (between the dashed lines) show a preferred negative
drift. Also in the 21-cm data a negative drift direction was detected for
the short period feature, but the long period feature shows a positive
drift. The latter could be because the 92-cm observation was much
longer or because this feature only shows a preferred drift at high
frequencies. There is no evidence for a $P_3\simeq2P_0$ modulation at
this frequency. \\
{\bf {B1933+16}$^\ast$}: this pulsar shows subpulse modulation over the
whole $P_3$ range (Fig. \ref{B1933+16}). \cite{bac73} found 
that there is no preferred subpulse drift direction at 430 MHz, however
regular drifting with $P_3\simeq2.2P_0$ has been reported by
\cite{ohs77} at 430 MHz. We found a preferred positive drift direction in a
broad feature near the $P_3=2P_0$ alias border in the 21-cm
data, but in the 92-cm data there is a clear preferred negative drift direction
with a much larger $P_3$.\\
{\bf {B1944+17}$^\ast$}: this pulsar shows a clear drift feature in its
2DFS (Fig. \ref{B1944+17}) and the drifting can clearly be seen by eye
in the pulse-stack. The feature is broadened because this pulsar shows
drift-mode changes (\citealt{dchr86} at both 430 and 1420 MHz). The
$P_3=13P_0$ `$A$'-mode drift is visible in the 2DFS at 0.075 cpp and the
$P_3=6.4P_0$ `$B$'-mode drift (0.16 cpp) does not appear in the
spectrum as a peak, although the centroid of the power in the 2DFS is
offset from the vertical axis up to at least 0.2 cpp.  There is also
evidence for a broad feature in a different alias mode at
$P_3\simeq20P_0$, although it appears to be weaker than in the 21-cm
data. It could be that the zero drift `$C$'-mode (\citealt{dchr86}) is
a drift mode for which the drift direction is changing continuously. The
modulation index is higher in the 92-cm data, possibly because there
were longer nulls in that data-set.\\
{\bf  {B1953+50}$^\ast$}: this pulsar showed a very clear broad drift
feature in its 2DFS at 21 cm. At this frequency there is also a
preferred positive drift direction (Fig. \ref{B1953+50}), confirming
the discovery at 21 cm, but the drifting subpulses are much less
pronounced. The measured values for $P_3$ are slightly different in
the two observations, but the observations are too short to state that
this is really significant. The feature is reaching the $P_0/P_3=0$
alias border and it is double peaked, suggesting that the drift
direction changes during the observation. This is probably also the
case at 21 cm, although it is less clear.\\
{\bf {B2016+28}$^\ast$}: the 2DFSs show a very broad drifting feature
(Fig. \ref{B2016+28}), which is caused by drift mode changes
(e.g. \citealt{ohs77a} at both 430 and 1720 MHz). Interestingly, at
this frequency (382 MHz), there is no sign of the strong
$P_3\simeq20P_0$ drift mode as seen in the leading component in the
21-cm data.  Also \citealt{ohs77a} concluded that the drift-rate
distribution is frequency dependent.
Also interestingly, the $P_3$ values are significantly different in
the two components. The drift bands can be seen by eye in the
pulse-stack. The differences between the two frequencies is really a
frequency dependence of the drift phenomenon, because the observations
at the two frequencies were recorded simultaneously. The pulse profile has also evolved significantly with frequency.\\
{\bf {B2020+28}$^\ast$}: the two components of the pulse profile show a
broad feature close to the alias border. A strong even-odd modulation
was also reported by \cite{brc75} at 430 MHz and by \cite{nuwk82} at
1.4 GHz and the trailing component was shown to have a preferred drift
direction. The leading component was shown to have a opposite
preferred drift direction in the 21-cm data, which is confirmed in the
92-cm data (Fig. \ref{B2020+28}). There is no evidence that the
feature extends over the alias border, although the feature is not
clearly separated from the alias border.\\
{\bf {B2021+51}$^\ast$}: the two components have a broad drift feature
with a preferred drift direction (Fig. \ref{B2021+51}), consistent
with e.g. \cite{ohs77} at 1720 MHz and the 21-cm data. The drift-rate
changes by a large factor during the observation, which was also
observed by \cite{ohs77}. They also suggested  that perhaps
the apparent drift direction changes sporadically.  There is no clear
evidence from the fluctuation spectra of the observation that the alias
mode is changing. The 92-cm observation was much shorter than the
21-cm observation, but nevertheless the drift feature appears to be
much weaker at 92 cm. Also the $P_3$ value seems to be significantly
different at the two frequencies. Because the observations at the two
frequencies were not simultaneously recorded it is not possible to
prove that these differences are really because a frequency dependence
of the drifting phenomenon rather than a time-dependence. Notice that
the pulse profile and the modulation index profiles are also very
different at the two frequencies.\\
{\bf {B2044+15}$^\ast$}: especially the 2DFS of the trailing component
of the pulse profile convincingly shows a broad drifting feature
(Fig. \ref{B2044+15}), which was also the case at 21 cm. These
observations confirm the drifting subpulses found by \cite{bac81} at
430 MHz. It is not clear if the feature is connected to the
$P_0/P_3=0$ alias border, because the $S/N$ of both observations was
low.\\
{\bf {B2045$-$16}$^\ast$}: subpulses with a negative drift have been
reported for the outer components and the the fluctuation feature is
observed to be broad with $P_3$ values between 2 and 3$P_0$
(\citealt{os77b} at 1720 MHz, \citealt{nuwk82} at 1.4 and 2.7 GHz
and \citealt{th71} at low frequencies). In our observation all three
components show negative drifting and the 2DFSs of the leading and
trailing component are plotted in Fig. \ref{B2045-16}. The middle
component shows a $P_2 = -\degrees{100}\pm50$ with $P_3 = 4\pm2P_0$
feature.  At 21 cm positive drifting was observed in the trailing
component and no drifting in the other components and the
drift-feature was classified as coherent. As noted in paper I,
the 21-cm observation was very short and therefore it is very well
possible that the results at 21 cm are caused by a sporadic event
rather than a systematic behavior of the drift phenomenon.  The middle
component  also shows a $P_3=32\pm2P_0$ feature without a
preferred drift direction at 92 cm, but because the
observation at 92 cm was also relatively short, it is not clear if this is
significant.  \\
{\bf  {B2110+27}$^\ast$}: this pulsar shows drifting over the whole
$P_3$ range in its 2DFS (Fig. \ref{B2110+27}), confirming the drifting
discovered at 21 cm.  Also at 92 cm, especially the lower part
of the 2DFS is double peaked, which could suggest that the
alias mode is constantly changing during the observation. In the
pulse-stack drifting is visible directly. The drift bands are
probably distorted by nulls, causing the drift feature in the 2DFS to
be extended over the whole $P_3$ range. There is also a strong
$P_3\simeq2P_0$ modulation. Some single drift bands with positive
drifting can be seen in the pulse-stack as well as a $P_3\simeq2P_0$
flickering. Drift bands with an apparent opposite drift direction can
be seen as well, although they also show the $P_3\simeq2P_0$
flickering. Nulling has not been reported in the literature before for this pulsar.\\
{\bf  {B2111+46}$^\ast$}: it has been reported that this pulsar shows
subpulse drift with a positive and negative drift direction, but
without either dominating (\citealt{tmh75} at 400 MHz). We find that
there is systematic drift in the leading component
(Fig. \ref{B2111+46}), which confirms our detection in the 21-cm
data. The feature is not well separated from the $P_3=2P_0$ alias
border. Notice also the high modulation index at 92 cm, especially in
the outer components.\\
{\bf  {B2148+63}$^\ast$}: at 21 cm the 2DFS of this pulsar showed a
broad, triple, well separated features at the $P_3=2P_0$ alias border,
which is most likely caused by apparent drift direction changes. The
2DFS of the 92-cm data shows a similar feature (Fig. \ref{B2148+63}),
confirming both the preferred drift direction and the apparent drift
direction changes discovered at 21 cm. The drift feature appears to be
much weaker at 92 cm, which is surprising because this observation has
a higher $S/N$ and contains more pulses.\\
{\bf  {B2154+40}$^\ast$}: the very broad drift feature discovered at
21 cm is confirmed in this observation (Fig. \ref{B2154+40}) and at
this frequency the drift feature is also mainly produced by the leading
component of the pulse profile and is possibly extended toward the
alias borders. The observations at the two frequencies have comparable
$S/N$ and about the same number of recorded pulses, but nevertheless
the feature is strongest at 92-cm.\\
{\bf  {B2255+58}$^\ast$}: the drift feature discovered at 21 cm is
confirmed (Fig. \ref{B2255+58}). At 21 cm the feature in the 2DFS was
clearly horizontally split which might be because of a subpulse phase
step. At this frequency there is no sign of a similar feature, which
could be related to the profile shape evolution that causes the pulse
profile to become double at higher frequencies.\\
{\bf {B2319+60}$^\ast$}: this pulsar shows a clear drift component in
the 2DFS of the trailing component of the pulse profile and a less
clear drift component in the middle and leading component. In
Fig. \ref{B2319+60} the 2DFS of the leading and trailing component are
shown. For the feature in the central component $P_2 =
{\degrees{130}}\pm80$ and $P_3 = 8\pm4P_0$ is measured.  It has been
found that this pulsar is a drift mode changer and that the allowed
drift mode transitions follow certain rules (\citealt{wf81} at 1415
MHz). In the 2DFS of the trailing component there is not much evidence
for a stable $P_3=8P_0$ `$A$' drift mode or a stable $P_3=4P_0$ `$B$'
drift mode, which could be because the observation is too
short to fully characterize the different drift-modes. Nevertheless
the preferred drift direction is detected in the trailing component.
The leading component seems to show mainly the $P_3=3P_0$ abnormal
mode, which is consistent with the results of \cite{wf81} who have
shown that the abnormal mode is related to a profile mode change in
which the leading component dominates the total emission. There is
also a low-frequency excess, which is similar to the $P_3 \simeq 130P_0$
feature found in the 21-cm data probably related to the nulls (\citealt{rit76}) or mode
changes. Notice the high modulation index, especially in the outer components.\\
{\bf  {B2351+61}$^\ast$}: the power in the 2DFS of this pulsar peaks
toward low frequencies and the centroid is significantly offset from
the vertical axis (Fig. \ref{B2351+61}), confirming the preferred
drift direction discovered in the 21-cm data.\\

\subsubsection{New drifters (classes ``Dif'' and ``Dif$^\ast$'')}

{\bf {J0421$-$0345}}: the 2DFS of this pulsar (Fig. \ref{J0421-0345})
shows a strong and clear detection of a drift feature that appears to
be strongest in the trailing component. This pulsar was not observed
at 21 cm.\\
{\bf {B0942$-$13}$^\ast$}: the 2DFS of this pulsar (Fig. \ref{B0942-13})
shows a weak, but clear detection of a drift feature with a very short
$P_2$ of only 2.5 ms. There is a hint that the feature is extended
toward the $P_3=2P_0$ alias border and changes alias mode (similar to PSR
B2148+63).  This pulsar was not observed at 21 cm.\\
{\bf {B1607$-$13}$^\ast$}: there is a clear detection of drifting
subpulses with a negative drift direction (Fig. \ref{B1607-13}). The
drifting subpulses are visible in the pulse-stack and the drifting
subpulses are also confirmed in another observation. The drift feature
is vertically extended, probably because of drift mode-changes. This
pulsar was not observed at 21 cm.\\
{\bf {J1652+2651}$^\ast$}: the 2DFS of this pulsar shows a weak drift
feature (Fig. \ref{J1652+2651}). The same feature is present in both
the first and second halve of the data making us confident in the
significance of this feature. This pulsar was not observed at 21 cm.\\
{\bf {B1700$-$18}}: there is clear detection of subpulses with a
negative drift direction and there is also a long period fluctuation
that appears to have an opposite preferred drift direction
(Fig. \ref{B1700-18}). The drifting subpulses can be  seen clearly
by eye in the pulse-stack. 
This pulsar was not observed at 21 cm. \\
{\bf {B1718$-$02}}: the 2DFS shows a clear broad drift feature
(Fig. \ref{B1718-02}). There is no evidence that the feature is
extended to the $P_0/P_3= 0$ alias border. This pulsar was not
observed at 21 cm.\\
{\bf {B1818$-$04}$^\ast$}: the 2DFS shows a very broad feature with a
preferred drift direction (Fig. \ref{B1818-04}), which was not significant
in the 21-cm data. The same preferred drift direction is detected in
the first, middle and last part of the observation, making us
confident in the significance. Interestingly, the 2DFS does not show a
sign of a double peaked distribution, as reported for the 21-cm data,
which could be related to the profile evolution. There is also a
longitude stationary $P_3=22\pm3$ feature, which did not appear in the
21-cm data. It has been reported that the subpulse modulation is not
well organized (\citealt{th71} and \cite{tmh75} both at 400 MHz).\\
{\bf {B1839+56}$^\ast$}: the 2DFS of this pulsar shows a low-frequency
modulation with a preferred positive drift direction
(Fig. \ref{B1839+56}). Consistent results are obtained for the first
and second half separately. Also the low-frequency excess in the 21-cm
data seems to show the same positive preferred drift direction, but
because of the relatively low number of pulses we were not confident in
the significance.\\
{\bf {B1905+39}$^\ast$}: the 2DFS of both the leading and trailing part
of the profile shows a drift feature (Fig. \ref{B1905+39}), although
it appears to be strongest in the trailing component. These features were not
discovered in the 21-cm data, not surprisingly because of the lower
$S/N$. The low frequency excess that was found in the 21-cm data is
confirmed at this frequency ($P_3=27\pm8P_0$) and possibly also shows
a preferred positive drift direction.\\
{\bf {B2053+21}$^\ast$}: both components of this pulsar show in the LRFS
a low-frequency modulation (Fig. \ref{B2053+21}). The power in the 2DFS of
the trailing component is clearly offset from the vertical axis and
shows a positive drift direction. The trailing components also show a
short term ($P_3 = 2.3\pm0.2P_0$) fluctuation, which interestingly
enough does not appear in the leading component. 
This pulsar was not observed at 21 cm.\\
{\bf {B2217+47}$^\ast$}: there is a (weak) preferred negative drift
direction detected in the feature in the 2DFS of this pulsar
(Fig. \ref{B2217+47}), which exists in both halfs of the observation.
This was not found in the 21-cm data, possibly because the observation
contained less pulses. At 147 MHz a flat spectrum was reported by
\citealt{th71}.
The 2DFS shows also two vertical bands smeared over the whole $P_3$
range. This modulation is, as observed at 21 cm, primarily generated
in the trailing part of the pulse profile. The interpretation is that
there is a quasiperiodic intensity modulation in the pulses with a
period of about 2.5 ms, but there is not much correlation in the
positions of the subpulses from pulse to pulse. This quasi-periodicity
is clearly visible in the single-pulses.\\

\subsection{Longitude stationary drifters (class ``Lon'')}

{\bf {B0756$-$15}}: a low-frequency modulation was found in the 21-cm
data. However this observation shows that the low-frequency excess is
in fact a longitude stationary feature with a relative long $P_3$
(Fig. \ref{B0756-15}). The feature in the 21-cm data is probably very
similar, although with a lower number of recorded pulses the
distinction between a low-frequency excess was harder to make.  \\
{\bf {B1541+09}}: a long period feature without preferred drift
direction is detected in the LRFS and 2DFS of this pulsar
(Fig. \ref{B1541+09}). This is consistent with the low-frequency
excess, mode changes and the organized, but short, drifts in both
directions reported by \citealt{now96} at 430 MHz. In the 21-cm data
no periodicity in the subpulse fluctuations could be measured,
possibly because of a too low $S/N$ and/or the drastic profile
evolution. The horizontal bands in the spectra are fluctuation
frequencies that were excluded because of RFI.\\
{\bf {B1732$-$07}}: a clear long period feature without preferred drift
direction is detected in the LRFS and 2DFS of this pulsar
(Fig. \ref{B1732-07}). This was not found in the 21-cm data, possibly
because of the lower $S/N$.\\
{\bf {B1737+13}}: especially the leading component shows a broad
modulation feature with $P_3\simeq10P_0$ (Fig. \ref{B1737+13}),
consistent with the the $P_3$=11-14$P_0$ longitude stationary subpulse
modulation without drifting reported by \cite{rws88} at 1412 MHz. No
drifting has been detected for this pulsar by \cite{bac81} at 430 MHz
and no fluctuation features were detected at 21 cm, most likely
because of the much lower $S/N$.\\
{\bf {B1811+40}}: there is a short period fluctuation in the spectra of
this pulsar (Fig. \ref{B1811+40}). The LRFS shows that this flickering
appears mainly in the leading part of the profile, which is also
clearly visible in the pulse-stack. It seems that the alias border
might be crossed constantly. At 21 cm the spectra are remarkably
featureless compared with the 92-cm result.\\
{\bf {B1846$-$06}}: the 2DFS of this pulsar shows a broad fluctuation
feature for which no significant preferred drift direction could be
detected (Fig. \ref{B1846-06}). The spectra look similar to that
observed at 21 cm.\\
{\bf {B1859+01}}: there is a broad feature without preferred drift
direction detected (Fig. \ref{B1859+01}). This pulsar was not observed
at 21 cm.\\
{\bf {B1907+10}}: a very clear long period feature is visible in the
LRFS and 2DFS (Fig. \ref{B1907+10}). This was not found to be
significant in the 21-cm data, although it is possibly there as
well (but weaker).\\
{\bf {B1910+20}}: the leading component of the pulse profile shows a
strong and vertically narrow feature in the LRFS (Fig. \ref{B1910+20})
and no significant preferred drift direction could be detected. This
feature appears also in the leading component of the 21-cm data,
although weaker.\\
{\bf {B1914+13}}: the spectra show a long period feature without a
preferred drift direction (Fig. \ref{B1914+13}). This modulation can
be seen by eye in the pulse-stack. At 21 cm this pulsar was also
observed to show a low-frequency excess, but it was not recognized as
a longitude stationary subpulse modulation because the modulation
appears to be weaker at high frequencies. \\
{\bf {B1946+35}}: the LRFS and 2DFS of this pulsar show a strong
low-frequency feature (Fig. \ref{B1946+35}). No significant offset
from the vertical axis has been detected in the 2DFS, confirming the
longitude stationary feature that was detected also at 21 cm.  At
21 cm a $P_3=33\pm2$ was measured, so the $P_3$'s appear to be different
at the two frequencies. However the observations were relatively
short, so possibly they were too short to obtain a good estimate for
the average $P_3$ value.\\
{\bf {B2327$-$20}}: a low-frequency excess has been measured at
21 cm. However at this frequency the feature appears to be
quasi-periodic (Fig. \ref{B2327-20}). This could be because the 92-cm
observation was longer and has a higher $S/N$. This feature is most
likely because of the nulls (\citealt{big92a}), which can clearly be
seen by eye in the pulse-stack. The the two vertical bands at $\pm100$
cpp (equals to $\pm\degrees{3.5}$) in the 2DFS are most likely related
to the component separation.\\

\subsection{Unconfirmed known drifters}

{\bf {B0037+56}}: the LRFS of this pulsar shows a low-frequency excess
(Fig. \ref{B0037+56}). The clear preferred drift direction discovered in
the 21-cm data is not confirmed, possibly because of a lower $S/N$ and
number of recorded pulses. \\
{\bf {B0138+59}}: a broad drift feature was found close to the
horizontal axis in the 21-cm data. In the 92-cm observation there is
no evidence for an offset from the vertical axis, but the
low-frequency excess is confirmed (Fig. \ref{B0138+59}). The $S/N$
should have been enough to confirm the drifting at 92 cm. It could be
that the drifting subpulses are very irregular and a longer
observation is required at 92 cm to confirm the drift feature.\\
{\bf {B0523+11}}: a weak drift feature has been discovered in the 21-cm
data, which could not be confirmed in the 92-cm data
(Fig. \ref{B0523+11}). Nevertheless, the 2DFS shows a hint that there
is a drift feature at this frequency as well, so the non-detection
could possibly be explained by a slightly lower $S/N$ and number of
pulses in the 92-cm data.  No drifting subpulses have been found at
430 MHz by \cite{bac81}.\\
{\bf {B0540+23}}: sporadic bursts of both positive and negative drift
have been reported by \cite{ash82} at 400 MHz and by \cite{now91} at
430 MHz. No preferred drift direction is detected in the 2DFS of this
pulsar (Fig. \ref{B0540+23}), consistent with the 21-cm data. At this
frequency we see short drift  bands with different
drift directions in the pulse-stack confirming the reported drifting
in the literature.  This is similar to the drifting subpulses reported in paper I.
Because this pulsar does not show a preferred drift
direction in our observations, this pulsar is not classified as a
drifter in our papers. This pulsar shows a clear low-frequency excess
at both frequencies. Notice the high modulation index, especially at
92 cm, caused by occasional bright pulses. \\
{\bf {B0609+37}}: the very clear and narrow drift feature discovered at
21 cm is not confirmed (Fig. \ref{B0609+37}). The $S/N$ of this
observation is a about twice as low, but that does not seem to
completely explain why the feature is not observed at this
frequency. With more than 10,000 pulses in this observation a
mode-change also does not seems to be a likely explanation. Although
an observation with a higher $S/N$ may reveal a drift feature at
92 cm, it seems that it must be less coherent than the feature found
in the 21-cm data to explain the current non-detection.\\
{\bf {B0611+22}}: our observation does not show any fluctuation features
in the spectra of this pulsar (Fig. \ref{B0611+22}), something that
has also been reported by \cite{brc75} at 430 MHz and is consistent
with the results  of paper I. Drifting with a periodicity $P_3=50-100P_0$
has been reported by \cite{fb80} at 430 MHz, who have analyzed
successive integrated pulse profiles. It is not clear if this kind of
drifting is related to subpulse drifting, because in their analysis
the subpulses are not directly measured. \\
{\bf {B1822$-$09}}: drifting subpulses and a correlation in the subpulse
modulation between the main- and interpulse have been observed for
this pulsar (e.g. \citealt{fwm81}).
The drifting subpulses as seen in the 21-cm data are not
confirmed. Possibly the $S/N$ of our observation is not good enough to
detect the very broad and weak drift feature, but it could also be
related to the profile evolution from a double morphology at high
frequencies to a more single morphology at low frequencies. As far as
we know drifting subpulses have only been reported at high frequencies
(\citealt{fwm81} at 1.620 GHz., \citealt{fw82} at 2.695 GHz,
\citealt{gjk+94} at 1.408 GHz -- 10.55 MHz). \\
{\bf {B1844$-$04}}: the narrow drift feature that was discovered in the
21-cm data is not confirmed at 92 cm (Fig. \ref{B1844-04}). The $S/N$
and the number of pulses are comparable at both frequencies. Because
of the relative low number of pulses it is possible that in the two
observations different drift-modes were dominating. Another
possibility is that it is related to the frequency dependence of the
profile morphology, because at 21 cm the drift-feature seems to be originating
from the trailing component and at 92 cm only one component is
observed. \\
{\bf {B1845$-$01}}: the drift feature reported in the 21-cm data and by
\cite{hw87} at 1414 MHz is not visible (Fig. \ref{B1845-01}), probably
because of a lower $S/N$ and the severe interstellar scattering. \\
{\bf {B1911$-$04}}: the LRFS of this pulsar (Fig. \ref{B1911-04}) shows
a low-frequency modulation (which is generated mainly in the trailing
part of the pulse profile), consistent with the 21-cm result. However,
at this frequency there is no evidence for a preferred drift
direction. Possibly this is related with the profile evolution from a
triple profile at 21 cm to a single profile at 92 cm.\\
{\bf {B1952+29}}: this pulsar was discovered to have drifting subpulses
in the 21-cm survey. There is an intensity modulation feature with the
same $P_3$ value at 92 cm (Fig. \ref{B1952+29}). Because the feature
in the 2DFS is not significantly offset from the vertical axis the
feature is classified as longitude stationary subpulse modulation at
this frequency. It is not clear if the modulation is really longitude
stationary at this frequency or if the drift direction is constantly
changing. The most clear detection of coherent drifting subpulses was
in the trailing part of the pulse profile at 21 cm.  Notice also that
the profile has evolved significantly with frequency. The component
which showed coherent drifting subpulses at 21 cm is at least much
weaker at 92 cm, but it is also possible that this component is not
visible at all.\\
{\bf {B2000+40}}: despite the presence of RFI, a clear drift feature was
discovered in the 21-cm data with a periodicity $P_3=0.4$ cpp. In the
92-cm data there is not much evidence for a similar feature
(Fig. \ref{B2000+40}). Maybe the non-detection at 92 cm is related to
the profile evolution with frequency in combination with the
non-linearity of the drift bands as observed at 21 cm. \\
{\bf {B2053+36}}: a broad clear drift feature was discovered at 21 cm,
which is not confirmed in the 92-cm data (Fig. \ref{B2053+36}). This
is surprising because the $S/N$ and the number of pulses should be
enough to detect the presence of drifting subpulses. Notice that also
the modulation index is much lower at 92 cm. Subpulse modulation
without a preferred drift direction has been reported for this pulsar
at 430 MHz by \cite{bac81}. The disappearance of the drift feature at
92 cm could be related to the frequency evolution of the profile.\\
{\bf {B2324+60}}: this pulsar showed a broad drift feature in its 2DFS
at the alias border in the 21-cm data and some drift bands could be
seen in the pulse-stack.  There was also a strong low-frequency
feature detected. Unfortunately, the $S/N$ of the 92-cm data is
insufficient to confirm any of this (Fig. \ref{B2324+60}).  \\
{\bf {J2346$-$0609}}: the 2DFS of the trailing component of the pulse
profile has a drift feature close to the $P_3=2P_0$ alias border in
the 21-cm data. There is probably some short period periodicity
modulation present in the trailing component in the 92-cm data,
although it is very weak (Fig. \ref{J2346-0609}). Drifting
subpulses could possibly be confirmed at 92 cm in an observation with a higher
$S/N$.  There is a strong sharp longitude stationary low-frequency
modulation in the leading component, which was also found in the 21-cm
data. This pulsar is not included in the statistics because it is just
too weak to be part of the source-list.\\

\subsection{Pulsars with low-frequency modulation}

{\bf {B0011+47}}: a low-frequency excess was found in the 21-cm data,
which is confirmed in the 92-cm data (Fig. \ref{B0011+47}). Notice the
high modulation index. It seems that the low-frequency excess and the
high modulation index are both caused by the burst like behavior of
the emission. This pulsar is not a known nuller.\\
{\bf {B0355+54}}: the spectra of this pulsar (Fig. \ref{B0355+54}) show
a low-frequency excess, consistent with the results from the 21-cm
data. The explanation for the low-frequency excess and the high
modulation index is that the emission is burst like.\\
{\bf {B0402+61}}: the power in the spectra of this pulsar is dominated
by the modulation in the leading component and is concentrated toward
long periodicities at 92 cm (Fig. \ref{B0402+61}). At 21 cm the
modulation of the trailing component dominated and a long period
feature without preferred drift direction was found. The 2DFS of the
trailing component shows a hint of a similar feature at 92 cm, but
this is much weaker and not significant.\\
{\bf {B0740$-$28}}: the LRFS of this pulsar shows low-frequency
modulation (Fig. \ref{B0740-28}), which is stronger than was found in
the 21-cm data. \\
{\bf {B1600$-$27}}: the low-frequency excess (Fig. \ref{B1600-27}) was
not observed at 21 cm, not unlikely because the $S/N$ of this
observation is higher.\\
{\bf {B1620$-$09}}: the observation is contaminated by RFI, but
nevertheless there appears a weak low-frequency excess in its
LRFS. This pulsar was not observed at 21 cm.\\
{\bf {B1649$-$23}}: this pulsar shows short period modulation as well as
long period modulation in its trailing component
(Fig. \ref{B1649-23}). Similar features are possibly also seen in the
21-cm data, although it was not found to be convincingly detected.\\
{\bf {B1706$-$16}}: this pulsar has a strong low-frequency excess
(Fig. \ref{B1706-16}), which was also found at 21 cm. The
low-frequency modulation is most likely caused by nulling, which can
clearly be seen in the pulse-stack. No nulling has previously been reported for
this pulsar.\\
{\bf {B1758$-$03}}: the low-frequency excess peaks at $P_3\simeq60P_0$,
which is probably caused by the burst like behavior of the
emission. This pulsar is not a known nuller and was not observed at
21 cm.\\
{\bf {J1808$-$0813}}: there is a clear low-frequency excess
(Fig. \ref{J1808-0813}), which was also found in the 21-cm data. The
feature peaks at $P_3=150P_0$, while at 21 cm the feature peaked at
$P_3=90P_0$. Because the observations are relatively short, this
difference is not significant. The low-frequency excess is probably
related to nulling, which can be seen by eye in the pulse stack at
21 cm. This pulsar is not known as a nuller.\\
{\bf {B1826$-$17}}: the LRFS shows a (very weak) detection of a low
frequency excess (Fig. \ref{B1826-17}), consistent with the 21-cm
data. The profile is scatter broadened at 92 cm.\\
{\bf {B1842+14}}: subpulse modulation without a drift direction has been
detected by \cite{bac81} at 430 MHz, which is consistent with the
spectra of our observation which do not reveal any preferred
periodicity or drift direction (Fig. \ref{B1842+14}). The side panels
of the LRFS and 2DFS shows that the power is increasing towards the
horizontal axis, indicating that the pulsar shows a low-frequency
intensity modulation. In the 21-cm data only the modulation index could be
measured.\\
{\bf {B1848+12}}: this pulsar has a strong low-frequency excess
(Fig. \ref{B1848+12}), probably related to nulling. This pulsar was
not observed at 21 cm and is not known to be a nuller.\\
{\bf {B1907+00}}: the spectra show a low frequency excess
(Fig. \ref{B1907+00}). This was not found in the 21-cm data, possibly
because of a lower $S/N$.\\
{\bf {B2003$-$08}}: there is a low-frequency excess observed in the
all components (Fig. \ref{B2003-08}), which is probably
also present at 21 cm.\\
{\bf {B2106+44}}: the spectra of this pulsar were found to peak towards
low frequencies at 21 cm, which is confirmed at this frequency
(Fig. \ref{B2106+44}). There is no evidence for a $P_3\simeq20P_0$
(longitude stationary) feature as found in the 21-cm data. This could
be related to the lower $S/N$ at 92 cm.\\

\subsection{Pulsars with a flat fluctuation spectrum}

\begin{center}
\begin{tabular}[htb]{|l|l|l|l|}
\hline
{B0447$-$12} & {B0450$-$18} & {B0458+46} & {J0520$-$2553}\\
{B0531+21} & {B0559$-$05} & {B0626+24} & {B0643+80}\\
{B0727$-$18} & {B0906$-$17} & {B1254$-$10} & {B1322+83}\\
{J1603--2531} & {J1654$-$2713} & {B1717$-$16} & {B1726$-$00}\\
{J1732$-$1930} &  {B1745$-$12} & {B1749$-$28} & {J1758+3030}\\
{B1804$-$08} & {B1821+05} & {B1831$-$04} & {B1831$-$03}\\
{B1839+09}  & {B1845$-$19} & {B1851$-$14} & {B1900$-$06}\\
{B1902$-$01} & {B1907-03} & {B1907+02} &  {B1914+09}\\
{B1915+13} & {B1920+21} & {B1929+20} & {B1937$-$26}\\
{B1940$-$12} & {B1943$-$29} & {B1946$-$25} & {B2002+31}\\
{B2022+50} & {B2027+37} & {B2113+14} & {B2224+65}\\
{B2227+61} & {B2306+55} & {B2334+61} & \\
\hline
\end{tabular}
\end{center}
{\bf B0450$-$18}: the three components (of which the outer ones are
plotted in Fig. \ref{B0450-18}) show flat featureless spectra,
consistent with paper I. \\
{\bf B0531+21}: both the 2DFS of the main- and interpulse of the Crab
pulsar do not show any sign of drifting subpulses
(Fig. \ref{B0531+21}). A very large modulation index  is
measured ($m=5$), which is caused by its giant pulses
(\citealt{sr68}). Notice that the modulation index is lower at low
frequencies, most likely because the giant pulses are much more
scatter-broadened at low frequencies. Notice also that the precursor
(that peaks at a pulse longitude of \degrees{65}) has a much lower
modulation index, which is consistent with the interpretation that its
emission is the ``normal'' emission of the pulsar (\citealt{psk+06}).  This
pulsar is not included in the statistics because its emission is
clearly dominated by a different type of emission than that of the
``normal'' pulsars.\\
{\bf B0643+80}: the spectra are featureless
(Fig. \ref{B0643+80}). This pulsar was not observed at 21 cm. A
``burst'' in the central component has been reported by \cite{mms98}
at 102.5 MHz. In a 22 minute observation (1120 pulses) they find that central
component of the profile was nine times brighter than usual, but because
no single pulses were recorded, it was not clear how long the burst
lasted. If it was caused by a single pulse, it must have been $10^4$
times brighter than the average. We do not see any evidence for strong
pulses in our 92-cm observation. \\
{\bf B0727$-$18}: the spectra are featureless (Fig. \ref{B0727-18}).
Notice the high modulation index, which are caused by bright single
pulses which are intermittently emitted. This pulsar was not observed
at 21 cm.\\
{\bf J1603$-$2531}: the spectra are featureless
(Fig. \ref{J1603-2531}). Notice the high modulation index, which is
probably caused by a few bright single pulses that appear above the
noise-level.  This pulsar was not observed at 21 cm.\\
{\bf B1749$-$28}: this pulsar shows flat featureless spectra
(Fig. \ref{B1749-28}), consistent with  paper I and the
results of \citealt{th71}.\\
{\bf B1804$-$08}: a low-frequency excess was reported for the 21-cm
data, which is not confirmed at 92 cm (Fig. \ref{B1804-08}), possibly
because of the lower $S/N$ or the drastic profile evolution.\\
{\bf B1821+05}: subpulse modulation without a drift direction has been
reported by \cite{bac81} at 430 MHz. No clear features are seen in the
2DFS (Fig. \ref{B1821+05}), possibly because of the low $S/N$ of our
observation. At 21 cm there was no detection of a modulation index.\\
{\bf B1839+09}: subpulse modulation without any drift direction has been
detected by \cite{bac81} at 430 MHz. We also observe subpulse
modulation without any sign of drifting subpulses
(Fig. \ref{B1839+09}). The spectra of the 21-cm data were
featureless.\\
{\bf B1915+13}: no features are seen in the spectra of this pulsar by
\cite{brc75} at 430 MHz, which is also the case for our observation
(Fig. \ref{B1915+13}). In the 21-cm data the modulation index was not
measured, probably because of a too low $S/N$.\\
{\bf B1920+21}: the 2DFS and LRFS of this pulsar shows three sharp
features with 0.19, 0.24 and 0.28 cpp (Fig. \ref{B1920+21}). This is
most likely because the single pulses ``flicker''. The 0.24 cpp
feature is in the first half of the data and the other two in the last
half. The observation is too short to state that there exist
a discrete number of flickering-frequencies that are switched on and
off similar to a mode-change rather than that a continuous range of
flickering-frequencies that can be excited. At 21 cm the spectra are
featureless, probably because of a lower $S/N$.\\
{\bf B1937$-$26}: a broad feature was detected at the $P_3=2P_0$ alias
border at 21 cm, but the spectra are flat at 92 cm
(Fig. \ref{B1937-26}). The two observations have a comparable $S/N$.\\
{\bf B2022+50}: only the 2DFS of the mainpulse of this pulsar is
plotted, because no modulation is measured for the interpulse
(Fig. \ref{B2022+50}). A flat spectrum was found for the mainpulse in
the 21-cm data.\\
{\bf B2113+14}: the spectra are flat (Fig. \ref{B2113+14}). Subpulse
modulation without preferred drift-direction and nulling has been
reported by \cite{bac81}.  This pulsar was not observed at 21 cm.\\
{\bf B2334+61}: notice the very high modulation index of this pulsar
(Fig. \ref{B2334+61}). This is because this pulsar appears to
intermittently emit bright single pulses (or it is off most of the
time except a few single pulses). In the 21-cm data the modulation
index is lower.\\

\subsection{Pulsars without a measured modulation index}

\begin{center}
\begin{tabular}[htb]{|l|l|l|l|}
\hline
{B0410+69} & {B1552$-$23} & {B1648$-$17} & {B1657$-$13}\\
{B1709$-$15} & {B1732$-$02} & {J1744$-$2335} & {B1756$-$22}\\
{J1759$-$2922} & {B1821$-$19} & {B1822+00} & {J1823$-$0154}\\
{J1835$-$1106} & {J1837-0045} & {J1848$-$1414} & {J1852$-$2610}\\
{B1859+03} & {B1900+05} & {B1913+10} & {B1924+16}\\
{J2005$-$0020} & {B2011+38} & {B2148+52} & {J2248-0101}\\
\hline
\end{tabular}
\end{center}
{\bf B1648$-$17}: the vertical line in the LRFS at pulse longitude
\degrees{218} is most likely because of spikes of RFI in the data
(Fig. \ref{B1648-17}). Because we do not believe the measured
modulation is caused by subpulse modulation, this pulsar is not
included in the statistics. This pulsar was not observed at 21 cm.\\
{\bf B1821$-$19}: a low-frequency excess was found in the 21-cm data,
which is not confirmed at this frequency (Fig. \ref{B1821-19}). The
profile is severely scatter-broadened. It is likely that the measured
modulation index is a result of statistical fluctuations of the very
small available off-pulse intensity level (baseline) that is
subtracted from the pulses. Because we do not
believe the measured modulation is caused by subpulse modulation, this
pulsar is not included in the statistics.\\
{\bf B1822+00}: no modulation index could be measured
(Fig. \ref{B1822+00}), which is, with an upperlimit of 0.4, unusual.
This pulsar was not observed at 21 cm.\\
{\bf B1859+03}: a flat spectrum was measured for this pulsar in the
21-cm data.  Although we also measure a flat spectrum in the 92-cm
data (Fig. \ref{B1859+03}), it is probably related to the severely
scatter-broadened profile.  Therefore this pulsar is, similar to
pulsar B1821$-$19, not included in the statistics.\\
{\bf B2011+38}: no modulation index could be measured for this pulsar
(Fig. \ref{B2011+38}), probably because of a too low $S/N$. In the
21-cm data a long period longitude stationary feature has been
found.\\

\section{\label{SctCompTwofreq}Individual sources at the two frequencies}

One of the aims of this paper is to discuss the frequency dependence
of the drifting phenomenon. Is the drifting phenomenon broadband, that
is, if the pulsar shows drifting subpulses at one frequency, does it
also have drifting subpulses at the other frequency? A related
question is if drifting subpulses are more regular at a certain
frequency. Other questions, such as how similar the values of $P_3$
are at the two frequencies, will be adressed in the next section.  In
this section a comparison is made between the results on the
individual sources at the two frequencies.  This will summarise some
aspects of the more detailed description in the previous section.

In total there are {\NrPulsarsTwoFreq} pulsars which were observed at
both 92 and 21 cm. Considering the pulsars that have the most clear
and regular drifting subpulses: \NrPulsarsTwoFreqAndCoherent of them
are observed to have coherent drifting subpulses at at least one
frequency. Of these pulsars \NrPulsarsTwoFreqAndCoherentTwo were
classified at both frequencies as coherent drifters, but
\NrPulsarsTwoFreqAndCoherentOne pulsars were classified
differently. In two cases the different classification is related to
different dominating drift-modes in the two observations (PSRs
B2319+60 and B1819$-$22). This could be because the
observations at the two frequencies were recorded at different times,
which means that one of the drift-modes could just have been missed in
one of the observations. Another explanation could be that some
drift-modes are only visible at certain frequencies, while at the
same time no drifting subpulses are observable at other
frequencies. An example of the latter is PSR B0031$-$07 (\citealt{smk05,sms+07}).

Also PSRs B0149$-$16 and B2045$-$16 are observed to have drifting
subpulses at both frequencies, but most likely because of a higher
$S/N$ the drift-feature is found to be broader at 92 cm. The drifting
subpulses of PSR B1918+19 were only in the observation at one
frequency, which could just be because of a lack of $S/N$ at one
frequency, although this pulsar is also known to be a drift-mode
changer. In the case of PSR B0105+65 the drifting subpulses appear to
only be observable at one frequency, but it must be noted that
subpulse modulation with the same $P_3$ is observed at the other
frequency. This is also the case for PSR B1730$-$22, which moreover
has a severe profile evolution. In the case of PSR B2315+21 the
drifting subpulses are only observed in a very narrow pulse longitude
range, so again profile evolution could be responsible for the
non-detection at the other frequency. For B1844$-$04 and B2000+40 it
cannot be ruled out that the observations were too short to catch the
drift-mode (if they would turn out to have drift-mode changes), but it
could also be related to the profile evolution. The most convincing
evidence that drifting subpulses can in some cases only appear at one
frequency is found for PSR B0609+37, which does not have detectable
drifting subpulses in a long observation with only a slightly lower
$S/N$ at 92 cm.

It seems that the majority of the pulsars with coherent drifting
subpulses have similar drifting subpulses at both frequencies, at
least when the $S/N$ of the observations are similar. In most cases
when drifting subpulses are only observed at one frequency the profile
evolution could be responsible. For the diffuse drifters the picture
is similar. In total there are \NrPulsarsTwoFreqAndDiffuse pulsars
which are classified as a diffuse drifter at at least one frequency
and the majority of them ({\NrPulsarsTwoFreqAndDiffuseTwo}) are
identically classified at both frequencies. Moreover,
\NrPulsarsDiffuseCoherent of them were classified as coherent drifters
at one frequency (as discussed above) and \NrPulsarsDiffuseLong pulsar
is classified as a pulsar with longitude stationary subpulse
modulation. The remaining \NrPulsarsDiffuseNothing pulsars are only
found to have drifting subpulses at one frequency. This could in many
cases just be a $S/N$ issue, although interstellar scattering or
profile evolution could also be responsible in some cases.

It is clear that at least a large fraction of the pulsars with
drifting subpulses at one frequency have drifting subpulses at the
other frequency as well. It is difficult to say what
the exact fraction of pulsars is that have drifting subpulses at both
frequencies, because in many cases where drifting subpulses were
found at only one frequency it could be related to the $S/N$ and the
length of the observation. Longer observations are therefore required
for these pulsars to confirm with more confidence that they really
show drifting subpulses at only one frequency.

The drifting subpulse phenomenon appears to be in general a broadband
phenomenon in the sense that the chance is high to find drifting
subpulses at both frequencies. However it is also clear that the
details of the drifting subpulses can be different at both
frequencies. This is clear for PSR B2016+28, the only pulsar in our
sample which was observed simultaneously at both frequencies with the
WSRT, but also for instance PSRs B1642$-$03 and B2021+51 appear to
have clearly different drifting subpulse characteristics at the two
frequencies.

Because the subpulse modulation patterns are in general similar at
different frequencies it is possible to use the LRFS to identify which
component in the pulse profile corresponds to which at the other
frequency. Especially when the frequency evolution of the profile is
drastic, this can be a powerful tool to track components between
different frequencies. Two examples of this application of the LRFS
are PSRs B1857$-$26 and B1919+21.

A number of pulsars are found which intermittently emit single pulses
that have pulse energies of about, or even brighter than, ten times
the average pulse energy. These pulses are not true giant pulses
(\citealt{sr68}), because they have widths comparable with the width of
the pulse profile. Nevertheless the emission of these \NrBursters
pulsars (PSRs B0011+47, B0355+54, B0540+23, B0727$-$18, J1603$-$2531,
B1917+00, B2319+60 and B2334+61) could potentially be different from
that of normal pulsars and be related to that of PSR B0656+14 and the
RRATs (\citealt{wsr+06,wws+06}).

Another side result of our single pulse analysis is that we have
discovered nulls in a number of sources which were not known to null
before. These \NrNullers sources are: PSRs B0011+47, B1706$-$16,
B1738$-$08, B1758$-$03, J1808$-$0813, B1819$-$22, B1848+12 and
B2110+27. Also one more pulsar has been discovered that shows a pulse
phase step: PSR J0459$-$0210.  This means that this pulsar has highly
non-linear driftbands with an almost discontinuous step in
subpulse-phase. This is possibly caused by destructive interference
between two drifting subpulse patterns (\citealt{esv03}).

\section{\label{Statistics92}Statistics}

\subsection{The numbers}

This survey has revealed another {\NrNewDrifters} new
drifters to go with those discovered in the 21-cm survey. 
For \NrDrifters pulsars out of the \NrPulsars that are included in the
92-cm source-list drifting subpulses are detected, which is about one
detection per three pulsars. This number is similar to that found at
21 cm, which is maybe not that surprising because {\NrPulsarsTwoFreq}
of these pulsars were also observed in the 21-cm survey. This means
that a considerable fraction of the source-lists at both frequencies
are overlapping, and as discussed in the previous section the chance
of detecting drifting subpulses at both frequencies is high.
As discussed in paper I, it is not surprising that drifting is not
always seen, even if drifting would be an intrinsic property of the
emission mechanism.

\begin{figure}[t]
\begin{center}
\rotatebox{270}{\resizebox{!}{0.94\hsize}{\includegraphics[angle=0]{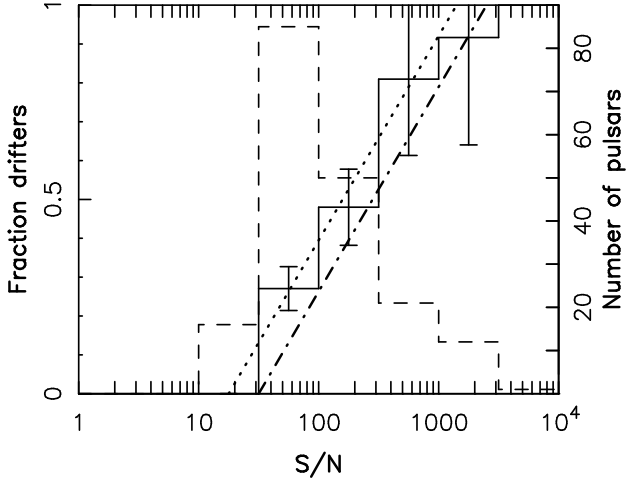}}}
\end{center}
\caption{\label{s2n_hist92} The solid line shows the fraction of
pulsars we observe to exhibit the drifting phenomenon versus the
measured $S/N$ ratio of the observation. The dashed histogram shows
the total number of pulsars observed in each $S/N$ ratio bin.  The
root-mean-square (RMS) is calculated as an estimate for the error (if
the bin contains more than one observation). The dotted line is a fit
for the $S/N$ dependence of the chance to detect drifting subpulses
and the dash-dotted line is that of paper I.}
\end{figure}

In Fig. \ref{s2n_hist92} the fraction of pulsars that show drifting
subpulses is plotted versus the integrated $S/N$ of the
observation. It can be seen that the chance of detecting drifting
subpulses is  correlated with the $S/N$. The dash-dotted line
is the correlation found for the 21-cm survey, which is similar to the
observed relation at 92 cm (dotted line). High $S/N$ observations
could in principle be biased toward well studied or long period
pulsars. However, very similar correlations are found when
observations containing more than 3000 pulses or the pulsars with
periods longer than 0.5 seconds are excluded. Therefore this
correlation seems to be a real observational bias in the sense that
drifting subpulses can only be significantly detected in a low $S/N$
observation when the drifting is reasonably coherent (e.g. PSR
J1650$-$1654).

If there is any significant difference in the fraction of drifters
found at the two different frequencies, it would be that at 92 cm the
chance of detecting drifting subpulses for a given $S/N$ is somewhat
higher than at 21 cm (see Fig. \ref{s2n_hist92}). This could in
principle be because drifting subpulses are more likely to exist at
low frequencies, or they may be more regular or more intense at low
frequencies. From the observations there is not much evidence that
there are more coherent drifters at 92 cm than at 21 cm, but we will
argue that there is some additional evidence that it is easier to
detect drifting subpulses at 92 cm (Sect. \ref{SctDisc}).

To reduce the effect of low $S/N$ observations on the statistics, the
observations with the lowest $S/N$'s were excluded from further
analysis. Because it turns out that the correlation between the chance
to detect drifting subpulses and the $S/N$ ratio of the observation is
very similar at both frequencies, the same threshold as in paper I is
applied. Only observations with a $S/N$ higher than 100 are included
in the statistics. Of the \NrPulsars observed pulsars \NrPulsarsSNR
have a high enough $S/N$ to be included in the statistics, which is
slightly lower than at 21 cm. Of these pulsars \NrDriftersSNR
({\DriftPercentageSNR}\%) show drifting subpulses and \NrCandidatesSNR
({\CandidatePercentageSNR}\%) show longitude stationary
modulation. These fractions are slightly higher than those found at 21
cm, so our conclusion that at least half of the pulsars have drifting
subpulses seems to be justified also at 92 cm.  

Scatter-broadening of the pulses is very likely to reduce the
probability to detect drifting subpulses when the amount of broadening
is comparable or even larger than the subpulse separation $P_2$. The
amount of scatter broadening increases rapidly with decreasing
observing frequency $f$. For instance the scattering time is expected
to be proportional with $f^{-4}$ in the thin screen model
(\citealt{sch68}), so scattering was expected to potentially be a
severe problem at 92 cm. However it turns out that of the sources that
have enough $S/N$ to be included in the statistics, only PSRs
B1821$-$19 and B1859+03 are severely scatter broadened. So by
excluding these two pulsars from further analysis the statistics are
not much affected by the effect of scatter-broadening.

\subsection{The drifting phenomenon and the $P$-$\dot P$ diagram}

\begin{figure*}[tb]
\begin{center}
\rotatebox{270}{\resizebox{0.57\hsize}{!}{\includegraphics[angle=0]{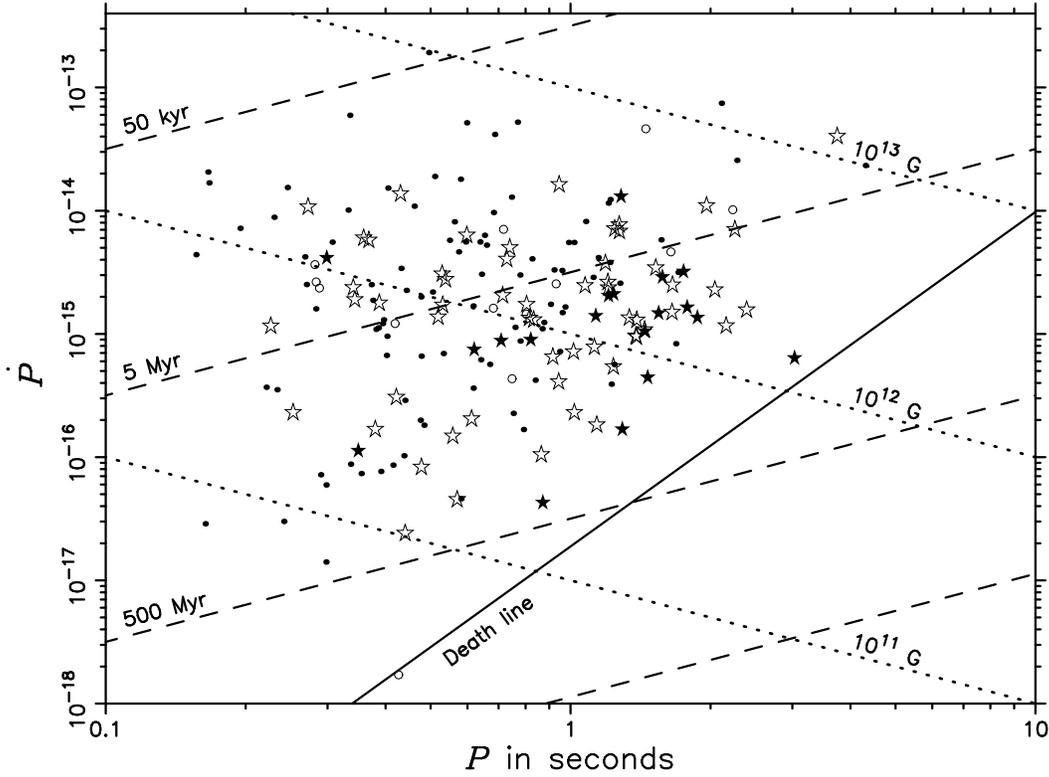}}}
\caption{\label{PPdot92}The $P$-$\dot P$ diagram of the analysed pulsars  at 92 cm
(including the low $S/N$ observations). The pulsars without drifting
subpulses are the dots, the diffuse (Dif and Dif$^\ast$) drifters are
the open stars, the coherent drifters are the filled stars and the
pulsars showing longitude stationary subpulse modulation are the open
circles. Lines of equal surface magnetic field strength and
characteristic ages are plotted, as well as a death line
(\citealt{cr93a}).}
\end{center}
\end{figure*}

\begin{figure*}[t]
\begin{center}
\rotatebox{270}{\resizebox{!}{0.45\hsize}{\includegraphics[angle=0]{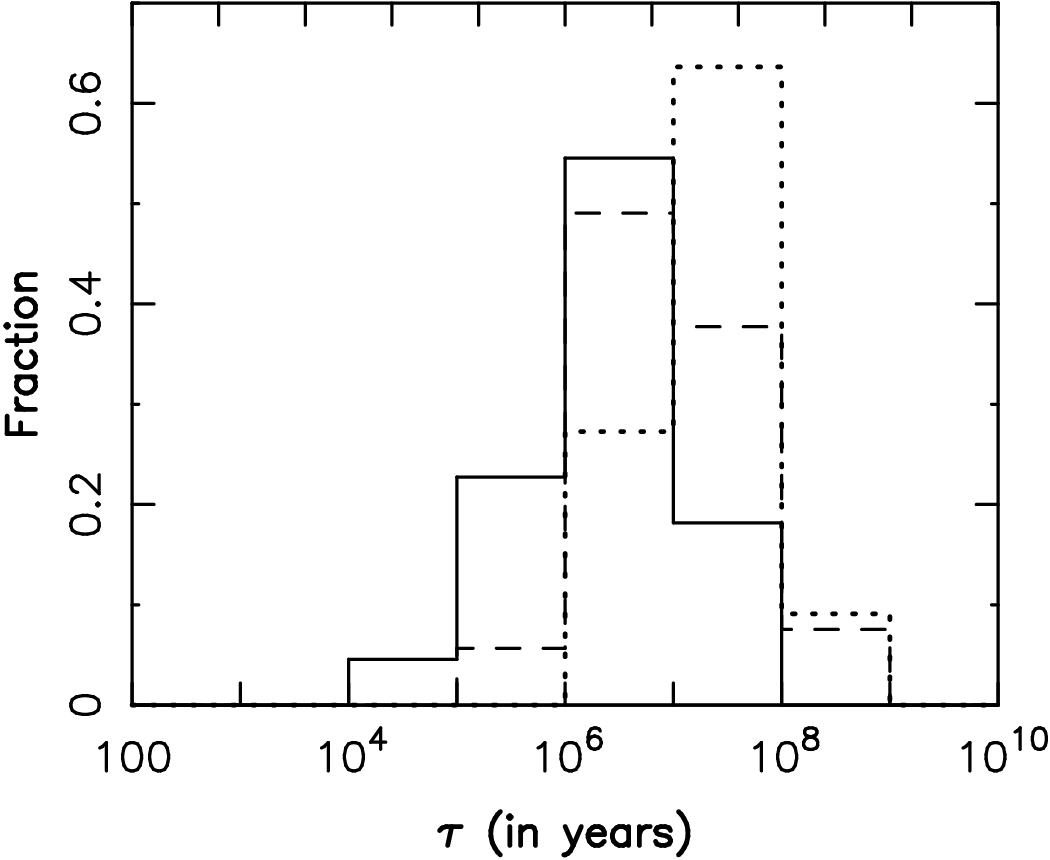}}}\hspace{0.05\hsize}
\rotatebox{270}{\resizebox{!}{0.45\hsize}{\includegraphics[angle=0]{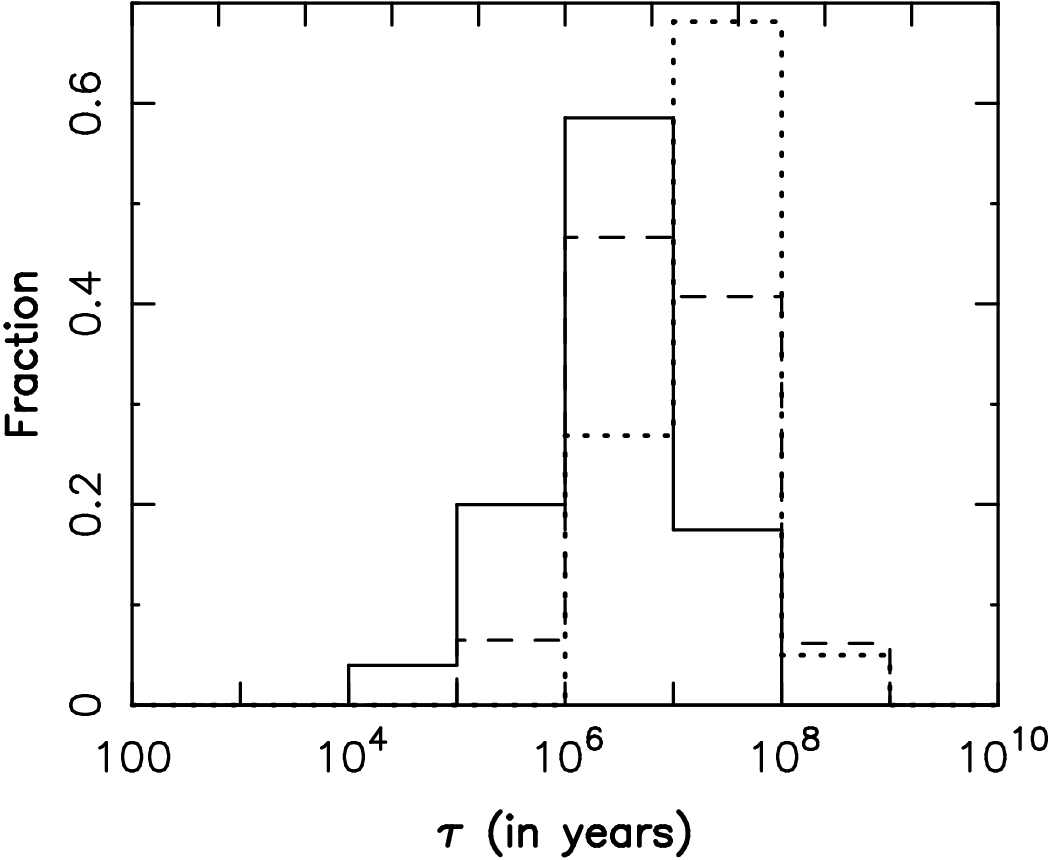}}}
\end{center}
\caption{\label{age_hist92}The left panel shows the histogram of the
characteristic ages of the analyzed pulsars with a S/N $\geq100$.  The
solid line is the age distribution of pulsars without drifting
subpulses, the dashed line shows all the drifters and the dotted line
shows the coherent drifters. The right panel shows the ``$S/N$ versus
age bias'' corrected histogram.}
\end{figure*}

Many of the important physical parameters of pulsars are derived from
the pulse period and its time derivative $\dot{P}$ (spin-down
parameter). It is therefore useful to look for correlations in the
$P-\dot{P}$ diagram. All the pulsars analysed at 92 cm are plotted in
Fig. \ref{PPdot92} and the coherent drifters, diffuse drifters and
pulsars with a longitude stationary subpulse modulation have different
symbols. This figure looks qualitatively similar to that in paper I
and this may not be that surprising as a considerable
fraction of the source-lists at both frequencies are overlapping.

As seen at 21 cm, the different groups of pulsars occupy a large
fraction of the $P-\dot{P}$ diagram and are clearly overlapping each
other. This implies that there is no strict correlation between the
drifting phenomenon and $P$, $\dot{P}$ or any combination of them.
Nevertheless, there appears to be a weak trend that pulsars with
drifting subpulses, especially the coherent drifters, are located
closer to the death-line. This would confirm the correlation reported
in paper I. 

To confirm this trend the characteristic age distributions (where
$\tau=\frac{1}{2}P/\dot{P}$) are plotted in the left panel of
Fig. \ref{age_hist92}. Indeed the pulsars with drifting subpulses are
on average older, as has also been reported by \cite{ash82}. Moreover
the pulsars with coherent drifting subpulses are even older. This
correlation suggests that there is an evolutionary trend that the
youngest pulsars have the most disordered subpulses and that the
subpulses become more and more organized into drifting subpulses as
the pulsar ages. We will come back to the interpretation of this trend
in the summary and conclusions (Sect. \ref{SumAndConclusions}).

To determine the significance of this trend the KS-test
(Kolmogorov-Smirnov test) is used, which calculates how likely it is
that two distributions are drawn from an identical
distribution.  There is only a \AgeDriftNonDriftperc
chance that the age distribution of the pulsars with and without
drifting subpulses are the same.  This is too high to state that this
trend is significant.
 There is only a \AgeCohNondriftperc chance that the age
distribution of the coherent drifters and the pulsars without drifting
subpulses are the same, which therefore appear to be significantly
different.  There is only a \AgeCohNoCohperc
chance that the age distribution of the drifters with and without
coherent drifting subpulses are the same.

The same trend is found when the low $S/N$ pulsars are included in the
statistics. We checked if the trend could be explained by a difference
in the $S/N$ of the observations in different age bins. As noted in
Sect. \ref{SctSourceList}, such bias can be expected if both the age
of the pulsar and $T_\mathrm{sky}$ are correlated with the galactic
latitude.  The $S/N$ versus age bias corrected age distributions are
plotted in the right panel of Fig. \ref{age_hist92} and they are
qualitatively the same as the left panel, giving us confidence in the
correlation found (the procedure used is described in paper I).

Because the drifting subpulse phenomenon appears to be broadband, it
is useful to combine the sample of pulsars at both frequencies to
increase the significance of the statistics. For the sources that were
observed at two frequencies the observation with the highest $S/N$ was
used. According to the KS-test there is only a
\AgeDriftNonDriftpercCombined chance that the age distribution of the
pulsars with and without drifting subpulses are the same and this
chance is \AgeCohNoCohpercCombined for the age distribution of the
coherent drifters and the other pulsars.  The difference between the
age distribution of the pulsars with and without drifting subpulses is
significant, while the different age distribution of the coherent
drifters and the other drifters is not significant. Nevertheless the
data at both frequencies is consistent with each other.  Because the
sample of pulsars at the two frequencies are partially overlapping,
the 92-cm data cannot independently confirm the trend observed at 21
cm.

Another interesting quantity that can be derived from the $P-\dot{P}$
diagram is the surface magnetic field strength
($B_s=10^{12}\sqrt{10^{15}P\dot P}$ Gauss).  There is a \BNonDriftperc
chance that the magnetic field strength distribution of the pulsars
with and without drifting subpulses are the same and this chance is
{\BNonCohpers} for the magnetic field strength distribution of the
pulsars with coherent drifting subpulses and those without drifting
subpulses.  This means that, as was found in paper I, there is no
evidence that the surface magnetic field strength distribution is
significantly different for the pulsars with and without drifting
subpulses.

\subsection{The drifting phenomenon and the modulation index}

It is clear that the modulation index is a parameter that is closely
related with the drifting phenomenon, because drifting subpulses imply
an intensity modulation. However, it is somewhat arbitrary how {\em
the} modulation should be defined, because the longitude-resolved
modulation index is in most cases highly dependent on pulse
longitude. Following paper I (and \citealt{jg03}) the statistics are
calculated using the minimum in the modulation index profile. The
motivation is that the minimum in the modulation index profile is
often found at the pulse longitude of a maximum in the intensity
profile. This means that for instance the longitude-averaged
modulation index will depend on the $S/N$ of the observation, because
a higher $S/N$ allows the detection of a modulation index farther away
from the peaks in the pulse profile where the modulation index tends
to be higher.

\begin{figure}[tb]
\begin{center}
\rotatebox{270}{\resizebox{!}{0.94\hsize}{\includegraphics[angle=0]{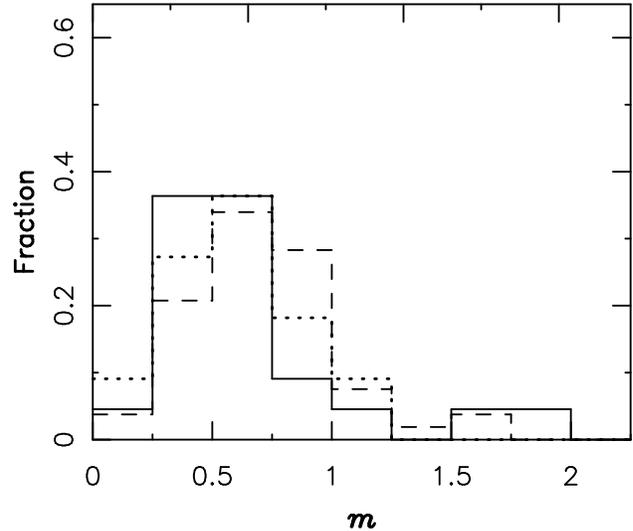}}}
\end{center}
\caption{\label{mod_hist92} The modulation index distribution of the
pulsars without drifting subpulses (solid line), with drifting
subpulses (dashed line) and with coherent drifting subpulses (dotted
line). Only observations with a $S/N\geq 100$ are included this
plot.}
\end{figure}

The modulation index distributions are shown in
Fig. \ref{mod_hist92} for the pulsars with and without drifting
subpulses and those with coherently drifting subpulses. 
There is no indication that the distributions are significantly
different, although the modulation index of the pulsars with drifting
subpulses tend to be higher. Interesting, in the 21-cm data there was
an indication that the pulsars with coherent drifting subpulses
appeared to have a lower modulation index (however not statistically
significant). We will come back to this point in 
Sect. \ref{SctDisc}.

\begin{figure}[tb]
\begin{center}
\rotatebox{270}{\resizebox{!}{0.94\hsize}{\includegraphics[angle=0]{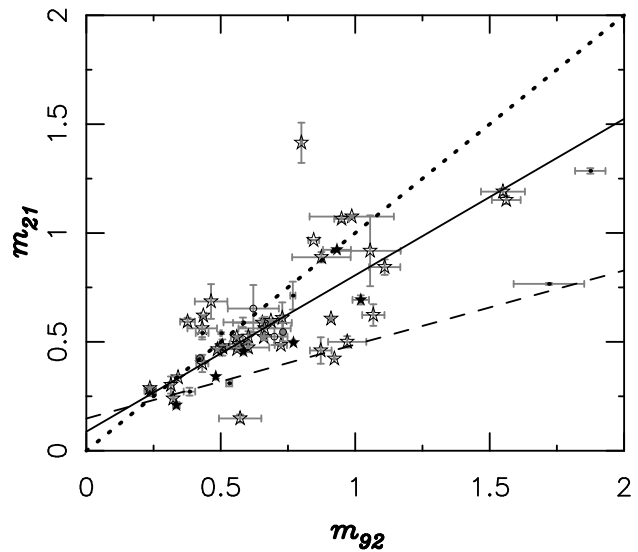}}}
\end{center}
\caption{\label{mod_mod}The modulation index measured at 21 cm versus
that measured at 92 cm. The points would lie on the dotted line when
the modulation index is frequency independent. Only observations with
a $S/N\geq 100$ are included this plot. The dashed and solid line are
straight line fits through the data-points with and without weighting
for the errorbars. The symbols are identical to those in
Fig. \ref{PPdot92}.}
\end{figure}

It is also interesting to compare the measured modulation indices at
92 and 21 cm. In Fig. \ref{mod_mod} the minimum modulation index of
each pulsar measured at 21 cm are plotted versus the minimum
modulation index measured at 92 cm. Only the pulsars that are observed
at both frequencies with a $S/N\geq 100$ are included in this
plot. Moreover, all the observations which had a modulation index at
one frequency that was below the detection threshold at the other
frequency were not included. This means that all the pulsars included
in this plot had the $S/N$ to measure an identical minimum in the
modulation index profile at both frequencies.

In Fig. \ref{mod_mod} it can be seen that the modulation indices
measured at the two frequencies are not independent, but they are
clearly correlated.  If the modulation index was independent of
observing frequency the points would be scattered symmetrically around
the dotted line $m_{21}=m_{92}$.  However, there are more pulsars with
a higher modulation index at 92 cm than visa versa. It is also
interesting that there are more pulsars which have large ($>1$)
modulation indices at 92 cm than at 21 cm.

To confirm the trend, and to check its significance, a straight line
fit was made through the data. This is done by minimizing the $\chi^2$
using the methethod described by \cite{ptv+92}, which incorporates the
measurement errors of both coordinates. The dashed line shows the best
fit through the data-points by weighting them by their measurement
errors (dashed line: $m_{21}=0.34\pm0.03 \times m_{92} +
0.15\pm0.03$). However, including the measurement errors heavily
biases the correlation to a few high $S/N$ observations, such as PSR
B0329+54 with $m_{92}=0.9$ and $m_{21}=0.4$.  To avoid the best fit
being dominated by a few high $S/N$ observations, the fitting is also
done by weighting the data-points equally. This fit confirms that the
modulation index at 92 cm is on average higher than at 21 cm (solid
line: $m_{21}=0.72\pm0.08 \times m_{92} + 0.09\pm0.06$), although the
scatter around the correlation is large.

\subsection{\label{SectComplexity}Complexity parameter}

One way to try to distinguish between emission theories is to try to
find correlations between the modulation index and the so-called
complexity parameters. The idea is that the modulation index will be
low when there are many overlapping subpulses in a single pulse,
because then the intensity will also vary less from pulse to
pulse. The number of subpulses per pulse depends, for instance, for
the sparking gap model on the number of sparks on the polar gap, which
can be quantified by a complexity parameter (\citealt{gs00}). In the
case of the sparking gap model the complexity parameter is the polar
cap radius $r_p$ divided by the gap height $h$.  The
complexity parameter is a model dependent function of $P$ and
$\dot{P}$ and the modulation index should have an anti-correlation
with this function (\citealt{jg03}).

Following \cite{jg03} and paper I the correlations between the
measured modulation indices and four complexity parameters are
calculated. These parameters correspond to the sparking gap model
($a_1$; \citealt{gs00}), continuous current outflow instabilities
($a_2$; \citealt{as79}; \citealt{ha01a}), surface magnetohydrodynamic
wave instabilities ($a_3$; \citealt{lou01e}) and outer magnetospheric
instabilities ($a_4$; \citealt{jg03}) respectively.  As noted in paper
I, $a_2$ is also proportional to the acceleration parameter, the total
current outflow from the polar cap, the square root of the spin down
energy loss rate and roughly to the circulation time of the sparks
expressed in pulse periods. Physically, $a_3$ and $a_4$ are
proportional to the magnetic field strength at the surface and at the
light cylinder respectively.

\begin{figure*}[tb]
\begin{center}
\rotatebox{270}{\resizebox{!}{0.7\hsize}{\includegraphics[angle=0]{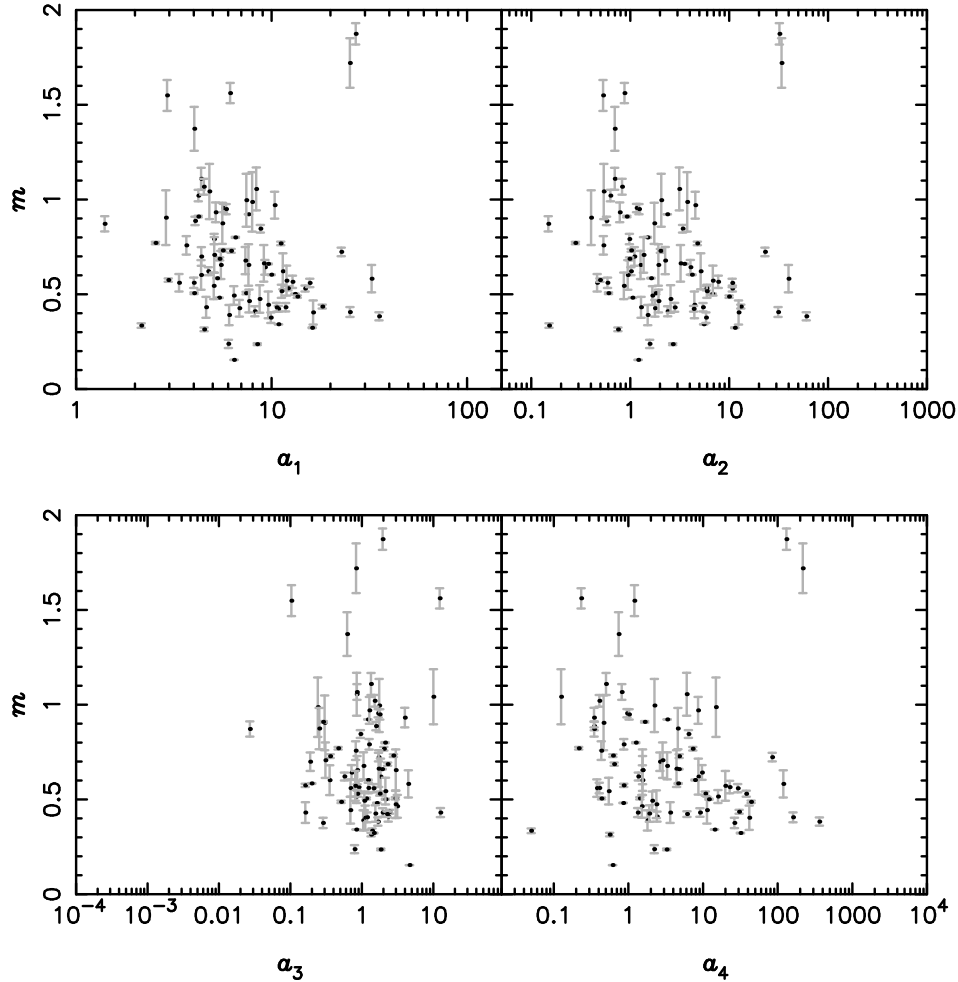}}}
\end{center}
\caption{\label{Complexity92}The modulation index for all analyzed
pulsars with a $S/N\geq 100$ versus the four complexity parameters as
described in the text.}
\end{figure*}

\begin{figure}[tb]
\begin{center}
\rotatebox{270}{\resizebox{!}{0.99\hsize}{\includegraphics[angle=0]{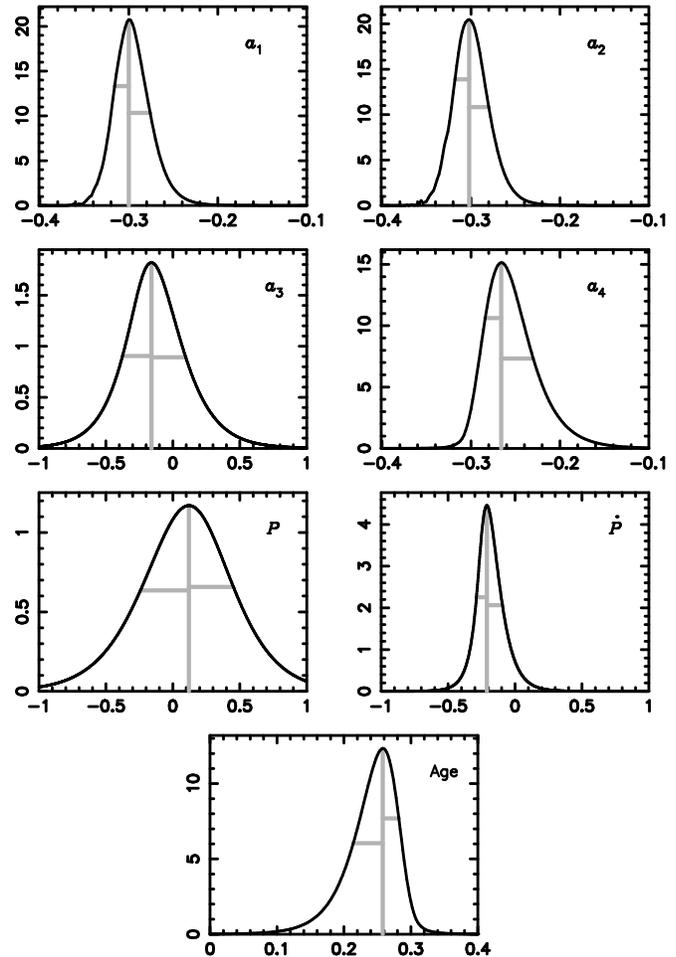}}}
\end{center}
\caption{\label{Probability92}The probability functions $P(\rho)$ of the
correlation coefficient $\rho$ for the four complexity parameters, the
pulse period, the spin-down parameter and the characteristic age of
the pulsar. The position of the maximum and the 1-$\sigma$ widths are
indicated by the gray lines.}
\end{figure}

 Because the modulation index appears to be frequency dependent, we
did not combine the sample of pulsars at the two frequencies. In
Fig. \ref{Complexity92} the modulation index is plotted versus the
four complexity parameters for the 92-cm data. In contrast to the
corresponding plot in paper I, only pulsars with a $S/N\geq 100$ are
included. This means that the measurement uncertainties are relatively
small on the data-points, which makes any correlation found more
convincing. There is a hint of an anti-correlation for each of the
four complexity parameters, although the correlation with $a_3$
appears to be the weakest. To determine the significance of these
trends we followed the procedure of paper I.  This involves the
calculation of the Spearman rank-ordered correlation coefficient
$\rho$ and its significance parameter $\Delta$ in a Monte Carlo
approach. This results in a probability distribution $P(\rho)$, which
is plotted in Fig. \ref{Probability92}.  In Table
\ref{Probability92Table} the measured location and widths of the peaks
of the probability functions are tabulated. For details of the method
used we refer to paper I\footnote{The normalization of the probability
functions in Fig. 11 of paper I is incorrect, but the values in Table
1 and therefore all the conclusions of paper I are unaltered.}.

\begin{table}[!tb]
\begin{center}
{
\begin{tabular}{llr@{$\;$}lr@{$\;$}l}
\hline
\hline
Parameter & Function & \multicolumn{2}{c}{$\rho$ ($S/N\geq 100$)} & \multicolumn{2}{c}{$\rho$ (all)}\\
\hline
& & & & \\[-8pt]
$a_1=r_\mathrm{p}/h$ & $5\frac{\displaystyle(\dot P/10^{-15})^{2/7}}{\displaystyle (P/1\mathrm{s})^{-9/14}}$ & $-0.30\!$&$^{+0.02}_{-0.02}$ & $-0.30\!$&$^{+0.02}_{-0.01}$\\[10pt]
$a_2\propto B_\mathrm{s}/P^2$ & $\sqrt{\dot PP^{-3}}$                    & $-0.30\!$&$^{+0.02}_{-0.02}$ & $-0.31\!$&$^{+0.02}_{-0.02}$\\[3pt]
$a_3\propto B_\mathrm{s}$ & $\sqrt{P\dot P}$                             & $-0.2\!$&$^{+0.3}_{-0.2}$    & $0.0\!$&$^{+0.5}_{-0.5}$\\[3pt]
$a_4\propto B_\mathrm{lc}$ & $\sqrt{\dot P/P^{-5}}$                      & $-0.27\!$&$^{+0.04}_{-0.02}$ & $-0.29\!$&$^{+0.02}_{-0.01}$\\[3pt]
$P$ & $P$                                                                & $0.1\!$& $^{+0.3}_{-0.4}$    & $0.21\!$&$^{+0.01}_{-0.03}$\\[3pt]
$\dot P$ & $\dot P$                                                      & $-0.21\!$&$^{+0.12}_{-0.08}$  & $-0.2\!$&$^{+0.1}_{-0.1}$\\[3pt]
Age & $\frac{1}{2}P/\dot{P}$                                             & $0.26\!$&$^{+0.02}_{-0.04}$  & $0.22\!$&$^{+0.03}_{-0.02}$\\[3pt]
\hline
\end{tabular}
}
\end{center}
\caption{\label{Probability92Table}The correlation coefficients $\rho$
and their significance as derived from Fig. \ref{Probability92} for all the pulsars (right column) and for only pulsars with a $S/N\geq100$.}
\end{table}

Based on a sample of 12 pulsars \cite{jg03} found the highest
anti-correlation for the sparking gap model ($a_1$) and that the
surface magnetohydrodynamic wave instabilities ($a_3$) is unlikely.
In the 21-cm data the modulation index was found to be both
consistent with being anti-correlated, as well as being uncorrelated,
with any of the four complexity parameters. It was therefore not
expected to find any strong correlations in the 92-cm data. However,
at this frequency the modulation index appears to be much more
strongly anti-correlated with all four complexity parameters. All four
complexity parameters are consistent with being anti-correlated with
the data and therefore none of the corresponding models can be ruled
out.  However, it must be noted the data provide little support for
the surface magnetohydrodynamic wave instabilities, as only marginal
anti-correlation is seen for $a_3$.

To further investigate the significance of these correlations we
excluded all the pulsars with $m>1$, as they are possibly forming a
separate group. None of the correlations seems to be
influenced much by excluding these points. It turns out that including
all the low $S/N$ pulsars also does not make much difference, except
that the significance of the positive correlation between the
modulation index and $P$ becomes stronger.

It should be noted that the modulation index is also likely to be
affected by the viewing geometry. Better results are expected when
only pulsars are included that are known to emit core emission
(\citealt{jg03}). Because not all the pulsars in our source-list have
an established classification into core or conal emission we have
included all pulsars with a measured modulation index in our
analysis. However, the modulation index is chosen to be the minimum in
the longitude-resolved modulation index (as in \citealt{jg03}), which
should give the best estimate for the modulation index of the core
component if present.

A difficulty in the interpretation of the correlations between the
modulation index and the complexity parameters is that it is assumed
that the (longitude-resolved) modulation arises solely because of the
movement of the subbeams. However the reality is in many cases more
difficult, because the subpulses also have intrinsic fluctuating
intensities. This additional component in the modulation index could
in principle have a very different (or none) correlation with $P$ and
$\dot{P}$ than that of the complexity parameter. It
would therefore be important to disentangle these different effects
on the modulation index. Also it should  also be noted
that the whole concept of an anti-correlation between the modulation
index with the complexity parameter only  is predicted
when there is a significant broadening in the mapping from the polar
gap to the radiation beam pattern. If this is not the case the
subpulses cannot overlap, hence there is no anti-correlation expected.

It is intriguing that the correlations with the modulation index are
much more constrained at 92 cm than at 21 cm. There are no obvious
observational biases we can think of that could explain this. For
instance, excluding the ms-pulsars from the 21-cm data (which were
excluded in this 92-cm survey) does not lead to different results. If
only the sources are considered which were observed with a $S/N\geq
100$ at both frequencies the results are again very similar, although
the significance of the correlations go down which is probably simply
related to the decreased number of data-points.  Therefore there is no
reason to believe these correlations are not real, although they are
weak with a large scatter (Fig. \ref{Complexity92}).

\subsection{\label{PropDriftSct}Properties of drift behavior}

\begin{figure}[tb]
\begin{center}
\rotatebox{270}{\resizebox{!}{0.94\hsize}{\includegraphics[angle=0]{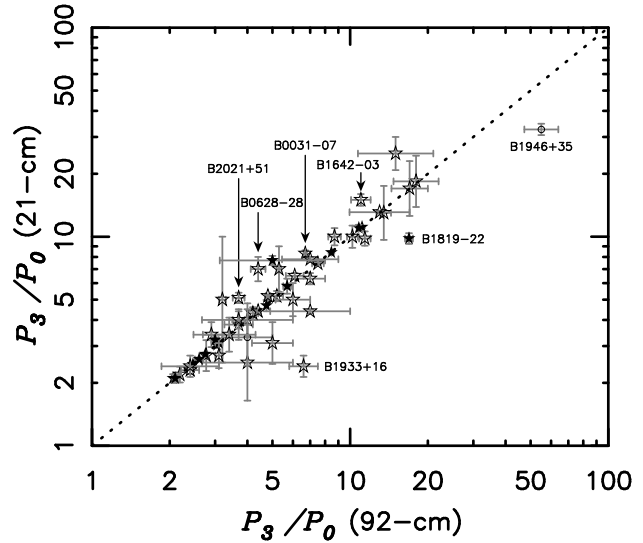}}}
\end{center}
\caption{\label{p3p3} The measured value of the vertical drift band
separation $P_3$ compared at the two frequencies. Also the low-$S/N$
observations are included.  The diffuse (Dif and Dif$^\ast$) drifters
are the open stars, the coherent drifters are the filled stars and the
pulsars with longitude stationary subpulse modulation are the open
circles. The dotted line shows the expected correlation when $P_3$ is
independent of frequency.}
\end{figure}

It is generally thought that the value of $P_3$ is independent of the
observing frequency, while the value of $P_2$ could vary
(e.g. \citealt{es03b}). All pulsars with a measured $P_3$ at two
frequencies are compared in Fig. \ref{p3p3} to test  the
absence of a dependency on the observing frequency of $P_3$. The
correlation is indeed  extremely tight 
and it is important to note that this correlation also applies for the
pulsars with a diffuse drift feature. This correlation confirms the
report for nine pulsars by \cite{nuwk82}.  Moreover, many points that
do not fall on the  correlation can be explained. In
the case of PSRs B0031$-$07 and B1819$-$22 drift-modes with different
$P_3$ values dominate at the two frequencies. For others, such as PSRs
B1738$-$08 and B1946+35 it seems  not unlikely that
longer observations will reveal that the $P_3$ values are consistent
at the two frequencies.  The different values for $P_3$ of
B0628$-$28 at the two frequencies are because in the 21-cm observation
the two differing values of $P_3$ in the leading and trailing halves
of the profile were not separated.  There seems to be only three
cases in which there is evidence for frequency dependence of $P_3$:
PSRs B1642$-$03, B1933+16 and B2021+51. It would be interesting to
have simultaneous observations at multiple frequencies to
 really proof if they exhibit a frequency dependent
drifting subpulse pattern.

If the pulse profile is double peaked and there are drifting subpulse
then the $P_3$ values measured for both components are in many cases
very similar.  This can be expected when the drifting
subpulses in both components share a common physical origin.
Nevertheless there are also some clear exceptions to this rule, for
example PSRs B0628$-$28, B0751+32, B1508+55, J1901$-$0906 and
B2016+28. Therefore this behavior cannot be explained
in the framework of the sparking gap model by a cut of the line of
sight through a single carousel of subbeams.

\begin{figure}[tb]
\begin{center}
\rotatebox{270}{\resizebox{!}{0.94\hsize}{\includegraphics[angle=0]{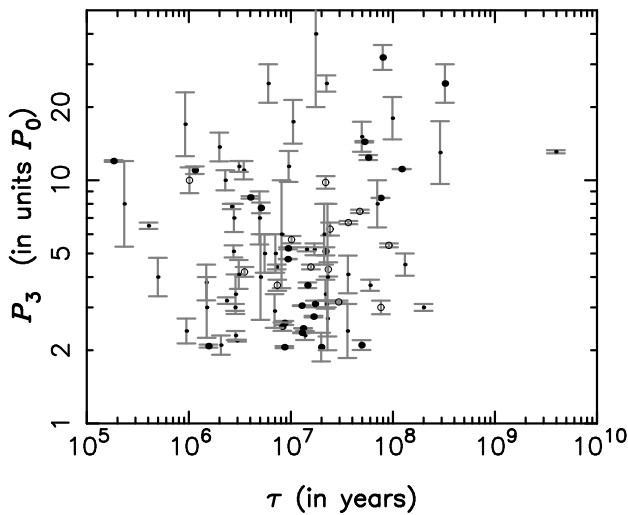}}}
\end{center}
\caption{\label{p3age92}The measured values of the vertical drift band
separation $P_3$ versus the pulsar age of all the pulsars showing the
drifting phenomenon. The coherent drifters are the filled circles, the
Dif drifters (with drift feature clearly separated from the alias
borders) are the open circles and the Dif$^\ast$ drifters are the
small dots. This figure contains both 92 and 21-cm observations. }
\end{figure}

In the 21-cm data there was no obvious correlation between $P_3$ and
the pulsar age contrary to reports in the past
(\citealt{wol80,ash82,ran86}). Because we find that $P_3$ is, at least
in the majority of the pulsars, highly correlated between the two
frequencies, the observations at the two frequencies are combined in
Fig. \ref{p3age92}. If the pulsar was observed at both frequencies,
the observation with the highest $S/N$ was used. It can be seen that
there is, as was found for the 21-cm data, no evidence for any
correlation between $P_3$ and the characteristic age. Possibly, such a
correlation would be most pronounced for pulsars with dominating conal
emission. Because no distinction in emission type has been made, the
correlation could be less precise. However the absence of a
correlation with age in this large sample of pulsars with drifting
subpulses (\NrDriftersCombinedSample) suggests that if such a
correlation would exist, it must be a very weak correlation. Also, the
evidence for a pulsar subpopulation located close to the $P_3=2P_0$
Nyquist limit (\citealt{wri03,ran86}) is weak. No correlations were
reported in paper I between $P_3$ and the magnetic field strength
(contrary to what was reported by \citealt{wol80} and
\citealt{ash82}), the pulse period (consistent with \citealt{wol80},
contrary to the tendency reported by \citealt{bac73}).  In the
combined sample of 21 and 92-cm observations there is also no evidence
for any of these correlations with $P_3$.

While $P_3$ appears to be highly independent of the observing
frequency, the value of $P_2$ is expected, and found, to vary with
frequency (e.g. \citealt{es03b}) and measuring a value for $P_2$ can
be far from trivial (e.g. \citealt{es03c}). Moreover, $P_2$ is only a
meaningful parameter if the drift bands are linear. Also, when the
pulsar only occasionally shows drifting subpulses the value measured
for $P_2$ will depend on the fraction of time the pulsar spent in the
drift-mode. This is because when the pulsar does not show drifting
subpulses most of the time, there will be more power in the 2DFS along
the vertical axis, which will make $P_2$ larger. Nevertheless the
measured values at the two frequencies appear to be roughly correlated
(not plotted). As reported for the 21-cm data, there
is no correlation between the drift direction and the pulsar spin-down
(contrary to what was reported by \citealt{rl75,bac81,ash82}) and
there is no difference between the number of pulsars with positive and
negative drifting subpulses.

\section{\label{SctDisc}The quasi-steady emission component}

\cite{nuwk82} concluded, based on a multi-frequency study of nine
pulsars, that it is a common feature of pulsars to have a quasi-steady
component in their emission which becomes stronger with increasing
frequency. This confirmed the idea of \cite{cor75}, who found that the
drifting subpulse signal of PSR B1919+21 becomes relatively weaker at
higher observing frequencies. A similar effect is observed for PSR
B0320+39 by \cite{iks82} at 102 and 406 MHz and \cite{es03c} at 328
and 1380 MHz. Other examples are PSR B0809+74 (\citealt{es03c} at 328
and 1380 MHz) and PSR B0031$-$07 (\citealt{sms+07} at 243, 607, 840
and 4850 MHz). This effect can also be seen very clearly by comparing
the side-panels of the spectra of PSR B0809+74 in this paper and those
in paper I.

A pure sinusoidal drifting subpulse signal would have a modulation
index $m = 1/\sqrt{2}$ and subpulse patterns with different waveforms
or drift band shapes will generally have larger modulation
indices. 
However, a considerable fraction of the pulsars with drifting
subpulses have a modulation index lower than 0.75 at 92 cm (see
Fig. \ref{mod_hist92}) and this fraction appears to be even larger in
the 21-cm data (see Fig. 8 of paper I). This confirms the idea that
there is a quasi-steady component in the emission of pulsars which is
relatively strong at higher frequencies. This interpretation is also
consistent with
Fig. \ref{mod_mod}, which shows the trend that the modulation index
is, on average, lower at 21 cm.

This picture may also explain why the chance of detecting drifting
subpulses is higher at low frequencies, independent of $S/N$
(Fig. \ref{s2n_hist92}). This again suggests that the drifting subpulse
signal is relatively stronger at lower frequencies. The drifting
subpulse signal appears to be, on average, stronger for older pulsars
(see Fig. 4 and the corresponding plot in paper I). It is therefore
expected that older pulsars have a higher modulation index and this is
confirmed by the positive correlation between the modulation index
and the characteristic age (Fig. \ref{Probability92}).

So why was the correlation between the modulation index and the age of
the pulsar not found for the 21-cm data? This can possibly be
explained because the quasi-steady component of the emission is not
entirely unmodulated, which means that there is a limit on how low the
modulation index can become by increasing the strength of the
quasi-steady component. Therefore the modulation index of especially
the pulsars with a strong drifting subpulse signal (which tend to be
older) will be suppressed by the strengthening of the quasi-steady
component with frequencies. This means that the (positive) correlation
between the modulation index and age is expected to be weaker at high
frequencies, as is observed.  

There are a few pulsars known which show an almost discontinuous step
in subpulse-phase in the middle of the profile. The modulation index
is very low at the pulse longitude of the step, which can be
understood in terms of destructive interference between two drifting
subpulse patterns (\citealt{esv03}). As pointed out in paper I, if
especially the coherent drifters can have highly non-linear
driftbands, then those pulsars will have an on average lower
modulation index. This trend was indeed found in the 21-cm data,
although it was not proved to be significant. Both PSRs B0320+39 and
PSR B0809+74 show such phase-steps at high frequencies
(\citealt{es03c}) and also the phase step found for PSR B2255+58 in
the 21-cm data is not present at low-frequencies. If phase-steps are
in general more likely to occur at high frequencies, than it is
expected that the modulation indices of the coherent drifters are only
found to be on average low at high frequencies. Indeed, there is not
much evidence for such a correlation in the 92-cm data
(Fig. \ref{mod_hist92}).

It must be noted that the interpretations made in this section are
based on correlations that are only weak and often not proved to be
significant. Nevertheless it is encouraging that it is possible to
explain the trends with concepts that are, at least from an
observational point of view, not unfounded.

Observations at higher frequencies are thought to correspond to lower
emission heights and this would imply that the emission beam is
narrower at higher frequencies (radius-to-frequency mapping;
\citealt{cor78}). If the components of pulse profiles are therefore
more overlapping at higher frequencies, then also the subpulses could
be expected to overlap more and this would imply a lower modulation
index (i.e. a stronger quasi-steady component) at higher
frequencies. This effect could also be related to refraction of plasma
waves caused by plasma density gradients in the pulsar
magnetosphere. The effect of refraction is stronger at higher
frequencies (e.g. \citealt{pet00,wsv+03,fq04}).  Additionally, if
refraction smears out the subpulses in time, the
subpulses will overlap more. Refraction could therefore be (partially)
responsible for the increasing quasi-steady component at higher
frequencies.

\section{\label{SumAndConclusions}Summary and conclusions}

One of the main points of paper I was that drifting subpulses are at
the very least a common phenomenon for radio pulsars, if not an
intrinsic property of the emission mechanism. This conclusion is
 further supported by the 92-cm data. For
{\NrNewDrifters} pulsars drifting subpulses are discovered for the
first time in the 92-cm data and {\NrConfirmedNewDrifters} of the new
drifters found in the 21-cm data are confirmed. Drifting subpulses are
detected in total for \NrDrifters pulsars out of the \NrPulsars and it
is estimated that at least half of the total population of pulsars
will show drifting subpulses when observations with high enough $S/N$
would be available. This fraction is similar to that found at 21 cm,
which is not surprising because a considerable fraction of the
source-lists at both frequencies are overlapping ({\NrPulsarsTwoFreq}
pulsars) and we find that the chance of detecting drifting subpulses
at both frequencies is high, indicating that the drifting phenomenon
is in general broadband. The fraction of pulsars with drifting
subpulses is possibly even slightly higher at 92 cm than at 21 cm and
the majority of the pulsars show subpulse modulation.

It is not expected that all pulsars show drifting subpulses.  This
could simply be because some of the observations had a lower than
expected $S/N$, because of unfortunate interstellar scintillation, a
higher sky temperature than the average over the whole sky, severe RFI
during the observation, or because the signal of the pulsar was weaker
than expected from the flux quoted in the database used.  But also if
line of sight cuts the magnetic pole centrally, longitude stationary
subpulse modulation is expected. Refractive distortion in the pulsar
magnetosphere or nulling will disrupt the drift bands, making it
difficult or even impossible to detect drifting subpulses. Also the
$P_3$ values of some pulsars could be very large, or the fraction of
time the pulsar spends in a drift mode could be very
short. There are also pulsars (such as B0540+23) for
which sporadic, regular bursts of drifting subpulses can be seen by
eye in the pulse-stack, but because the apparent drift direction and
rate is very different each time they are not found to have a
significant preferred drift direction. In order to understand drifting
subpulses better, it will be important to not only study the pulsars
with highly regular drifting subpulses, but also those systems that
show very variable drifting subpulses properties.

Because drifting subpulses are at the very least a common
phenomenon for radio pulsars, which implies that the physical
conditions required for the emission mechanism and the drifting
mechanism to work are similar. This is consistent with the absence of
a correlation with the surface magnetic field strength. 

The measured values of $P_3$ at the two frequencies are
highly correlated and the same is true for pulsars with drifting
subpulses and double peaked intensity profiles. These correlations are
expected when the drifting subpulses share a common physical origin,
such as expected for instance in the framework of the sparking gap
model. Therefore it will be a great challenge for theoretical models
to explain the pulsars that deviate from these correlations.

A correlation between $P_3$ and other pulsar parameters is expected if 
the drift rate depends on any physical parameters of the pulsar and the
strongest correlation is expected to be found when $P_2$ is
 identical for different pulsars. Such a correlation
would be a very important observational restriction on pulsar emission
models. However, there are no such correlations found. As explained in
paper I this could suggest that many pulsars in our sample are aliased
or that $P_2$ is highly variable from pulsars to pulsar.

In the sparking gap model, drifting subpulses are expected for conal
emission and therefore for pulsars with an on average higher
modulation index. 
As noted in  paper I, drifting subpulses are found for pulsars
classified as ``core single stars'' (for instance PSR B2255+58;
\citealt{ran93b}). This stresses the importance of being unbiased on
pulsar type when studying the drifting phenomenon.

Our sample of pulsars is not biased on pulsar type or any particular
pulsar characteristics, which allows us to do meaningful statistics on
the drifting phenomenon. There is a weak trend found that pulsars with
drifting subpulses are on average older, especially the coherent
drifters, confirming the trend found at 21 cm. This correlation
suggests that there is an evolutionary trend such that the youngest
pulsars have the most disordered subpulses and that the subpulses
become more and more organized into drifting subpulses when the pulsar
ages. This trend does not necessarily imply a direct link between
the true age of the pulsar and the drifting phenomenon. It could well
be that the characteristic age is correlated to some other physical
quantity, which is correlated with the drifting subpulse mechanism.
This could be some electrodynamical quantity in the polar gap of the
pulsar, but it could also be explained by the evolution of the angle
between the magnetic axis and the rotation axis or the evolution of
the pulse morphology. In the non-radial pulsations model
(\citealt{cr04}) this trend can also be explained, because the
appearance of narrow drifting subpulses is favored in pulsars with an
aligned magnetic axis.  As discussed in paper I, this trend cannot
explained by nulling. 

The coherent drifters appear to have low modulation indices at high
frequencies.  This trend can be explained if especially the coherent
drifters show subpulse-phase steps, which are related to low
modulation indices and appear to be more common at high frequencies.

The modulation indices measured at the two frequencies are clearly
correlated, although they tend to be higher at low
frequencies. The modulation index also appears to be
correlated with three of the four complexity parameters and with the
characteristic age of the pulsar. Such correlations were not found in
the 21-cm data.  This is consistent with the picture in
which the radio emission can be divided into a drifting subpulse
signal plus a  quasi-steady component which becomes
stronger at high observing frequencies.

\begin{acknowledgements}
The authors would like to express their gratitude to
the staff of the WSRT for their flexibility with scheduling and
assisting with the observations. The Westerbork Synthesis Radio
Telescope is operated by the ASTRON (Netherlands Foundation for
Research in Astronomy) with support from NWO.
\end{acknowledgements}

\clearpage

\begin{table*}[!ht]
\begin{center}
\begin{tabular}{lc|rc|rr|r@{}lrr@{$\;$}lr@{}l|l}
\hline
\hline
Pulsar & Class  & $P_0$ (s) & $\dot P$ & Pulses & $S/N$\hspace{1.0mm} &  $m$ & & $m_\mathrm{thresh}$ & $P_2$ & (deg)& $P_3$ & $\;(P_0)$&  Figure\\
\hline
B0011+47 &    & 1.2407 & $ 5.6\times10^{-16}$ & 2565 & 88  & 1.51$\pm$&0.04 & 0.35 & & & & &  \ref{B0011+47}\\
B0031$-$07 & Dif  & 0.9430 & $ 4.1\times10^{-16}$ & 7666 & 887  & 1.4$\pm$&0.1 & 0.12 & $ -21.9\!$&$^{+0.2}_{-0.1}$ & $ 6.7$&$\pm0.1$ &  \ref{B0031-07}\\
B0037+56 &    & 1.1182 & $ 2.9\times10^{-15}$ & 1555 & 48  & 0.73$\pm$&0.05 & 0.47 & & & & &  \ref{B0037+56}\\
B0105+65 & Coh  & 1.2837 & $ 1.3\times10^{-14}$ & 1363 & 99  & 0.49$\pm$&0.04 & 0.33 & $ -13\!$&$^{+2}_{-4}$ & $ 2.08$&$\pm0.03$ &  \ref{B0105+65}\\
B0136+57 & Dif$^\ast$  & 0.2725 & $ 1.1\times10^{-14}$ & 6303 & 201  & 0.72$\pm$&0.02 & 0.24 & $ -35\!$&$^{+5}_{-15}$ & $ 6.1$&$\pm0.5$ &  \ref{B0136+57}\\
B0138+59 &    & 1.2229 & $ 3.9\times10^{-16}$ & 1904 & 147  & 0.56$\pm$&0.05 & 0.27 & & & & &  \ref{B0138+59}\\
B0148$-$06 & Coh  & 1.4647 & $ 4.4\times10^{-16}$ & 1180 & 77  & 0.9$\pm$&0.1 & 0.44 & $ -14.0\!$&$^{+0.6}_{-0.5}$ & $ 14.4$&$\pm0.1$ &  \ref{B0148-06}\\
 &  &    & & &  & & &  & $ -36\!$&$^{+6}_{-2}$ & $ 14.51$&$\pm0.04$ &  \\
B0149$-$16 & Dif  & 0.8327 & $ 1.3\times10^{-15}$ & 1026 & 190  & 0.39$\pm$&0.05 & 0.17 & $ -13\!$&$^{+2}_{-4}$ & $ 5.7$&$\pm0.2$ &  \ref{B0149-16}\\
 &  &    & & &  & & &  & $ -14\!$&$^{+4}_{-10}$ & $ 5.6$&$\pm0.4$ &  \\
B0301+19 & Dif$^\ast$  & 1.3876 & $ 1.3\times10^{-15}$ & 2541 & 221  & 1.11$\pm$&0.06 & 0.18 & $ -25\!$&$^{+4}_{-3}$ & $ 5.2$&$\pm0.3$ &  \ref{B0301+19}\\
 &  &    & & &  & & &  & $ -25\!$&$^{+6}_{-12}$ & $ 4.5$&$\pm0.4$ &  \\
B0320+39 & Coh  & 3.0321 & $ 6.4\times10^{-16}$ & 3545 & 477  & 0.33$\pm$&0.01 & 0.11 & $ 6.4\!$&$^{+0.2}_{-0.3}$ & $ 8.46$&$\pm0.01$ &  \ref{B0320+39}\\
B0329+54 & Dif$^\ast$  & 0.7145 & $ 2.0\times10^{-15}$ & 16379 & 8005  & 0.9222$\pm$&0.0001 & 0.002 & $ 300\!$&$^{+150}_{-50}$ & $ 4$&$\pm2$ &  \ref{B0329+54}\\
 &  &    & & &  & & &  & $ 60\!$&$^{+8}_{-2}$ & $ 6$&$\pm1$ &  \\
B0355+54 &    & 0.1564 & $ 4.4\times10^{-15}$ & 18897 & 230  & 1.7$\pm$&0.1 & 0.26 & & & & &  \ref{B0355+54}\\
B0402+61 &    & 0.5946 & $ 5.6\times10^{-15}$ & 8995 & 181  & 0.62$\pm$&0.09 & 0.30 & & & & &  \ref{B0402+61}\\
B0410+69 &    & 0.3907 & $ 7.7\times10^{-17}$ & 2965 & 50  & & & 0.42 & & & & &  \ref{B0410+69}\\
J0421$-$0345 & Dif  & 2.1613 & $ 1.2\times10^{-15}$ & 1090 & 51  & 0.91$\pm$&0.03 & 0.36 & $ 5\!$&$^{+3}_{-1}$ & $ 3.16$&$\pm0.04$ &  \ref{J0421-0345}\\
B0447$-$12 &    & 0.4380 & $ 1.0\times10^{-16}$ & 4010 & 60  & 0.7$\pm$&0.1 & 0.62 & & & & &  \ref{B0447-12}\\
B0450+55 & Dif$^\ast$  & 0.3407 & $ 2.4\times10^{-15}$ & 5161 & 124  & 0.57$\pm$&0.03 & 0.30 & $ -170\!$&$^{+15}_{-100}$ & $ 8.7$&$\pm0.5$ &  \ref{B0450+55}\\
B0450$-$18 &    & 0.5489 & $ 5.8\times10^{-15}$ & 1560 & 273  & 0.53$\pm$&0.01 & 0.15 & & & & &  \ref{B0450-18}\\
B0458+46 &    & 0.6386 & $ 5.6\times10^{-15}$ & 2750 & 53  & 0.66$\pm$&0.08 & 0.60 & & & & &  \ref{B0458+46}\\
J0459$-$0210 & Coh  & 1.1331 & $ 1.4\times10^{-15}$ & 3123 & 129  & 0.79$\pm$&0.03 & 0.24 & $ -9\!$&$^{+1}_{-3}$ & $ 2.36$&$\pm0.01$ &  \ref{J0459-0210}\\
J0520$-$2553 &    & 0.2416 & $ 3.0\times10^{-17}$ & 7275 & 38  & 0.69$\pm$&0.09 & 0.61 & & & & &  \ref{J0520-2553}\\
B0523+11 &    & 0.3544 & $ 7.4\times10^{-17}$ & 4948 & 142  & 0.43$\pm$&0.05 & 0.26 & & & & &  \ref{B0523+11}\\
B0525+21 & Dif$^\ast$  & 3.7455 & $ 4.0\times10^{-14}$ & 949 & 333  & 1.56$\pm$&0.05 & 0.12 & $ -70\!$&$^{+10}_{-10}$ & $ 3.7$&$\pm0.5$ &  \ref{B0525+21}\\
B0531+21 &    & 0.0331 & $ 4.2\times10^{-13}$ & 106150 & 983  & 0.36$\pm$&0.02 & 0.21 & & & & &  \ref{B0531+21}\\
B0540+23 &    & 0.2460 & $ 1.5\times10^{-14}$ & 3481 & 111  & 1.87$\pm$&0.06 & 0.31 & & & & &  \ref{B0540+23}\\
B0559$-$05 &    & 0.3960 & $ 1.3\times10^{-15}$ & 2168 & 58  & 0.6$\pm$&0.1 & 0.54 & & & & &  \ref{B0559-05}\\
B0609+37 &    & 0.2980 & $ 5.9\times10^{-17}$ & 11926 & 64  & 0.62$\pm$&0.07 & 0.53 & & & & &  \ref{B0609+37}\\
B0611+22 &    & 0.3350 & $ 5.9\times10^{-14}$ & 2504 & 112  & 0.58$\pm$&0.07 & 0.22 & & & & &  \ref{B0611+22}\\
B0626+24 &    & 0.4766 & $ 2.0\times10^{-15}$ & 1791 & 89  & 0.43$\pm$&0.04 & 0.30 & & & & &  \ref{B0626+24}\\
B0628$-$28 & Dif$^\ast$  & 1.2444 & $ 7.1\times10^{-15}$ & 1397 & 566  & 0.7$\pm$&0.1 & 0.09 & ?& & $ 10$&$\pm3$ &  \ref{B0628-28}\\
 &  &    & & &  & & &  & $ 70\!$&$^{+80}_{-20}$ & $ 4.4$&$\pm0.3$ &  \\
B0643+80 &    & 1.2144 & $ 3.8\times10^{-15}$ & 2749 & 42  & 0.53$\pm$&0.05 & 0.49 & & & & &  \ref{B0643+80}\\
B0727$-$18 &    & 0.5102 & $ 1.9\times10^{-14}$ & 2822 & 54  & 0.93$\pm$&0.05 & 0.46 & & & & &  \ref{B0727-18}\\
B0740$-$28 &    & 0.1668 & $ 1.7\times10^{-14}$ & 4317 & 336  & 0.38$\pm$&0.02 & 0.15 & & & & &  \ref{B0740-28}\\
B0751+32 & Dif$^\ast$  & 1.4423 & $ 1.1\times10^{-15}$ & 2445 & 84  & 1.2$\pm$&0.1 & 0.28 & $ -90\!$&$^{+20}_{-30}$ & $ 6$&$\pm2$ &  \ref{B0751+32}\\
 &  &    & & &  & & &  & ?& & $ 2.5$&$\pm0.7$ &  \\
B0756$-$15 & Lon  & 0.6823 & $ 1.6\times10^{-15}$ & 5212 & 101  & 0.68$\pm$&0.07 & 0.39 & ?& & $ 20$&$\pm5$ &  \ref{B0756-15}\\
B0809+74 & Coh  & 1.2922 & $ 1.7\times10^{-16}$ & 15984 & 3107  & 0.77$\pm$&0.01 & 0.04 & $ -13.2\!$&$^{+0.1}_{-0.7}$ & $ 11.12$&$\pm0.01$ &  \ref{B0809+74}\\
B0818$-$13 & Coh  & 1.2381 & $ 2.1\times10^{-15}$ & 8686 & 1935  & 0.482$\pm$&0.001 & 0.04 & $ -5.1\!$&$^{+0.1}_{-0.6}$ & $ 4.74$&$\pm0.01$ &  \ref{B0818-13}\\
B0820+02 & Dif$^\ast$  & 0.8649 & $ 1.0\times10^{-16}$ & 15905 & 903  & 0.9$\pm$&0.1 & 0.11 & $ 10\!$&$^{+2}_{-1}$ & $ 4.5$&$\pm0.5$ &  \ref{B0820+02}\\
B0823+26 & Dif$^\ast$  & 0.5307 & $ 1.7\times10^{-15}$ & 1617 & 865  & 0.85$\pm$&0.02 & 0.03 & $ 70\!$&$^{+10}_{-12}$ & $ 5.3$&$\pm0.1$ &  \ref{B0823+26}\\
B0834+06 & Dif$^\ast$  & 1.2738 & $ 6.8\times10^{-15}$ & 2314 & 898  & 0.51$\pm$&0.01 & 0.08 & $ -20\!$&$^{+5}_{-35}$ & $ 2.19$&$\pm0.02$ &  \ref{B0834+06}\\
 &  &    & & &  & & &  & ?& & $ 2.17$&$\pm0.02$ &  \\
B0906$-$17 &    & 0.4016 & $ 6.7\times10^{-16}$ & 2132 & 86  & 0.79$\pm$&0.03 & 0.39 & & & & &  \ref{B0906-17}\\
B0919+06 & Dif$^\ast$  & 0.4306 & $ 1.4\times10^{-14}$ & 1988 & 292  & 0.44$\pm$&0.01 & 0.15 & $ -200\!$&$^{+150}_{-200}$ & $ 4$&$\pm2$ &  \ref{B0919+06}\\
B0942$-$13 & Dif$^\ast$  & 0.5703 & $ 4.5\times10^{-17}$ & 1503 & 237  & 0.57$\pm$&0.01 & 0.14 & $ -1.5\!$&$^{+0.1}_{-6}$ & $ 3.0$&$\pm0.1$ &  \ref{B0942-13}\\
B0950+08 & Dif$^\ast$  & 0.2531 & $ 2.3\times10^{-16}$ & 16395 & 343  & 1.0$\pm$&0.2 & 0.26 & $ -500\!$&$^{+100}_{-300}$ & $ 6$&$\pm3$ &  \ref{B0950+08}\\
\end{tabular}
\end{center}
\caption{
\label{Table92}
The details of all the analysed pulsars. The classification of the pulsar in the second column, where ``Coh'' is a coherent drifter, ``Dif'' and ``Dif$^\ast$'' are diffuse drifters with or without drift features which are clearly separated from the alias borders and ``Lon'' are pulsars showing longitude stationary subpulse modulation. The next columns are the pulse period, its dimensionless time derivative, the number of pulses in the observation, the signal to noise ratio, the minimum in the longitude resolved modulation index, the minimum detectable modulation index, the horizontal and vertical driftband separation and the figure number.}
\end{table*}
\begin{table*}[!ht]
\begin{center}
\begin{tabular}{lc|rc|rr|r@{}lrr@{$\;$}lr@{}l|l}
\hline
\hline
Pulsar & Class  & $P_0$ (s) & $\dot P$ & Pulses & $S/N$\hspace{2.5mm} &  $m$& & $m_\mathrm{thresh}$ & $P_2$ & (deg)& $P_3$ & $\;(P_0)$ &  Figure\\
\hline
B1039$-$19 & Dif  & 1.3864 & $ 9.4\times10^{-16}$ & 1258 & 70  & 0.8$\pm$&0.1 & 0.46 & $ 50\!$&$^{+300}_{-20}$ & $ 4.2$&$\pm0.2$ &  \ref{B1039-19}\\
 &  &    & & &  & & &  & $ 12\!$&$^{+20}_{-1}$ & $ 4.2$&$\pm0.2$ &  \\
B1112+50 & Dif$^\ast$  & 1.6564 & $ 2.5\times10^{-15}$ & 1055 & 55  & 2.1$\pm$&0.1 & 0.52 & $ 40\!$&$^{-10}_{+20}$ & $ 9$&$\pm5$ &  \ref{B1112+50}\\
B1133+16 & Dif$^\ast$  & 1.1879 & $ 3.7\times10^{-15}$ & 3999 & 2170  & 0.800$\pm$&0.003 & 0.04 & $ 170\!$&$^{+60}_{-50}$ & $ 4$&$\pm2$ &  \ref{B1133+16}\\
 &  &    & & &  & & &  & $ 120\!$&$^{+50}_{-20}$ & $ 30$&$\pm8$ &  \\
B1237+25 & Dif$^\ast$  & 1.3824 & $ 9.6\times10^{-16}$ & 5219 & 851  & 0.505$\pm$&0.005 & 0.05 & $ -33\!$&$^{+4}_{-1}$ & $ 2.77$&$\pm0.01$ &  \ref{B1237+25}\\
 &  &    & & &  & & &  & $ 37\!$&$^{+10}_{-2}$ & $ 2.77$&$\pm0.04$ &  \\
B1254$-$10 &    & 0.6173 & $ 3.6\times10^{-16}$ & 2811 & 58  & 0.91$\pm$&0.07 & 0.49 & & & & &  \ref{B1254-10}\\
B1322+83 &    & 0.6700 & $ 5.7\times10^{-16}$ & 2590 & 50  & 0.58$\pm$&0.08 & 0.58 & & & & &  \ref{B1322+83}\\
B1508+55 & Dif$^\ast$  & 0.7397 & $ 5.0\times10^{-15}$ & 10486 & 1949  & 0.660$\pm$&0.002 & 0.05 & $ -18\!$&$^{+2}_{-12}$ & $ 3.2$&$\pm0.1$ &  \ref{B1508+55}\\
 &  &    & & &  & & &  & $ -35\!$&$^{+2}_{-7}$ & $ 2.4$&$\pm0.1$ &  \\
B1530+27 & Dif$^\ast$  & 1.1248 & $ 7.8\times10^{-16}$ & 1543 & 65  & 1.50$\pm$&0.07 & 0.47 & $ -40\!$&$^{+10}_{-25}$ & $ 4$&$\pm4$ &  \ref{B1530+27}\\
B1540$-$06 & Coh  & 0.7091 & $ 8.8\times10^{-16}$ & 5130 & 588  & 0.24$\pm$&0.02 & 0.09 & $ 6.0\!$&$^{+3}_{-0.5}$ & $ 3.05$&$\pm0.01$ &  \ref{B1540-06}\\
B1541+09 & Lon  & 0.7484 & $ 4.3\times10^{-16}$ & 2349 & 131  & 0.62$\pm$&0.02 & 0.17 & ?& & $ 17$&$\pm4$ &  \ref{B1541+09}\\
B1552$-$23 &    & 0.5326 & $ 6.9\times10^{-16}$ & 4962 & 21  & & & 1.73 & & & & &  \ref{B1552-23}\\
B1600$-$27 &    & 0.7783 & $ 3.0\times10^{-15}$ & 1101 & 70  & 1.21$\pm$&0.05 & 0.39 & & & & &  \ref{B1600-27}\\
J1603$-$2531 &    & 0.2831 & $ 1.6\times10^{-15}$ & 12552 & 34  & 1.4$\pm$&0.1 & 0.99 & & & & &  \ref{J1603-2531}\\
B1604$-$00 & Dif$^\ast$  & 0.4218 & $ 3.1\times10^{-16}$ & 2225 & 361  & 0.73$\pm$&0.01 & 0.11 & $ 100\!$&$^{+10}_{-15}$ & $ 2.9$&$\pm0.5$ &  \ref{B1604-00}\\
B1607$-$13 & Dif$^\ast$  & 1.0184 & $ 2.3\times10^{-16}$ & 1710 & 37  & 1.2$\pm$&0.1 & 0.85 & $ -20\!$&$^{+5}_{-20}$ & $ 8$&$\pm2$ &  \ref{B1607-13}\\
B1612+07 & Dif$^\ast$  & 1.2068 & $ 2.4\times10^{-15}$ & 1455 & 113  & 0.96$\pm$&0.02 & 0.21 & $ -60\!$&$^{+40}_{-60}$ & $ 6$&$\pm4$ &  \ref{B1612+07}\\
B1620$-$09 &    & 1.2764 & $ 2.6\times10^{-15}$ & 1377 & 56  & 0.74$\pm$&0.05 & 0.40 & & & & &  \ref{B1620-09}\\
B1633+24 &    & 0.4905 & $ 1.2\times10^{-16}$ & 4806 & 74  & 0.65$\pm$&0.06 & 0.48 & $ -25\!$&$^{+8}_{-40}$ & $ 2.2$&$\pm0.1$ &  \ref{B1633+24}\\
B1642$-$03 & Dif$^\ast$  & 0.3877 & $ 1.8\times10^{-15}$ & 14587 & 2192  & 0.3413$\pm$&0.0005 & 0.02 & $ -70\!$&$^{+15}_{-30}$ & $ 11$&$\pm1$ &  \ref{B1642-03}\\
B1648$-$17 &    & 0.9734 & $ 3.0\times10^{-15}$ & 1804 & 59  &  & & 0.41 & & & & &  \ref{B1648-17}\\
B1649$-$23 &    & 1.7037 & $ 3.2\times10^{-15}$ & 1030 & 65  & 0.94$\pm$&0.04 & 0.41 & & & & &  \ref{B1649-23}\\
J1650$-$1654 & Coh  & 1.7496 & $ 3.2\times10^{-15}$ & 1002 & 33  & 0.9$\pm$&0.2 & 0.67 & $ 15\!$&$^{+8}_{-4}$ & $ 2.6$&$\pm0.1$ &  \ref{J1650-1654}\\
J1652+2651 & Dif$^\ast$  & 0.9158 & $ 6.5\times10^{-16}$ & 3875 & 78  & 0.7$\pm$&0.1 & 0.47 & $ -40\!$&$^{+10}_{-80}$ & $ 25$&$\pm2$ &  \ref{J1652+2651}\\
J1654$-$2713 &    & 0.7918 & $ 1.7\times10^{-16}$ & 4480 & 31  & 0.9$\pm$&0.1 & 0.72 & & & & &  \ref{J1654-2713}\\
B1657$-$13 &    & 0.6410 & $ 6.2\times10^{-16}$ & 2743 & 13  &  && 1.91 & & & & &  \ref{B1657-13}\\
B1700$-$18 & Dif  & 0.8043 & $ 1.7\times10^{-15}$ & 1067 & 72  & 1.24$\pm$&0.05 & 0.39 & $ -9\!$&$^{+1}_{-3}$ & $ 3.7$&$\pm0.2$ &  \ref{B1700-18}\\
B1702$-$19 & Coh  & 0.2990 & $ 4.1\times10^{-15}$ & 11885 & 115  & 0.40$\pm$&0.06 & 0.38 & $ -90\!$&$^{+40}_{-50}$ & $ 10.8$&$\pm0.2$ &  \ref{B1702-19}\\
B1706$-$16 &    & 0.6531 & $ 6.3\times10^{-15}$ & 1314 & 228  & 0.77$\pm$&0.01 & 0.13 & & & & &  \ref{B1706-16}\\
B1709$-$15 &    & 0.8688 & $ 1.1\times10^{-15}$ & 2024 & 32  &  && 0.56 & & & & &  \ref{B1709-15}\\
B1717$-$16 &    & 1.5656 & $ 5.8\times10^{-15}$ & 2252 & 94  & 0.65$\pm$&0.03 & 0.34 & & & & &  \ref{B1717-16}\\
B1717$-$29 & Coh  & 0.6204 & $ 7.5\times10^{-16}$ & 1380 & 54  & 1.0$\pm$&0.2 & 0.90 & $ -10.9\!$&$^{+0.4}_{-0.7}$ & $ 2.461$&$\pm0.001$ &  \ref{B1717-29}\\
B1718$-$02 & Dif  & 0.4777 & $ 8.3\times10^{-17}$ & 3665 & 61  & 1.2$\pm$&0.1 & 0.67 & $ 100\!$&$^{+60}_{-10}$ & $ 5.4$&$\pm0.1$ &  \ref{B1718-02}\\
B1726$-$00 &    & 0.3860 & $ 1.1\times10^{-15}$ & 9191 & 68  & 0.54$\pm$&0.08 & 0.51 & & & & &  \ref{B1726-00}\\
B1730$-$22 & Coh  & 0.8717 & $ 4.3\times10^{-17}$ & 3027 & 93  & 0.52$\pm$&0.03 & 0.33 & $ -25\!$&$^{+5}_{-50}$ & $ 25$&$\pm5$ &  \ref{B1730-22}\\
B1732$-$02 &    & 0.8394 & $ 4.2\times10^{-16}$ & 2093 & 14  &&  & 1.68 & & & & &  \ref{B1732-02}\\
B1732$-$07 & Lon  & 0.4193 & $ 1.2\times10^{-15}$ & 2047 & 121  & 0.64$\pm$&0.04 & 0.18 & ?& & $ 13$&$\pm2$ &  \ref{B1732-07}\\
J1732$-$1930 &    & 0.4838 & $ 1.8\times10^{-16}$ & 7345 & 77  & 0.7$\pm$&0.1 & 0.48 & & & & &  \ref{J1732-1930}\\
B1737+13 & Lon  & 0.8031 & $ 1.5\times10^{-15}$ & 2188 & 109  & 0.49$\pm$&0.05 & 0.35 & ?& & $ 10$&$\pm4$ &  \ref{B1737+13}\\
B1738$-$08 & Dif$^\ast$  & 2.0431 & $ 2.3\times10^{-15}$ & 1003 & 63  & 1.13$\pm$&0.07 & 0.41 & $ 80\!$&$^{+80}_{-10}$ & $ 4.8$&$\pm0.1$ &  \ref{B1738-08}\\
 &  &    & & &  & & &  & $ 15\!$&$^{+2}_{-3}$ & $ 4.7$&$\pm0.1$ &  \\
J1744$-$2335 &    & 1.6835 & $ 8.3\times10^{-16}$ & 2101 & 18  &  && 1.43 & & & & &  \ref{J1744-2335}\\
B1745$-$12 &    & 0.3941 & $ 1.2\times10^{-15}$ & 9015 & 111  & 0.44$\pm$&0.07 & 0.42 & & & & &  \ref{B1745-12}\\
B1749$-$28 &    & 0.5626 & $ 8.1\times10^{-15}$ & 5791 & 1087  & 0.501$\pm$&0.003 & 0.08 & & & & &  \ref{B1749-28}\\
B1753+52 & Dif$^\ast$  & 2.3914 & $ 1.6\times10^{-15}$ & 1474 & 36  & 1.4$\pm$&0.1 & 0.55 & $ 14\!$&$^{+8}_{-4}$ & $ 7$&$\pm1$ &  \ref{B1753+52}\\
B1756$-$22 &    & 0.4610 & $ 1.1\times10^{-14}$ & 5766 & 48  & & & 0.52 & & & & &  \ref{B1756-22}\\
J1758+3030 &    & 0.9473 & $ 7.2\times10^{-16}$ & 1855 & 45  & 1.0$\pm$&0.1 & 0.73 & & & & &  \ref{J1758+3030}\\
B1758$-$03 &    & 0.9215 & $ 3.3\times10^{-15}$ & 3855 & 194  & 1.0$\pm$&0.1 & 0.21 & & & & &  \ref{B1758-03}\\
J1759$-$2922 &    & 0.5744 & $ 4.6\times10^{-15}$ & 1494 & 12  & & & 1.53 & & & & &  \ref{J1759-2922}\\
B1804$-$08 &    & 0.1637 & $ 2.9\times10^{-17}$ & 5123 & 78  & 0.57$\pm$&0.03 & 0.34 & & & & &  \ref{B1804-08}\\
J1808$-$0813 &    & 0.8760 & $ 1.2\times10^{-15}$ & 4049 & 45  & 0.71$\pm$&0.07 & 0.61 & & & & &  \ref{J1808-0813}\\
B1811+40 & Lon  & 0.9311 & $ 2.5\times10^{-15}$ & 3811 & 213  & 0.43$\pm$&0.04 & 0.20 & ?& & $ 2.3$&$\pm0.4$ &  \ref{B1811+40}\\
\end{tabular}
\end{center}
\setcounter{table}{1}
\caption{continued.}
\end{table*}
\begin{table*}[!ht]
\begin{center}
\begin{tabular}{lc|rc|rr|r@{}lrr@{$\;$}lr@{}l|l}
\hline
\hline
Pulsar & Class  & $P_0$ (s) & $\dot P$ & Pulses & $S/N$\hspace{2.5mm} &  $m$& & $m_\mathrm{thresh}$ & $P_2$ & (deg)& $P_3$ & $\;(P_0)$ &  Figure\\
\hline
B1818$-$04 & Dif$^\ast$  & 0.5981 & $ 6.3\times10^{-15}$ & 4943 & 513  & 0.43$\pm$&0.02 & 0.07 & $ -300\!$&$^{+50}_{-100}$ & $ 3$&$\pm1$ &  \ref{B1818-04}\\
B1819$-$22 & Coh  & 1.8743 & $ 1.4\times10^{-15}$ & 1095 & 58  & 0.9$\pm$&0.2 & 0.47 & $ -17\!$&$^{+2}_{-6}$ & $ 16.9$&$\pm0.6$ &  \ref{B1819-22}\\
B1821+05 &    & 0.7529 & $ 2.3\times10^{-16}$ & 1137 & 60  & 0.70$\pm$&0.02 & 0.29 & & & & &  \ref{B1821+05}\\
B1821$-$19 &    & 0.1893 & $ 5.2\times10^{-15}$ & 4528 & 39  & & & 0.92 & & & & &  \ref{B1821-19}\\
B1822+00 &    & 0.7789 & $ 8.8\times10^{-16}$ & 4561 & 70  &  && 0.40 & & & & &  \ref{B1822+00}\\
B1822$-$09 &    & 0.7690 & $ 5.2\times10^{-14}$ & 3841 & 90  & 0.30$\pm$&0.01 & 0.12 & & & & &  \ref{B1822-09}\\
J1823$-$0154 &    & 0.7598 & $ 1.1\times10^{-15}$ & 1129 & 18  & & & 0.94 & & & & &  \ref{J1823-0154}\\
B1826$-$17 &    & 0.3071 & $ 5.6\times10^{-15}$ & 2781 & 62  & 0.62$\pm$&0.08 & 0.49 & & & & &  \ref{B1826-17}\\
B1831$-$03 &    & 0.6867 & $ 4.2\times10^{-14}$ & 1246 & 87  & 0.42$\pm$&0.03 & 0.27 & & & & &  \ref{B1831-03}\\
B1831$-$04 &    & 0.2901 & $ 7.2\times10^{-17}$ & 3099 & 17  & 0.6$\pm$&0.1 & 0.62 & & & & &  \ref{B1831-04}\\
J1835$-$1106 &    & 0.1659 & $ 2.1\times10^{-14}$ & 5052 & 22  & & & 1.33 & & & & &  \ref{J1835-1106}\\
J1837$-$0045 &    & 0.6170 & $ 1.7\times10^{-15}$ & 2361 & 14  & & & 1.48 & & & & &  \ref{J1837-0045}\\
B1839+09 &    & 0.3813 & $ 1.1\times10^{-15}$ & 2245 & 79  & 0.61$\pm$&0.06 & 0.37 & & & & &  \ref{B1839+09}\\
B1839+56 & Dif$^\ast$  & 1.6529 & $ 1.5\times10^{-15}$ & 1052 & 100  & 0.89$\pm$&0.02 & 0.24 & $ 120\!$&$^{+130}_{-20}$ & $ 40$&$\pm40$ &  \ref{B1839+56}\\
B1842+14 &    & 0.3755 & $ 1.9\times10^{-15}$ & 2287 & 164  & 0.52$\pm$&0.03 & 0.29 & & & & &  \ref{B1842+14}\\
B1844$-$04 &    & 0.5978 & $ 5.2\times10^{-14}$ & 1431 & 97  & 0.37$\pm$&0.05 & 0.18 & & & & &  \ref{B1844-04}\\
B1845$-$01 &    & 0.6594 & $ 5.3\times10^{-15}$ & 1295 & 29  &  && 1.36 & & & & &  \ref{B1845-01}\\
B1845$-$19 &    & 4.3082 & $ 2.3\times10^{-14}$ & 1029 & 117  & 1.0$\pm$&0.2 & 0.33 & & & & &  \ref{B1845-19}\\
B1846$-$06 & Lon  & 1.4513 & $ 4.6\times10^{-14}$ & 1002 & 78  & 1.11$\pm$&0.08 & 0.25 & ?& & $ 4$&$\pm1$ &  \ref{B1846-06}\\
B1848+12 &    & 1.2053 & $ 1.2\times10^{-14}$ & 2950 & 73  & 1.0$\pm$&0.1 & 0.40 & & & & &  \ref{B1848+12}\\
J1848$-$1414 &    & 0.2978 & $ 1.4\times10^{-17}$ & 4897 & 10 & &  & 3.04 & & & & &  \ref{J1848-1414}\\
B1851$-$14 &    & 1.1466 & $ 4.2\times10^{-15}$ & 3093 & 77  & 0.81$\pm$&0.04 & 0.39 & & & & &  \ref{B1851-14}\\
J1852$-$2610 &    & 0.3363 & $ 8.8\times10^{-17}$ & 5200 & 45  &&  & 0.65 & & & & &  \ref{J1852-2610}\\
B1857$-$26 & Dif  & 0.6122 & $ 2.0\times10^{-16}$ & 3196 & 297  & 0.60$\pm$&0.08 & 0.11 & $ 160\!$&$^{+100}_{-20}$ & $ 7.5$&$\pm0.4$ &  \ref{B1857-26}\\
 &  &    & & &  & & &  & $ 200\!$&$^{+90}_{-30}$ & $ 6.8$&$\pm0.2$ &  \\
B1859+01 & Lon  & 0.2882 & $ 2.4\times10^{-15}$ & 12337 & 96  & 0.65$\pm$&0.03 & 0.35 & ?& & $ 14$&$\pm1$ &  \ref{B1859+01}\\
B1859+03 &    & 0.6555 & $ 7.5\times10^{-15}$ & 5418 & 108  &  && 0.31 & & & & &  \ref{B1859+03}\\
B1900+01 & Dif$^\ast$  & 0.7293 & $ 4.0\times10^{-15}$ & 2395 & 152  & 0.7$\pm$&0.1 & 0.18 & $ 180\!$&$^{+20}_{-30}$ & $ 3.4$&$\pm0.9$ &  \ref{B1900+01}\\
B1900+05 &    & 0.7466 & $ 1.3\times10^{-14}$ & 4751 & 40  &  && 0.90 & & & & &  \ref{B1900+05}\\
B1900$-$06 &    & 0.4319 & $ 3.4\times10^{-15}$ & 1987 & 54  & 0.7$\pm$&0.1 & 0.54 & & & & &  \ref{B1900-06}\\
J1901$-$0906 & Coh  & 1.7819 & $ 1.6\times10^{-15}$ & 984 & 35  & 1.27$\pm$&0.08 & 0.38 & $ -50\!$&$^{+30}_{-50}$ & $ 5.1$&$\pm0.3$ &  \ref{J1901-0906}\\
 &  &    & & &  & & &  & $ -7\!$&$^{+4}_{-2}$ & $ 3.1$&$\pm0.1$ &  \\
B1902$-$01 &    & 0.6432 & $ 3.1\times10^{-15}$ & 2717 & 35  & 0.67$\pm$&0.08 & 0.60 & & & & &  \ref{B1902-01}\\
B1905+39 & Dif$^\ast$  & 1.2358 & $ 5.4\times10^{-16}$ & 1176 & 158  & 0.76$\pm$&0.05 & 0.25 & $ 40\!$&$^{+110}_{-7}$ & $ 4.1$&$\pm0.8$ &  \ref{B1905+39}\\
 &  &    & & &  & & &  & $ 17\!$&$^{+10}_{-4}$ & $ 4.3$&$\pm0.1$ &  \\
B1907+00 &    & 1.0169 & $ 5.5\times10^{-15}$ & 1139 & 90  & 0.60$\pm$&0.02 & 0.26 & & & & &  \ref{B1907+00}\\
B1907+02 &    & 0.9898 & $ 5.5\times10^{-15}$ & 1170 & 109  & 0.41$\pm$&0.03 & 0.20 & & & & &  \ref{B1907+02}\\
B1907+10 & Lon  & 0.2836 & $ 2.6\times10^{-15}$ & 3024 & 219  & 0.53$\pm$&0.02 & 0.21 & ?& & $ 13$&$\pm2$ &  \ref{B1907+10}\\
B1907$-$03 &    & 0.5046 & $ 2.2\times10^{-15}$ & 1701 & 55  & 0.35$\pm$&0.04 & 0.30 & & & & &  \ref{B1907-03}\\
B1910+20 & Lon  & 2.2330 & $ 1.0\times10^{-14}$ & 1095 & 40  & 0.93$\pm$&0.04 & 0.81 & ?& & $ 2.70$&$\pm0.04$ &  \ref{B1910+20}\\
B1911$-$04 &    & 0.8259 & $ 4.1\times10^{-15}$ & 1908 & 955  & 0.236$\pm$&0.005 & 0.07 & & & & &  \ref{B1911-04}\\
B1913+10 &    & 0.4045 & $ 1.5\times10^{-14}$ & 4333 & 16  & & & 1.87 & & & & &  \ref{B1913+10}\\
B1914+09 &    & 0.2703 & $ 2.5\times10^{-15}$ & 3105 & 52  & 0.58$\pm$&0.09 & 0.46 & & & & &  \ref{B1914+09}\\
B1914+13 & Lon  & 0.2818 & $ 3.6\times10^{-15}$ & 12615 & 54  & 1.0$\pm$&0.1 & 0.90 & ?& & $ 21$&$\pm5$ &  \ref{B1914+13}\\
B1915+13 &    & 0.1946 & $ 7.2\times10^{-15}$ & 3739 & 115  & 0.41$\pm$&0.03 & 0.26 & & & & &  \ref{B1915+13}\\
B1917+00 & Dif$^\ast$  & 1.2723 & $ 7.7\times10^{-15}$ & 1144 & 127  & 0.46$\pm$&0.06 & 0.15 & $ 90\!$&$^{+20}_{-20}$ & $ 7$&$\pm2$ &  \ref{B1917+00}\\
B1918+19 & Coh  & 0.8210 & $ 9.0\times10^{-16}$ & 1397 & 111  & 0.65$\pm$&0.07 & 0.28 & $ 20\!$&$^{+15}_{-5}$ & $ 3.7$&$\pm0.1$ &  \ref{B1918+19}\\
B1919+21 & Dif  & 1.3373 & $ 1.3\times10^{-15}$ & 466 & 730  & 0.31$\pm$&0.01 & 0.05 & $ -8.5\!$&$^{+0.5}_{-1.5}$ & $ 4.4$&$\pm0.1$ &  \ref{B1919+21}\\
 &  &    & & &  & & &  & $ -8\!$&$^{+1}_{-2}$ & $ 4.4$&$\pm0.1$ &  \\
B1920+21 &    & 1.0779 & $ 8.2\times10^{-15}$ & 1350 & 177  & 0.47$\pm$&0.07 & 0.15 & & & & &  \ref{B1920+21}\\
B1923+04 & Dif$^\ast$  & 1.0741 & $ 2.5\times10^{-15}$ & 1077 & 57  & 1.01$\pm$&0.07 & 0.43 & $ 80\!$&$^{+30}_{-30}$ & $ 2.9$&$\pm0.5$ &  \ref{B1923+04}\\
B1924+16 &    & 0.5798 & $ 1.8\times10^{-14}$ & 4580 & 24  &&  & 1.26 & & & & &  \ref{B1924+16}\\
B1929+10 & Dif$^\ast$  & 0.2265 & $ 1.2\times10^{-15}$ & 13044 & 1274  & 0.49$\pm$&0.01 & 0.05 & ?& & $ 11.4$&$\pm0.6$ &  \ref{B1929+10}\\
 &  &    & & &  & & &  & $ -350\!$&$^{+60}_{-70}$ & $ 5.0$&$\pm0.5$ &  \\
B1929+20 &    & 0.2682 & $ 4.2\times10^{-15}$ & 6535 & 43  & 0.7$\pm$&0.1 & 0.70 & & & & &  \ref{B1929+20}\\
\end{tabular}
\end{center}
\setcounter{table}{1}
\caption{continued.}
\end{table*}
\begin{table*}[!ht]
\begin{center}
\begin{tabular}{lc|rc|rr|r@{}lrr@{$\;$}lr@{}l|l}
\hline
\hline
Pulsar & Class  & $P_0$ (s) & $\dot P$ & Pulses & $S/N$\hspace{2.5mm} &  $m$& & $m_\mathrm{thresh}$ & $P_2$ & (deg)& $P_3$ & $\;(P_0)$ &  Figure\\
\hline
B1933+16 & Dif$^\ast$  & 0.3587 & $ 6.0\times10^{-15}$ & 4791 & 1041  & 0.323$\pm$&0.002 & 0.05 & $ -150\!$&$^{+30}_{-30}$ & $ 6.6$&$\pm0.9$ &  \ref{B1933+16}\\
B1937$-$26 &    & 0.4029 & $ 9.6\times10^{-16}$ & 2082 & 72  & 1.31$\pm$&0.05 & 0.45 & & & & &  \ref{B1937-26}\\
B1940$-$12 &    & 0.9724 & $ 1.7\times10^{-15}$ & 1190 & 81  & 0.87$\pm$&0.05 & 0.27 & & & & &  \ref{B1940-12}\\
B1943$-$29 &    & 0.9594 & $ 1.5\times10^{-15}$ & 1206 & 47  & 0.81$\pm$&0.06 & 0.35 & & & & &  \ref{B1943-29}\\
B1944+17 & Dif$^\ast$  & 0.4406 & $ 2.4\times10^{-17}$ & 3990 & 143  & 1.55$\pm$&0.08 & 0.28 & $ -30\!$&$^{+5}_{-10}$ & $ 13.5$&$\pm0.9$ &  \ref{B1944+17}\\
B1946+35 & Lon  & 0.7173 & $ 7.1\times10^{-15}$ & 2396 & 414  & 0.42$\pm$&0.02 & 0.10 & ?& & $ 55 $&$\pm 9$ &  \ref{B1946+35}\\
B1946$-$25 &    & 0.9576 & $ 3.3\times10^{-15}$ & 1834 & 38  & 0.91$\pm$&0.05 & 0.48 & & & & &  \ref{B1946-25}\\
B1952+29 & Lon  & 0.4267 & $ 1.7\times10^{-18}$ & 5755 & 127  & 0.87$\pm$&0.04 & 0.26 & ?& & $ 13$&$\pm4$ &  \ref{B1952+29}\\
B1953+50 & Dif$^\ast$  & 0.5189 & $ 1.4\times10^{-15}$ & 1650 & 153  & 1.1$\pm$&0.1 & 0.19 & $ 60\!$&$^{+30}_{-20}$ & $ 15$&$\pm6$ &  \ref{B1953+50}\\
B2000+40 &    & 0.9051 & $ 1.7\times10^{-15}$ & 1268 & 53  & 0.61$\pm$&0.07 & 0.49 & & & & &  \ref{B2000+40}\\
B2002+31 &    & 2.1113 & $ 7.5\times10^{-14}$ & 1114 & 104  & 0.43$\pm$&0.02 & 0.20 & & & & &  \ref{B2002+31}\\
B2003$-$08 &    & 0.5809 & $ 4.6\times10^{-17}$ & 4521 & 44  & 1.18$\pm$&0.09 & 0.87 & & & & &  \ref{B2003-08}\\
J2005$-$0020 &    & 2.2797 & $ 2.6\times10^{-14}$ & 1066 & 18  & & & 1.21 & & & & &  \ref{J2005-0020}\\
B2011+38 &    & 0.2302 & $ 8.9\times10^{-15}$ & 7597 & 33  & & & 1.47 & & & & &  \ref{B2011+38}\\
B2016+28 & Dif$^\ast$  & 0.5580 & $ 1.5\times10^{-16}$ & 25909 & 1461  & 0.910$\pm$&0.004 & 0.11 & $ -8.0\!$&$^{+0.2}_{-1.5}$ & $ 3.7$&$\pm0.2$ &  \ref{B2016+28}\\
 &  &    & & &  & & &  & $ -8.5\!$&$^{+0.2}_{-3}$ & $ 4.3$&$\pm0.2$ &  \\
B2020+28 & Dif$^\ast$  & 0.3434 & $ 1.9\times10^{-15}$ & 6857 & 415  & 0.57$\pm$&0.08 & 0.11 & $ -65\!$&$^{+5}_{-25}$ & $ 2.4$&$\pm0.2$ &  \ref{B2020+28}\\
 &  &    & & &  & & &  & $ 50\!$&$^{+10}_{-5}$ & $ 2.4$&$\pm0.2$ &  \\
B2021+51 & Dif$^\ast$  & 0.5292 & $ 3.1\times10^{-15}$ & 3471 & 331  & 0.97$\pm$&0.07 & 0.12 & $ 150\!$&$^{+150}_{-50}$ & $ 3$&$\pm1$ &  \ref{B2021+51}\\
 &  &    & & &  & & &  & $ 40\!$&$^{-5}_{+15}$ & $ 3.7$&$\pm0.2$ &  \\
B2022+50 &    & 0.3726 & $ 2.5\times10^{-15}$ & 6590 & 35  & 0.74$\pm$&0.07 & 0.60 & & & & &  \ref{B2022+50}\\
B2027+37 &    & 1.2168 & $ 1.2\times10^{-14}$ & 1444 & 54  & 0.75$\pm$&0.06 & 0.45 & & & & &  \ref{B2027+37}\\
B2043$-$04 & Coh  & 1.5469 & $ 1.5\times10^{-15}$ & 1521 & 147  & 1.02$\pm$&0.03 & 0.28 & $ 4.3\!$&$^{+0.5}_{-0.1}$ & $ 2.75$&$\pm0.02$ &  \ref{B2043-04}\\
B2044+15 & Dif$^\ast$  & 1.1383 & $ 1.8\times10^{-16}$ & 1540 & 70  & 0.56$\pm$&0.05 & 0.26 & $ -25\!$&$^{-10}_{+10}$ & $ 18$&$\pm4$ &  \ref{B2044+15}\\
 &  &    & & &  & & &  & $ -16\!$&$^{+3}_{-4}$ & $ 13$&$\pm1$ &  \\
B2045$-$16 & Dif$^\ast$  & 1.9616 & $ 1.1\times10^{-14}$ & 896 & 1969  & 0.154$\pm$&0.001 & 0.02 & $ -26\!$&$^{+4}_{-2}$ & $ 3.0$&$\pm0.1$ &  \ref{B2045-16}\\
 &  &    & & &  & & &  & $ -40\!$&$^{+2}_{-17}$ & $ 2.7$&$\pm0.1$ &  \\
B2053+21 & Dif$^\ast$  & 0.8152 & $ 1.3\times10^{-15}$ & 2497 & 76  & 0.54$\pm$&0.04 & 0.37 & ?& & $ 105$&$\pm20$ &  \ref{B2053+21}\\
 &  &    & & &  & & &  & $ 15\!$&$^{+5}_{-5}$ & $ 130$&$\pm7$ &  \\
B2053+36 &    & 0.2215 & $ 3.7\times10^{-16}$ & 14369 & 132  & 0.38$\pm$&0.03 & 0.25 & & & & &  \ref{B2053+36}\\
B2106+44 &    & 0.4149 & $ 8.6\times10^{-17}$ & 4216 & 102  & 0.70$\pm$&0.05 & 0.38 & & & & &  \ref{B2106+44}\\
B2110+27 & Dif$^\ast$  & 1.2029 & $ 2.6\times10^{-15}$ & 2041 & 273  & 0.95$\pm$&0.03 & 0.13 & $ 50\!$&$^{+10}_{-20}$ & $ 7$&$\pm3$ &  \ref{B2110+27}\\
B2111+46 & Dif$^\ast$  & 1.0147 & $ 7.1\times10^{-16}$ & 1136 & 550  & 1.07$\pm$&0.04 & 0.04 & $ -200\!$&$^{+10}_{-30}$ & $ 3.1$&$\pm0.1$ &  \ref{B2111+46}\\
B2113+14 &    & 0.4402 & $ 2.9\times10^{-16}$ & 5619 & 69  & 0.63$\pm$&0.07 & 0.39 & & & & &  \ref{B2113+14}\\
B2148+52 &    & 0.3322 & $ 1.0\times10^{-14}$ & 5265 & 48  & & & 0.49 & & & & &  \ref{B2148+52}\\
B2148+63 & Dif$^\ast$  & 0.3801 & $ 1.7\times10^{-16}$ & 9336 & 196  & 0.9$\pm$&0.1 & 0.28 & $ -40\!$&$^{+30}_{-90}$ & $ 2.4$&$\pm0.7$ &  \ref{B2148+63}\\
B2154+40 & Dif$^\ast$  & 1.5253 & $ 3.4\times10^{-15}$ & 1140 & 536  & 0.69$\pm$&0.01 & 0.07 & $ 100\!$&$^{+30}_{-10}$ & $ 5$&$\pm1$ &  \ref{B2154+40}\\
B2217+47 & Dif$^\ast$  & 0.5385 & $ 2.8\times10^{-15}$ & 4934 & 1214  & 0.603$\pm$&0.002 & 0.04 & $ -140\!$&$^{+10}_{-100}$ & $ 4.1$&$\pm0.6$ &  \ref{B2217+47}\\
B2224+65 &    & 0.6825 & $ 9.7\times10^{-15}$ & 2553 & 66  & 0.77$\pm$&0.04 & 0.37 & & & & &  \ref{B2224+65}\\
B2227+61 &    & 0.4431 & $ 2.3\times10^{-15}$ & 8007 & 76  & 0.64$\pm$&0.09 & 0.62 & & & & &  \ref{B2227+61}\\
J2248$-$0101 &    & 0.4772 & $ 6.6\times10^{-16}$ & 2423 & 25  & & & 1.35 & & & & &  \ref{J2248-0101}\\
B2255+58 & Dif$^\ast$  & 0.3682 & $ 5.8\times10^{-15}$ & 6705 & 183  & 0.56$\pm$&0.02 & 0.19 & $ 200\!$&$^{+150}_{-60}$ & $ 10.2$&$\pm0.5$ &  \ref{B2255+58}\\
J2302+6028 & Coh  & 1.2064 & $ 2.0\times10^{-15}$ & 2643 & 87  & 0.33$\pm$&0.05 & 0.30 & $ -40\!$&$^{+10}_{-40}$ & $ 5.25$&$\pm0.05$ &  \ref{J2302+6028}\\
B2303+30 & Coh  & 1.5759 & $ 2.9\times10^{-15}$ & 1685 & 200  & 0.54$\pm$&0.07 & 0.22 & $ 10.6\!$&$^{+0.8}_{-0.2}$ & $ 2.06$&$\pm0.02$ &  \ref{B2303+30}\\
B2306+55 &    & 0.4751 & $ 2.0\times10^{-16}$ & 3697 & 113  & 0.71$\pm$&0.09 & 0.36 & & & & &  \ref{B2306+55}\\
B2310+42 & Coh  & 0.3494 & $ 1.1\times10^{-16}$ & 5260 & 1518  & 0.585$\pm$&0.002 & 0.04 & $ 13\!$&$^{+4}_{-6}$ & $ 2.10$&$\pm0.05$ &  \ref{B2310+42}\\
 &  &    & & &  & & &  & $ 13\!$&$^{+1}_{-1}$ & $ 2.11$&$\pm0.03$ &  \\
B2315+21 & Coh  & 1.4447 & $ 1.0\times10^{-15}$ & 1214 & 165  & 0.56$\pm$&0.03 & 0.22 & $ -8\!$&$^{+2}_{-6}$ & $ 5.1$&$\pm0.2$ &  \ref{B2315+21}\\
B2319+60 & Dif$^\ast$  & 2.2565 & $ 7.0\times10^{-15}$ & 1568 & 104  & 0.93$\pm$&0.05 & 0.38 & $ 80\!$&$^{+30}_{-20}$ & $ 5$&$\pm3$ &  \ref{B2319+60}\\
 &  &    & & &  & & &  & $ 20\!$&$^{+6}_{-2}$ & $ 6$&$\pm2$ &  \\
B2324+60 &    & 0.2337 & $ 3.5\times10^{-16}$ & 7486 & 21  &  && 1.56 & & & & &  \ref{B2324+60}\\
B2327$-$20 & Lon  & 1.6436 & $ 4.6\times10^{-15}$ & 1603 & 647  & 0.73$\pm$&0.01 & 0.06 & ?& & $ 20$&$\pm10$ &  \ref{B2327-20}\\
B2334+61 &    & 0.4953 & $ 1.9\times10^{-13}$ & 3546 & 37  & 1.7$\pm$&0.2 & 1.10 & & & & &  \ref{B2334+61}\\
J2346$-$0609 &    & 1.1815 & $ 1.4\times10^{-15}$ & 2972 & 53  & 1.09$\pm$&0.08 & 0.53 & ?& & $ 100$&$\pm30$ &  \ref{J2346-0609}\\
B2351+61 & Dif$^\ast$  & 0.9448 & $ 1.6\times10^{-14}$ & 1859 & 79  & 1.33$\pm$&0.06 & 0.35 & $ 50\!$&$^{+20}_{-15}$ & $ 17$&$\pm3$ &  \ref{B2351+61}\\
\end{tabular}
\end{center}
\setcounter{table}{1}
\caption{continued.}
\end{table*}

\clearpage
\appendix

\onecolumn
\section{Figures}

{\Large Astro-ph version is missing 191 figures due to file size
restrictions. Please download appendices from:\\ {\tt
http://www.astron.nl/$\sim$stappers/wiki/doku.php?id=resources:publications}. }

\label{Figures_ref}

\begin{figure*}[hbt]
\caption{
\label{B0011+47}
\label{B0031-07}
\label{B0037+56}
\label{B0105+65}
\label{B0136+57}
\label{B0138+59}
\label{B0320+39}
\label{B0355+54}
\label{B0410+69}
\label{J0421-0345}
}
\caption{
\label{B0447-12}
\label{B0450+55}
\label{B0458+46}
\label{J0459-0210}
\label{J0520-2553}
\label{B0523+11}
\label{B0540+23}
\label{B0559-05}
\label{B0609+37}
\label{B0611+22}
}
\caption{
\label{B0626+24}
\label{B0643+80}
\label{B0727-18}
\label{B0740-28}
\label{B0756-15}
\label{B0809+74}
\label{B0818-13}
\label{B0820+02}
\label{B0823+26}
\label{B0906-17}
}
\caption{
\label{B0919+06}
\label{B0942-13}
\label{B0950+08}
\label{B1112+50}
\label{B1254-10}
\label{B1322+83}
\label{B1530+27}
\label{B1540-06}
\label{B1541+09}
\label{B1552-23}
}
\caption{
\label{B1600-27}
\label{J1603-2531}
\label{B1604-00}
\label{B1607-13}
\label{B1612+07}
\label{B1620-09}
\label{B1633+24}
\label{B1642-03}
\label{B1648-17}
\label{B1649-23}
}
\caption{
\label{J1650-1654}
\label{J1652+2651}
\label{J1654-2713}
\label{B1657-13}
\label{B1700-18}
\label{B1702-19}
\label{B1706-16}
\label{B1709-15}
\label{B1717-29}
\label{B1717-16}
}
\caption{
\label{B1718-02}
\label{B1726-00}
\label{B1730-22}
\label{J1732-1930}
\label{B1732-07}
\label{B1732-02}
\label{J1744-2335}
\label{B1745-12}
\label{B1749-28}
\label{B1756-22}
}
\caption{
\label{B1758-03}
\label{J1758+3030}
\label{J1759-2922}
\label{B1804-08}
\label{J1808-0813}
\label{B1811+40}
\label{B1818-04}
\label{B1819-22}
\label{B1821-19}
\label{B1821+05}
}
\caption{
\label{B1822-09}
\label{B1822+00}
\label{J1823-0154}
\label{B1826-17}
\label{B1831-04}
\label{B1831-03}
\label{J1835-1106}
\label{J1837-0045}
\label{B1839+09}
\label{B1839+56}
}
\caption{
\label{B1842+14}
\label{B1844-04}
\label{B1845-19}
\label{B1845-01}
\label{B1846-06}
\label{J1848-1414}
\label{B1848+12}
\label{B1851-14}
\label{J1852-2610}
\label{B1859+01}
}
\caption{
\label{B1859+03}
\label{B1900-06}
\label{B1900+01}
\label{B1900+05}
\label{B1902-01}
\label{B1907-03}
\label{B1907+00}
\label{B1907+02}
\label{B1907+10}
\label{B1910+20}
}
\caption{
\label{B1911-04}
\label{B1913+10}
\label{B1914+09}
\label{B1914+13}
\label{B1915+13}
\label{B1917+00}
\label{B1918+19}
\label{B1920+21}
\label{B1923+04}
\label{B1924+16}
}
\caption{
\label{B1929+10}
\label{B1929+20}
\label{B1933+16}
\label{B1937-26}
\label{B1940-12}
\label{B1943-29}
\label{B1944+17}
\label{B1946-25}
\label{B1946+35}
\label{B1952+29}
}
\caption{
\label{B1953+50}
\label{B2000+40}
\label{B2002+31}
\label{B2003-08}
\label{J2005-0020}
\label{B2011+38}
\label{B2022+50}
\label{B2027+37}
\label{B2043-04}
\label{B2053+36}
}
\caption{
\label{B2106+44}
\label{B2110+27}
\label{B2113+14}
\label{B2148+52}
\label{B2148+63}
\label{B2154+40}
\label{B2217+47}
\label{B2224+65}
\label{B2227+61}
\label{J2248-0101}
}
\caption{
\label{B2255+58}
\label{J2302+6028}
\label{B2303+30}
\label{B2306+55}
\label{B2324+60}
\label{B2327-20}
\label{B2334+61}
\label{B2351+61}
}

\caption{
\label{B0148-06}
\label{B0149-16}
\label{B0301+19}
\label{B0329+54}
\label{B0402+61}
}
\caption{
\label{B0450-18}
\label{B0525+21}
\label{B0531+21}
\label{B0628-28}
\label{B0751+32}
}
\caption{
\label{B0834+06}
\label{B1039-19}
\label{B1133+16}
\label{B1237+25}
\label{B1508+55}
}
\caption{
\label{B1737+13}
\label{B1738-08}
\label{B1753+52}
\label{B1857-26}
\label{J1901-0906}
}
\caption{
\label{B1905+39}
\label{B1919+21}
\label{B2016+28}
\label{B2020+28}
\label{B2021+51}
}
\caption{
\label{B2044+15}
\label{B2045-16}
\label{B2053+21}
\label{B2111+46}
\label{B2310+42}
}
\caption{
\label{B2315+21}
\label{B2319+60}
\label{J2346-0609}
}
\end{figure*}

\end{document}